\documentclass[a4paper,11pt]{article}
\pdfoutput=1 

\usepackage{jheppub} 
\usepackage{amsmath,amssymb,amsthm}
\usepackage{graphicx}
\usepackage{subcaption}
\usepackage[percent]{overpic}
\usepackage{color}
\usepackage{multirow}
\usepackage{slashed}
\usepackage{mathrsfs}
\usepackage{tikz}
\usepackage[all,cmtip,knot]{xy}

\newcommand{\R}{\mathbb{R}}
\newcommand{\C}{\mathbb{C}}
\newcommand{\xUDarrow}[1]{%
 {\left\updownarrow\vbox to #1{}\right.\kern-\nulldelimiterspace}
}
\renewcommand{\Re}{\mathrm{Re}}

\makeatletter
\newcommand\xLRarrow[2][]{%
  \ext@arrow 9999{\longleftrightarrowfill@}{#1}{#2}}
\newcommand\longleftrightarrowfill@{%
  \arrowfill@\leftarrow\relbar\rightarrow}
 \newcommand{\xdashleftrightarrow}[2][]{\ext@arrow 3359\leftrightarrowfill@@{#1}{#2}}
 \def\leftrightarrowfill@@{\arrowfill@@\leftarrow\relbar\rightarrow}
\def\arrowfill@@#1#2#3#4{%
  $\m@th\thickmuskip0mu\medmuskip\thickmuskip\thinmuskip\thickmuskip
   \relax#4#1
   \xleaders\hbox{$#4#2$}\hfill
   #3$%
}
\makeatother

\newtheorem{thm}{Theorem}[section]

\newtheorem{dfn}[thm]{Definition}

\def\be{\begin{equation}}
\def\ee{\end{equation}}
\def\IR{{\mathbb{R}}}
\def\IZ{{\mathbb{Z}}}
\def\IP{{\mathbb{P}}}
\def\IC{{\mathbb{C}}}
\def\IU{{\mathbb{U}}}

\def\CB{{\mathcal{B}}}
\def\CM{{\mathcal{M}}}
\def\CL{{\mathcal{L}}}

\def\CN{{\mathcal{N}}}
\def\CP{{\mathcal{P}}}

\def\fm{{\mathfrak{m}}}

\def\CZ{\mathcal{Z}}

\def\tCW{\widetilde{\mathcal{W}}}
\def\Li{\mathrm{Li}}
\def\gauge{\mathrm{g}}
\def\axial{\mathrm{a}}
\def\top{\mathrm{t}}
\def\vort{\vortex{g}}

\def\vort{\mathrm{vortex}}

\def\tot{{\rm{tot}}}

\def\upR{\xygraph{
!{0;/r1.0em/:}
[u(0.5)]
!{\xoverv=<}
}}

\def\upL{\xygraph{
!{0;/r1.0em/:}
[u(0.5)]
!{\xunderv=<}
}}

\def\upS{\xygraph{
!{0;/r1.0em/:}
[u(0.5)]
!{\huncross=>}
}}

\renewcommand{\(}{\left(}
\renewcommand{\)}{\right)}

\title{Multi-cover skeins, quivers, and 3d $\CN=2$ dualities}

\author[a,b]{Tobias Ekholm,}
\author[c,d]{Piotr Kucharski,}
\author[e,f]{and Pietro Longhi}

\affiliation[a]{Department of Mathematics, Uppsala University, Box 480, 751 06 Uppsala, Sweden}
\affiliation[b]{Institut Mittag-Leffler, Aurav 17, 182 60 Djursholm, Sweden}
\affiliation[c]{Walter Burke Institute for Theoretical Physics, California Institute of Technology, \\ Pasadena, CA 91125, USA}
\affiliation[d]{Faculty of Physics, University of Warsaw, ul. Pasteura 5, 02-093 Warsaw, Poland}
\affiliation[e]{Institute for Theoretical Physics, ETH Zurich, CH - 8093, Zurich, Switzerland}
\affiliation[f]{Mathematical Sciences Research Institute, 17 Gauss Way, Berkeley, CA 94720, USA}

\emailAdd{tobias.ekholm@math.uu.se, piotrek@caltech.edu, longhip@phys.ethz.ch}

\abstract{
The relation between open topological strings and representation theory of symmetric quivers is explored beyond the original setting of the knot-quiver correspondence.
Multiple cover generalizations of the skein relation for boundaries of holomorphic disks on a Lagrangian brane are observed to generate dual quiver descriptions of the geometry.
Embedding into M-theory, a large class of dualities of 3d $\CN=2$ theories associated to quivers is obtained. 
The multi-cover skein relation admits a compact formulation in terms of quantum torus algebras associated to the quiver and in this language the relations are similar to wall-crossing identities of  Kontsevich and Soibelman.
}

\begin{document} 

\maketitle

\section{Introduction}

There is an interesting relation between quivers and open topological strings that was first observed in applications to knot theory \cite{Kucharski:2017ogk, Kucharski:2017poe}. In \cite{Ekholm:2018eee} we discussed the underlying geometry and physics, in terms of counts of open holomorphic curves ending on a knot conormal $L_{K}$, and in terms of the 3d $\CN=2$ physics on an M5-brane wrapping $L_{K}\times S^1\times \IR^2$.

In the present paper we explore further aspects. We relate counts of open holomorphic curves, quivers, and certain 3d $\CN=2$ quantum field theories, in a way that takes
simple properties of one theory to highly nontrivial statements in the others. 
This leads to new results both on the mathematical and physical sides, including mechanisms for generating classes of distinct quivers (with different number of nodes) that determine the same physics, multi-cover skein relations, and {a large class of} 3d $\CN=2$ dualities. 
The results are not limited to the original knot theory setting of~\cite{Kucharski:2017ogk, Kucharski:2017poe} but give connections between quivers and open topological strings also in many other situations.

\subsection{Physics and geometry of the knots-quivers correspondence}
In order to introduce the main results of this paper, we first recall our previous work \cite{Ekholm:2018eee}.
The motivation and starting point was the observation in \cite{Kucharski:2017ogk, Kucharski:2017poe} that the generating series of the symmetrically colored HOMFLY-PT polynomials of a knot $K$ can be written as the partition function (motivic generating series) of a symmetric quiver. A~symmetric quiver $Q$ is a finite graph with a set of nodes connected by undirected edges.\footnote{Equivalently one can consider directed edges (arrows) with a condition that the number of arrows from vertex $i$ to $j$, $i\ne j$ is equal to the number of arrows from vertex $j$ to $i$. In this paper we switch between these two pictures: an undirected edge between two distinct vertices corresponds to a pair of arrows in opposite directions, whereas loops remain unchanged.}
In~\cite{Ekholm:2018eee} we found a geometric interpretation of the nodes of $Q$ as basic holomorphic disks ending on $L_K\approx S^1\times \IR^2$, the knot conormal Lagrangian in the resolved conifold associated to a~knot~$K$, see \cite{Ooguri:1999bv}. 
The number of edges between two nodes of $Q$ was identified with a version of the linking number between corresponding disk boundaries, defined via bounding chains as in \cite{Aganagic:2013jpa,Ekholm:2018iso}.

We showed that if one assumes that all holomorphic curves with boundary on $L_{K}$ are multiple covers of the basic holomorphic disks, then -- using the multiple cover formula for generic disks together with the definition of generalized holomorphic curves in \cite{Ekholm:2018iso} -- the wave function of $L_{K}$ counting generalized holomorphic curves agrees with the quiver partition function. 

For the corresponding physical setting, consider M-theory on the resolved conifold times $S^1\times \IR^4$ with an M5-brane on $L_K\times S^1\times\IR^2$. Then each basic holomorphic disk can be wrapped by an~M2-brane ending on the M5. 
The quiver representation theory computes the~spectrum of BPS M2-branes in terms of a finite set of basic BPS states that correspond to the M2-branes that are wrapped on the basic disks.
The geometric setup in M-theory has a~field-theoretic counterpart in the flat spacetime directions.
In \cite{Dimofte:2010tz}, it was observed that the spectrum of BPS M2-branes descends to the spectrum of BPS vortices in a 3d $\CN=2$ theory $T[L_K]$ on the M5 worldvolume in the transverse flat $S^1\times \IR^2$. 
The~quiver description of the vortex spectrum leads to a simple dual Lagrangian description for this theory, denoted by $T[Q_{K}]$.
This picture indicates that the whole spectrum of BPS vortices, or higher-genus holomorphic curves, can be generated completely by a finite set of linked basic genus-zero curves (disks).

\subsection{Multi-cover skein relations and quivers}\label{sec:introskeinquiv}
From the perspective of topological strings it is natural to view holomorphic curves in a~Calabi-Yau 3-fold with boundary on a Lagrangian $L$ as deforming Chern-Simons theory on $L$, see \cite{Witten:1992fb}. In \cite{ES} this perspective was used to give a new mathematical approach to open curve counts: 1-dimensional defects in Chern-Simons theory of $L$ are links in $L$ modulo isotopy and the~framed skein relation (the defining relation of the framed HOMFLY-PT polynomial). The~resulting module of 1-dimensional defects is called the \emph{framed skein module of $L$}. 

The central idea in \cite{ES} is to count holomorphic curves with boundary in $L$ by the elements represented by their boundaries in the framed skein module of $L$ and
a key point in that approach is to separate contributions of zero symplectic area curves from those of positive area curves, i.e., separate instanton contributions from perturbative contributions. More geometrically, in order to count holomorphic curves, one must take into account contributions from constant maps. The approach in \cite{ES} leading to the skein relation is to keep the~constants unperturbed, focus on curves without components of symplectic area zero (called bare), add the contributions to the counts from constants attached to a bare curve `by hand', and show that this separation of bare and constant curves does not change in generic 1-parameter families.

The total contribution of a bare curve comes from the first (non-multiple) part of the~well-known multiple cover formula for holomorphic curves: the~local contribution to the~open string or Gromov-Witten partition function from a curve of Euler characteristic~$\chi$, with generic normal bundle, and representing the homology class $a$ is
\begin{equation}\label{eq:fpformula}
\exp\left(\sum_{d>0}\frac{1}{d}\frac{a^{d}}{(q^{q}-q^{-d})^{\chi}}\right),
\quad q=e^{\frac12 g_{s}}.
\end{equation}    
The first term in the expansion of this formula says that at degree one the contribution is simply $(q-q^{-1})^{-\chi}$ and counting bare curves with this contribution one finds that the~count is indeed invariant in the framed skein. In other words the framed skein relation is a~`bifurcation' identity for bare curves: 
\be\label{eq:skein-relations-brane}
\upR =  \upL + (q-q^{-1}) \upS \,.
\ee

\begin{figure}[h!]
\begin{center}
\includegraphics[width=0.55\textwidth]{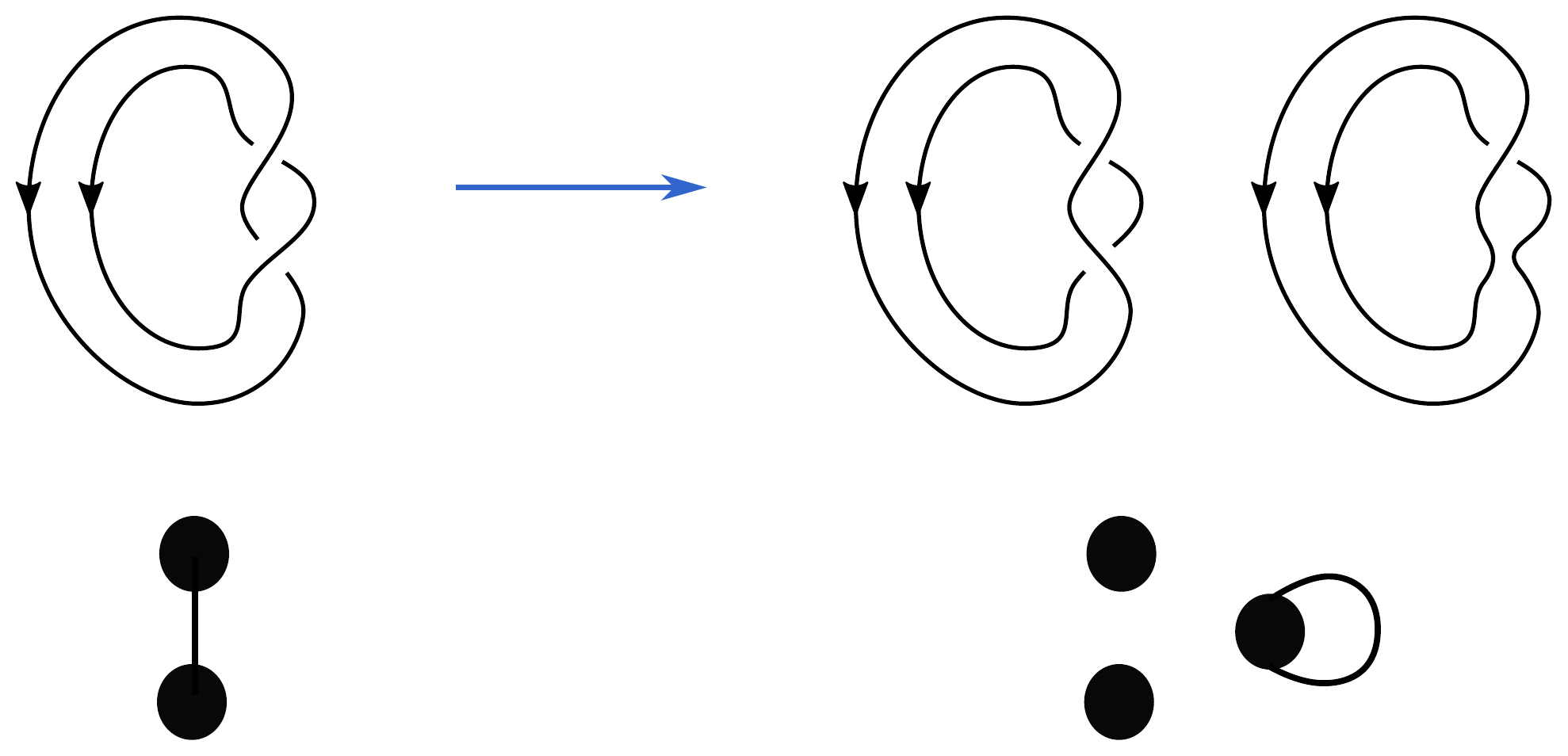}
\caption{Multi-cover skein relations on linking disks and dual quiver description.}
\label{fig:skein-relation-2-3-intro}
\end{center}
\end{figure}

From the holomorphic curve perspective this paper studies the same bifurcation, taking into account all multiple covers with constant curves attached. Our first result is that when the boundaries of two basic disks cross, they can be glued into a new disk and, taking the~multiple covers of this new disk into account, the partition function counting generalized holomorphic curves remains unchanged. 
This means that one can then use this bifurcation to trade two linked basic holomorphic disks for two unlinked basic disks plus a~new basic disk obtained from gluing them, see the upper part of Figure \ref{fig:skein-relation-2-3-intro}. 
Before unlinking, the glued disk is part of the boundstate spectrum. After unlinking the boundstate spectrum is trivial (because the disks do not link anymore) and the new disk should be a part of the new basic set. We call the invariance of generalized holomorphic curve counts under bifurcations of basic disks the~\emph{multi-cover skein relation}.

Interpreted in terms of quivers, 
the multi-cover skein relation changes an edge into an extra node (with a loop), as shown in the lower part of Figure \ref{fig:skein-relation-2-3-intro}. The invariance of the count of generalized curves then implies that the corresponding quiver partition function should also remain unchanged. 
We verify that this is indeed the case and observe that it extends to a large class of dualities on quivers generated by multi-cover skein relations in the dual geometric setting.  
We classify `quiver multi-cover skein moves' and prove that they leave the the partition function unchanged.


The multi-cover skein moves -- which are rather natural in the context of holomorphic curves -- give relations among quivers with different numbers of nodes which are nontrivial from the viewpoint of quiver representation theory.

Finally, from the holomorphic curve perspective it is natural to ask whether there are corresponding multi-cover skein formulas also for higher genus curves. In general the~answer is no, but the formulas for disks can be used to derive more involved formulas for higher genus curves. As an illustration, we discuss bifurcations for the annulus in Appendix~\ref{app:disk-annulus}.

\subsection{3d $\CN=2$ dualities}
A quiver $Q$ encodes the data of a~3d $\CN=2$ theory $T[Q]$, see \cite{Ekholm:2018eee}.
Each node corresponds to a~$U(1)$ gauge group with a charged chiral multiplet attached to it and arrows encode effective mixed Chern-Simons couplings. Then the partition function of quiver representation theory equals  {the partition function of $T[Q]$ on $\IR^2\times_q S^1$}.

Applying the multi-cover skein relation to the quiver $Q$ transforms it to a new quiver~$Q'$ with a different number of nodes and arrows. Consequently, the corresponding gauge theories $T[Q]$ and $T[Q']$ have different gauge groups, matter content, and couplings. Nevertheless, the {$\IR^2\times_q S^1$} partition functions of $T[Q]$ and $T[Q']$ must coincide since the dual quiver partition functions do (with suitable identifications of couplings). This {appears to} give a new class of dualities among 3d $\CN=2$ theories, generated by the quiver version of the multi-cover skein relation. A basic instance of this type of dualities is closely related to the well-known SQED-XYZ mirror duality \cite{Aharony:1997bx}.

\subsection{Quantum torus algebra and wall-crossing}
The geometric idea underlying the relation between quivers and open topological strings, that the whole BPS spectrum of open holomorphic curves is generated by a finite set of basic disks, is not evident from the standard form of quiver partition functions. Here the~BPS spectrum is encoded by motivic DT invariants, the exponents in the factorization of the quiver partition function
\be
	P^Q(\mathbf{x},q) = \prod_{\mathbf{d},j} 
	\Psi_q(
	q^j\mathbf{x}^{\mathbf{d}}
	)^{(-1)^{j}\Omega_{\mathbf{d},j}} \,,
\ee
where $\Psi_q$ is the quantum dilogarithm (see below for a definition), components of $\mathbf{x} = (x_1,\dots, x_m)$ are variables associated to quiver nodes that keep track of the charges of BPS states, and $\mathbf{d}=(d_1\dots d_m)$ is the dimension vector.

We introduce a new formalism, allowing us to write down the partition function as a~\emph{finite} product of basic contributions
\be\label{eq:quantum-P}
	\IP^{Q} = 
	\Psi_q( X_m )\cdot \Psi_q( X_{m-1} )\ \cdot\ \ldots\ \cdot\  \Psi_q( X_{1} )\,.
\ee
Now $X_i$ are \emph{non-commutative} variables, valued in a quantum torus algebra $X_i X_j = q^{\pm 2\mathrm{lk}(i,j)} X_j X_i$ obtained by a certain anti-symmetrization of the quiver linking matrix. The~new partition function $\IP^Q$ is therefore valued in the quantum torus algebra, and it reduces to $P^Q$ by an operation of \emph{normal ordering} which we define.

This new presentation of the quiver partition function has several nice features.
On the one hand, it makes manifest the fact that the whole spectrum is generated by multi-covers (quantum dilogarithms) of a finite set of basic disks (quiver nodes) through their interactions, encoded by the quiver linking matrix interpreted in the quantum torus algebra of the $X_i$. Here the quiver with $m$ nodes and no edges corresponds to commutative variables $X_{j}=x_{j}$ and the partition function is an actual product. To get non-commutative variables we introduce duals $y_{j}$ of $x_{j}$ with $y_{j}x_{j}=q^{\delta_{ij}}x_{j}y_{j}$ and define $X_{i}=x_{i}\prod_{j=1}^{m}y_{j}^{\mathrm{lk}(i,j)}$.  Normal ordering then corresponds to moving all $y_{j}$-variables to the right.
Thus, starting from the~product partition function and applying normal ordering, we see that the quantum torus algebra keeps track of how linking between basic disks propagates to counts of generalized holomorphic curves involving all their multi-covers and boundstates.

This formalism relates to work on wall-crossing by Kontsevich and Soibelman \cite{Kontsevich:2008fj}. We show in particular that dualities induced by quiver multi-cover skein relations reduce in simple cases to wall-crossing identities.
For example, the skein relation of Figure \ref{fig:skein-relation-2-3-intro} translates into the following equality of quantum partition functions
\be
	\Psi_q(X_2) \Psi_q(X_1) = \Psi_q(X_1) \Psi_q(-q \, X_1 X_2) \Psi_q(X_2) 
\ee
which is an instance of the pentagon identity.
More generally, quiver skein relations predicts many more involved identities for products of dilogarithms with arguments valued in quantum torus algebras. Although we collectively refer to these as `wall-crossing identities' (by analogy with the work of Kontsevich-Soibelman), we point out that they have a~somewhat different structure in general.

\subsection{Gauge theory on branes and quantum Lagrangian correspondences}

Geometrically \eqref{eq:quantum-P} can be understood as deformations of $U(1)$ Chern-Simons theory on a Lagrangian $L\approx S^{1}\times\R^{2}$. It is clear from the path integral that the quantization of $U(1)$ Chern-Simons theory gives a quantum torus, $x=e^{\xi}$, $y=e^{\eta}$, where $\eta=g_{s}\frac{\partial}{\partial\xi}$ and the equation for the wave function: $(1-y)\Psi(x)=0$, which means $\Psi=1$. 
Consider now instead $L$ with one holomorphic disk attached. This disk deforms the Chern-Simons theory and leads to the equation for quantum variables on a small torus surrounding it: $(1-x_{i}-y_{i})\Psi(x_{i})=0$. The global wave function is then obtained by $x=x_{i}$. Consider next attaching several disks which are mutually linked. Then the above implies that the wave function is a product of quantum dilogarithms of $X_{i}=e^{\xi_{i}}e^{\sum_{j}\mathrm{lk}(i,j)\eta_{j}}$, where the variables correspond to unlinked disks, and after normal ordering we get a function of local longitudes~$x_{i}$ that should be substituted by corresponding powers of the global longitude~$x$. 

In the terminology of \cite{Aganagic:2013jpa} this means that we build a D-model associated to the local tori surrounding the boundaries of the basic disks. The D-model is then an open topological string in $(\C^{\ast}\times\C^{\ast})^{m}$ with one factor and quantum torus coordinates $(x_{i},y_{i})=(e^{\xi_i},e^{\eta_i})$ for each $i=1,\dots,m$. In $(\C^{\ast}\times\C^{\ast})^{m}$ there is the space filling coistropic brane and a~Lagrangian brane which is simply a~product Lagrangian in the coordinates $(x_{i},y_{i})$ and a~product wave function $\Psi(x_{1},\dots,x_{m})=\prod_{i=1}^{m}\Psi_{q}(x_{i})$. We obtain the wave function and quantum curve for the composite system on all of $L_{K}$ by pushing the product Lagrangian through the~Lagrangian correspondence in $(\C^{\ast}\times\C^{\ast})^{m}\times(\C^{\ast}\times\C^{\ast})$, where the last factor with coordinates $(x,y)=(e^{\xi},e^{\eta})$ corresponds to the torus which is the ideal boundary of~$L_{K}$, determined by the linking of the disks via 
\be 
\xi=\xi_{1}-\sum_j\mathrm{lk}(1,j)\eta_{j}=\dots=\xi_{m}-\sum_j\mathrm{lk}(m,j)\eta_{j},\quad
\eta=\sum_j\eta_{j}.
\ee
At the full quantum level,
this corresponds to \eqref{eq:quantum-P}, which in the semi-classical limit (counting only disks) is closely related to the reasoning in the Atiyah-Floer conjecture, and here leads to a symplectic reduction formula for the disk potential of $L_{K}$.

\subsubsection*{Organization of the paper}
In Section \ref{sec:background} we  collect background on the relation between quivers and counts of holomorphic curves in the knot theory setting. Section \ref{sec:general-oGW-Q} describes how to generalize this correspondence to counts of holomorphic curves in more general toric Calabi-Yau threefolds with Lagrangian insertions.
In Section \ref{sec:skein-rel} we introduce quiver multi-cover skein relations, describing their form and proving invariance of partition functions.
Physical implications are studied in Section \ref{sec:skein-dualities}, where the relations are reformulated as dualities of 3d $\CN=2$ theories.
In Section \ref{sec:wall-crossing} we study connections with wall-crossing, show how the quantum torus algebra can be used to organize holomorphic curve counts, and present the multi-cover skein relation in this language.

\section*{Acknowledgements}
We would like to thank Tudor Dimofte, Andrew Neitzke, Sara Pasquetti, Du Pei, Marko Sto\v{s}i\'{c}, and Paul Wedrich for insightful discussions.
The work of T.E. is supported by 
the Knut and Alice Wallenberg Foundation and the Swedish Reserach Council.
The work of P.K. is supported by the Polish Ministry of Science and Higher
Education through its programme Mobility Plus (decision no. 1667/MOB/V/2017/0).
The work of P.L. is supported by a grant from the Swiss National Science foundation. 
He also acknowledges the support of the NCCR SwissMAP that is also funded by the Swiss National
Science foundation.
The work of P.L. is also supported by the National Science Foundation under Grant No. DMS-1440140 while the author is in residence at the Mathematical Sciences Research Institute in Berkeley, California, during the Fall 2019 semester.

\section{Background}\label{sec:background}
In this section we recall relevant aspects of the knots-quivers correspondence and of counts of open curves.

\subsection{Knots-quivers correspondence}\label{sub:Knots-quivers-correspondence}
If $K\subset S^{3}$ is a knot then its HOMFLY-PT polynomial $P^K(a,q)$ \cite{freyd1985,PT} is a 2-variable polynomial that is easily calculated from a knot diagram (a projection of $K$ with over/under information at crossings) via the skein relation. The polynomial is a knot invariant, 
i.e., invariant under isotopies and in particular independent of diagrammatic presentation. More generally, the colored HOMFLY-PT polynomials $P^K_{R}(a,q)$ are similar polynomial knot invariants depending also on a representation $R$ of the Lie algebra ${\mathfrak{u}}(N)$. Also the colored version admits a diagrammatic description: it is given by a linear combination of the~standard polynomial of certain satellite links of $K$. (In this setting, the original HOMFLY-PT corresponds to the standard representation.) In order to simplify the notation, we will write the HOMFLY-PT polynomial also when we refer to the more general colored version.

From the physical point of view, the HOMFLY-PT polynomial is the expectation value of the knot viewed as a Wilson line in $U(N)$ Chern-Simons gauge theory on $S^{3}$ \cite{witten1989} which then depends on a choice of representation $R$ for the Lie algebra ${\mathfrak{u}}(N)$. Here we will restrict attention to symmetric representations $R=S^{r}$ corresponding to Young diagrams with a~single row of $r$ boxes. For each $r$-box representation we get a polynomial $P_{r}^{K}(a,q)$ and we consider the \emph{HOMFLY-PT generating series} in the variable $x$:
\be\label{eq:HOMFLY-PT series}
	P^{K}(x,a,q)=\sum_{r=0}^{\infty}P_{r}^{K}(a,q)x^{r}\,.
\ee

In this setting, the Labastida-Mari\~{n}o-Ooguri-Vafa (LMOV) invariants \cite{Ooguri:1999bv,LM0004,LMV0010} are certain numbers 
assembled into the LMOV generating function:
\be 
N^{K}(x,a,q)=\sum_{r,i,j}N_{r,i,j}^{K}x^{r}a^{i}q^{j}
\ee
that gives the following expression for the HOMFLY-PT generating series
\begin{equation}\label{P^K=Exp}
P^{K}(x,a,q)=\mathrm{Exp}\left(\frac{N^{K}(x,a,q)}{1-q^{2}}\right)\,.
\end{equation}
$\textrm{Exp}$ is the plethystic exponential, if $f=\sum_{n}a_{n}t^{n}$, $a_{0}=0$ then
\be
\mathrm{Exp}\bigl(f\bigr)(t)=  
\exp\left(\sum_{k}\tfrac{1}{k}f(t^{k})\right) =\prod_{n}(1-t^{n})^{a_{n}}.
\ee
According to the LMOV conjecture \cite{Ooguri:1999bv,LM0004,LMV0010},  $N_{r,i,j}^{K}$ are integer numbers.

The knots-quivers (KQ) correspondence introduced in \cite{Kucharski:2017ogk,Kucharski:2017poe} and mentioned in the previous section provides a new approach to HOMFLY-PT polynomials and LMOV invariants as follows. 

A \emph{quiver} $Q$ is an oriented graph with a finite number of vertices connected by finitely many arrows (oriented edges). We denote the set of vertices by $Q_0$ and the set of arrows by $Q_1$. A \emph{dimension vector} for $Q$ is a vector in the integral lattice with basis $Q_{0}$, $\mathbf{d}\in \IZ Q_0$. We number the vertices of $Q$ by $1,2,\dots,m=|Q_{0}|$. A \emph{quiver representation with dimension vector $\mathbf{d}=(d_{1},\dots,d_{m})$} is the assignment of a vector space of dimension $d_i$ to the node $i\in Q_0$ and of a linear map $\gamma_{ij}\colon\IC^{d_i} \to \IC^{d_j}$ to each arrow from vertex $i$ to vertex $j$. The~\emph{adjacency matrix} of $Q$ is the $m\times m$ integer matrix with entries $C_{ij}$ equal to the~number of arrows from $i$ to $j$. A quiver is symmetric if its adjacency matrix is. 

Quiver representation theory studies moduli spaces of stable quiver representations (see e.g. \cite{kirillov2016quiver} for an introduction to this subject).
While explicit expressions for invariants describing those spaces are hard to find in general, they are quite well understood in
the~case of symmetric quivers \cite{Kontsevich:2008fj,KS1006,2011arXiv1103.2736E,MR1411,FR1512}. Important information about the moduli space
of representations of a symmetric quiver with trivial potential is encoded in the \emph{motivic generating series} defined as 
\be\label{eq:Efimov}
	P^{Q}(\mathbf{x},q)=\sum_{d_{1},\ldots,d_{m}\geq0}(-q)^{\sum_{1\leq i,j\leq m}C_{ij}d_{i}d_{j}}\prod_{i=1}^{m}\frac{x_{i}^{d_{i}}}{(q^{2};q^{2})_{d_{i}}} 
\ee
where the denominator is the so-called $q$-Pochhammer symbol
\be
	(z;q^2)_r  = \prod_{s=0}^{r-1} (1-z q^{2s}) \,.
\ee
Sometimes we will call $P^{Q}(\mathbf{x},q)$ the quiver partition function. We also point out that the~quiver representation theory involves the choice of an element, the potential, in the~path algebra of the quiver and that the trivial potential is the zero element.

Furthermore, for the quiver $Q$ there exist so called motivic Donaldson-Thomas (DT) invariants $\Omega_{\mathbf{d},s}^{Q}=\Omega_{(d_{1},...,d_{m}),s}^{Q}$.
They can be assembled into the DT generating function
\be\label{eq:DT-gen-function}
\Omega^{Q}(\mathbf{x},q)=\sum_{\mathbf{d},s}\Omega_{\mathbf{d},s}^{Q}\mathbf{x^d}q^{s}(-1)^{|\mathbf{d}|+s+1},\qquad \qquad \mathbf{x^d}=\prod_{i}x_{i}^{d_{i}}\,,
\ee
which is related to the motivic generating series in the following way
\begin{equation}\label{P^Q=Exp}
P^{Q}(\mathbf{x},q)=\textrm{Exp}\left(\frac{\Omega^{Q}(\mathbf{x},q)}{1-q^{2}}\right).
\end{equation}

The DT invariants have two geometric interpretations,
either as the intersection homology Betti numbers of the moduli space of all
semi-simple representations of $Q$ of dimension vector $\mathbf{d}$,
or as the Chow-Betti numbers of the moduli space of all simple representations
of $Q$ of dimension vector $\mathbf{d}$, see \cite{MR1411,FR1512}.
In \cite{2011arXiv1103.2736E} there is a proof that these invariants are positive integers.

The most basic version of the conjectured knot-quiver correspondence is the statement that for each knot $K$ there is a quiver $Q_{K}$ and integers $\{a_{i},q_{i}\}_{i\in {Q_{K}}_{0}}$, such that
\be\label{eq:KQ-corr-basic}
	\left.P^{Q_{K}}(\mathbf{x},q)\right|_{x_{i}=x a^{a_{i}}q^{q_{i}-C_{ii}}}=P^{K}(x,a,q) \,.
\ee
We call $x_{i}=x a^{a_{i}}q^{q_{i}-C_{ii}}$ the \emph{KQ change of variables}. In \cite{Kucharski:2017ogk,Kucharski:2017poe} there are also refined versions of the KQ correspondence. The correspondence on the level of LMOV and DT
invariants is obtained by substituting \eqref{P^K=Exp} and \eqref{P^Q=Exp} into \eqref{eq:KQ-corr-basic} 
\begin{equation}\label{eq:KQDT}
\left.\Omega^{Q_{K}}(\mathbf{x},q)\right|_{x_{i}=x a^{a_{i}}q^{q_{i}-C_{ii}}}=N^{K}(x,a,q)\,.
\end{equation}
Since DT invariants are integer, this equation implies the LMOV conjecture. 

We stress that the KQ correspondence is conjectural, and that it is currently not known how to construct the quiver $Q_{K}$ from a given knot $K$. Evidence for the conjecture includes checks on infinite families of torus and twist knots. A proof for 2-bridge knots appeared recently in \cite{Stosic:2017wno}, whereas \cite{PSS1802} explores the relation to combinatorics of counting paths. On the other hand \cite{Panfil:2018faz} proposes a relation between quivers and topological strings on various Calabi-Yau manifolds and \cite{Zhu:2019wew} contains many explicit formulas obtained in the context of LMOV invariants.

\subsection{Physics -- 3d $\mathcal{N}=2$ theories}\label{sub:Physics-of-KQ}

The physical intepretation of the KQ correspondence is a duality between two 3d $\mathcal{N}=2$ theories: one determined by the knot and the other by the quiver \cite{Ekholm:2018eee}.

The theory associated to the knot $K$ arises from the M-theory on the resolved conifold~$X$ with a single M5-brane wrapping the conormal Lagrangian of the knot $L_K$: 
\be\label{eq:M-theory-setup}
\begin{split}
	\text{space-time}: \quad& \IR^4 \times S^1 \times X \\
			& \cup \phantom{ \ \times S^1 \times \ \ } \cup\\
	\text{M5}: \quad & \IR^2\times S^1 \times L_K.
\end{split}
\ee

The compactification on $X$ leads to 3d $\mathcal{N}=2$ effective theory on $\IR^2\times S^1$, which we denote $T[L_{K}]$. The twisted superpotential of $T[L_K]$ is encoded by the combined large-color and $g_{s}\to 0$ limit of the HOMFLY-PT generating series. 
 The structure of the~theory $T[Q_K]$ can be read off from the analogous limit of the motivic generating series.
The exact form of the duality is given by the change of variables required by the~KQ~correspondence. It amounts to identifying the Fayet-Ilioupoulos couplings of $T[Q_K]$ with specific combinations of the physical fugacities in $T[L_{K}]$.
After this identification $T[Q_K]$ has the same moduli space of supersymmetric vacua as $T[L_K]$, by construction. Among the many dual descriptions of $T[L_K]$, the existence of a quiver $Q_K$ provides a~specific choice. 
The structure of 3d $\CN=2$ theories associated to quivers will be revisited in detail in Section \ref{sec:skein-dualities}.

We consider the duality between $T[L_K]$ and $T[Q_K]$ also from the perspective of the spectra of BPS vortices: BPS states of $T[L_K]$ are counted by LMOV invariants, BPS states of $T[Q_K]$ are counted by (quiver) DT invariants, and \eqref{eq:KQDT} is a manifestation of the duality between the two theories.

\subsection{Geometry -- holomorphic disks}\label{sub:Geometry-of-KQ}
In the previous secion we saw that $T[L_{K}]$ arises from M-theory as the effective theory on the surface of the M5-brane, and that its BPS particles originate from M2-branes ending on the M5. From the symplectic geometric point of view BPS states correspond to generalized holomorphic curves with boundary on the Lagrangian submanifold $L_{K}$. 

We recall the definition of generalized holomorphic curves in the resolved conifold $X$ with boundary on a knot conormal $L_{K}\subset X$ (as defined in \cite{Ekholm:2018iso,Ekholm:2018eee}) from the skeins on branes approach to open curve counts in \cite{ES}. The key observation in \cite{ES} is that the count of bare curves (i.e., curves without constant components) counted by the values of their boundaries in the skein module remains invariant under deformations. The count of such curves also requires the choice of a 4-chain $C_{K}$. Intersections of the interior of a holomorphic curve and the 4-chain contribute to the framing variable $a$ in the skein module. For generalized curves there is a single brane on $L_{K}$ and then $a=q$. When $a=q$ then the map from the~skein module to `homology class and linking' is well-defined and thus counting curves this way, less refined than the $U(1)$-skein, also remains invariant. In $L_{K}\approx S^{1}\times\R^{2}$ one can define such a map that depends on the choice of a framing of the torus at infinity. More precisely, one fixes bounding chains for the holomorphic curve boundaries that agree with multiples of the longitude at infinity and replace linking with intersections between curve boundaries and bounding chains. In \cite{Ekholm:2018iso} an explicit construction of such bounding chains and compatible 4-chain $C_{K}$ from a certain Morse function of $L_{K}$ was described.   

Consider now holomorphic disks with boundary in a multiple of the basic homology class. Such disks are generically embedded and for suitable representatives of the knot conormal can never be further decomposed under deformations. Assuming, in line with~\cite{Gopakumar:1998ii, Gopakumar:1998jq}, that all actual holomorphic curves with boundary on $L_{K}$ lie in neighborhoods of such holomorphic disks attached to the conormal, it would then follow that all generalized holomorphic curves are combinations of branched covers of the basic disks. Using the~multiple cover formula \eqref{eq:fpformula} the count of generalized curves then agrees with the quiver partition function with nodes at the basic disks and with arrows according to linking and additional contributions to the vertices given by $4$-chain intersections.

From this point of view, the theory $T[Q_{K}]$ can be thought of as changing the perspective and treating the basic holomorphic disks with a small tubular neighborhood at their boundaries as independent objects glued into (or attached to) the Lagrangian.

\section{Quiver description of open Gromov-Witten invariants}\label{sec:general-oGW-Q}

The geometric interpretation of the quiver nodes and edges in \cite{Ekholm:2018eee}, see also \cite{Panfil:2018faz}, indicates that the knots-quivers correspondence is a special instance of a more general phenomenon. 
There appears to be a quiver description not only of knot invariants, related to basic holomorphic disks on knot conormals in the resolved conifold, but more generally of BPS states in the open topological string for a larger class of Lagrangian branes in toric Calabi-Yau 3-folds, where both the physical and geometric underlying principles apply. In this section we expand on this viewpoint and 
discuss general features of the quiver description of BPS states of open topological strings.

We consider a Lagrangian brane $L$ with topology $S^{1}\times\IR^2$ inside a toric Calabi-Yau threefold $X$ and the partition function $\CZ^{\rm top}(X,L)$ of open topological strings in $X$ with boundaries on $L$ or in other words the generating function counting generalized holomorphic curves with boundary on $L$.

We observe that in many cases this partition function can be recast in the form of the partition function $\CZ^{\rm quiv}(Q)$ of a symmetric quiver $Q$ (such as (\ref{eq:Efimov})). The knot-quiver correspondence is the special case when $L$ is a knot conormal and $X$ the resolved conifold. In the case when $L$ is a toric brane and $X$ is a `strip geometry', this follows from results in \cite{Panfil:2018faz}. Here we propose that this picture is valid more generally.

Besides the identification of partition functions, the relation between $\CZ^{\rm top}(X,L)$ and $\CZ^{\rm quiv}(Q)$ suggests the existence of a configuration for $L\subset X$, where the whole spectrum of holomorphic curves counted comes from combinations of multiple covers of a finite set of basic holomorphic disks. Here each quiver node corresponds to a basic holomorphic disk in $X$ with boundary $\gamma$ along $L$, wrapping a certain number of times around $S^1$ and a~certain number of times around closed 2-cycles.
The disk boundaries have mutual linking numbers which can be viewed as intersections of the basic disk boundaries with bounding chains constructed from a  Morse flow on $L$. Using a 4-chain $C$ with $\partial C=2L$, as explained in~\cite{Ekholm:2018eee}, one defines also self-linking. These linking and self-linking numbers correspond to quiver arrows.
Any generalized holomorphic curve would then be a map from a worldsheet Riemann surface $\Sigma$ to a union of the basic disks. 
Linking of the basic disks gives linking on the boundary of such a map, which can then give rise to many formally connected  generalized curves. 

Such a decomposition of generalized holomorphic curves into basic disks induces a grading of the former, which corresponds precisely to the quiver dimension vector $(d_1,\dots ,d_m)$.
The relevant geometric data of all curves in the spectrum includes the homology classes of their boundaries (refined in this way by the dimension vector) and the relative homology classes of these curves in $(X,L)$, as well as the self-linking and intersections with the~4-chain. 
We observe that in many cases the following open-string/quiver relation holds:

\emph{
The spectrum of generalized holomorphic curves (holomorphic worldsheet instantons), with the above defined quantum numbers, is entirely encoded by a finite set of basic holomorphic disks as follows.
The disks correspond to the quiver nodes. The arrows of the~quiver and the values of the quiver variables of the disks are determined by their self-linking, mutual linking, 4-chain intersections and relative homology in $(X,L)$.
The quiver representation theory completely determines the full spectrum.}

It is an interesting problem to find conditions ensuring that the open-string/quiver relation holds. From the behavior of knot and link conormals one might speculate that it holds when the Lagrangian can be continuously deformed to a controlled cover of a special Lagrangian $S^{1}\times\R^{2}$.   
{For a quiver with many nodes we expect this to resemble roughly a multiple branched covering of the $\mathbb{C}^3$ toric brane, possibly with different framings on distinct sheets, and with basic disks arising as combinations of the basic disks on the underlying $S^{1}\times \R^{2}$-brane. The motivation behind this picture will become clear in Section \ref{sec:skein-higher}. Existence of such geometric configurations in the moduli space of Lagrangians is an open problem.}

When the open-string/quiver relation holds, the mirror curve of the system $(X,L)$ admits a `decomposition' into the quiver A-polynomials introduced in \cite{Ekholm:2018eee}.
At the quantum level, this translates into the statement that $\CZ^{\rm top}(x)$ admits a refinement to $P^Q(\mathbf{x})$ which is annihilated by the quantum version of the quiver A-polynomials, which we introduce below.

A less obvious consequence that follows from our previous work \cite{Ekholm:2018eee} is that the 3d $\CN=2$ low energy effective theory on an M5-brane wrapping $L$ is a theory of type $T[Q]$. These are abelian Chern-Simons matter theories with a very special structure. In particular, their BPS vortex spectrum coincides with the spectrum of open topological strings in the sense that the {$\IR^2\times_q S^1$} partition function of $T[Q]$ agrees exactly with $P^Q$, the motivic generating series of the quiver $Q$. 

Besides these direct consequences, there are others that give rise to new dualities. The rest of this paper is devoted to exploring these in more detail.

\section{Multi-cover skein relations and birth/death for quivers}\label{sec:skein-rel}
The open-string/quivers relation, where quiver nodes are identified with basic holomorphic disks and arrows encode linking (see Section \ref{sec:general-oGW-Q}), suggests a skein property for quivers.
More precisely, deforming $L$ may cause two basic disks to intersect and linking numbers to change. However the topological string partition function, as well as the disk potential, remain invariant. This follows from the invariance of curve counts in the $U(1)$-skein and projection to generalized holomorphic curves, as explained in Section \ref{sec:introskeinquiv}.
At instances where disk boundaries cross, the boundstates of the two disks also change, since their linking does. As we shall see below, previous bound states turn into contributions from a~new basic disk which is obtained by gluing the two crossing disks. This then means that there should be a~new quiver, with one extra node and with DT spectrum the same as the~previous one after a~suitable specialization of the quiver variables. We will study this in a~simple example in Section \ref{sec:simpleunlink} and prove the general relation in Section \ref{sec:Proof of invariance-unlink}.

Similarly, deformations of the Lagrangian $L$ may lead to birth/death bifurcations in the~moduli space of basic disks. Near this point there are two new basic disks of opposite sign. The partition function of covers of a negative disk is the inverse of the~partition function of the corresponding disk. It turns out that the partition function for a disk with self-linking of positive sign and 4-chain intersection of opposite sign equals that of a negative sign disk. This then leads to a stabilization operation on quivers where two canceling nodes are added. We study this in a simple example and the general case in Section \ref{sec:redundant}. 

As it turns out, orientations of moduli spaces play an important role in this study. More precisely, when disks cross, the local linking number changes from positive to negative or vice versa. The oriention sign of the glued disk depends on the orientation sign and to get quiver formulas for the direction where the joined disk would disappear for the natural orientation we use canceling disks and birth/deaths as just described.

In Section \ref{sec:equivalence-of-quivers} we collect these holomorphic disk bifurcations into a set of moves on quivers that leaves the partition function invariant.

\subsection{Simple unlinking}\label{sec:simpleunlink}
\begin{figure}[h!]
\begin{center}
\includegraphics[width=0.65\textwidth]{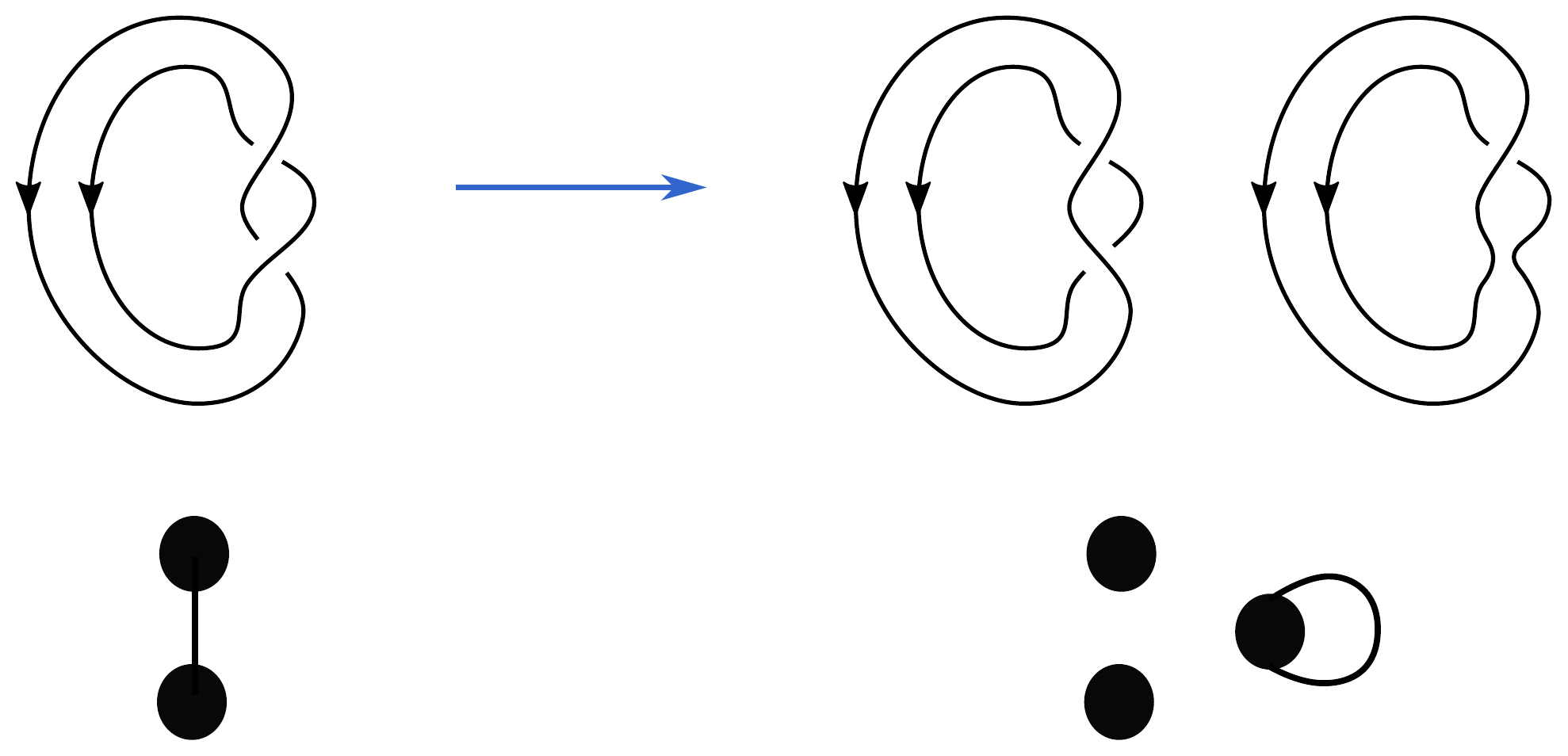}
\caption{The effect of disk boundary crossing on a simple quiver corresponding to two disks  linking once. Since we use only symmetric quivers, we simplify pictures and let an~unoriented line between nodes denote a pair of arrows in opposite directions.}
\label{fig:skein-relation-2-3}
\end{center}
\end{figure}

Let us consider two disks whose boundaries in $L$ link once, as in the left hand side of Figure~\ref{fig:skein-relation-2-3}. As the disk boundaries cross, the disks stay intact and end up in a new position with boundaries unlinked. There is also a new disk born. It is obtained by gluing the~two initial disks and its boundary has one self-crossing. Thus, after the crossing instant, the~configuration of the disk boundaries is as in the right hand side of Figure \ref{fig:skein-relation-2-3} where neither of the~old disks link with the new disk.  

\begin{figure}[h!]
	\begin{center}
		\includegraphics[width=0.4\textwidth]{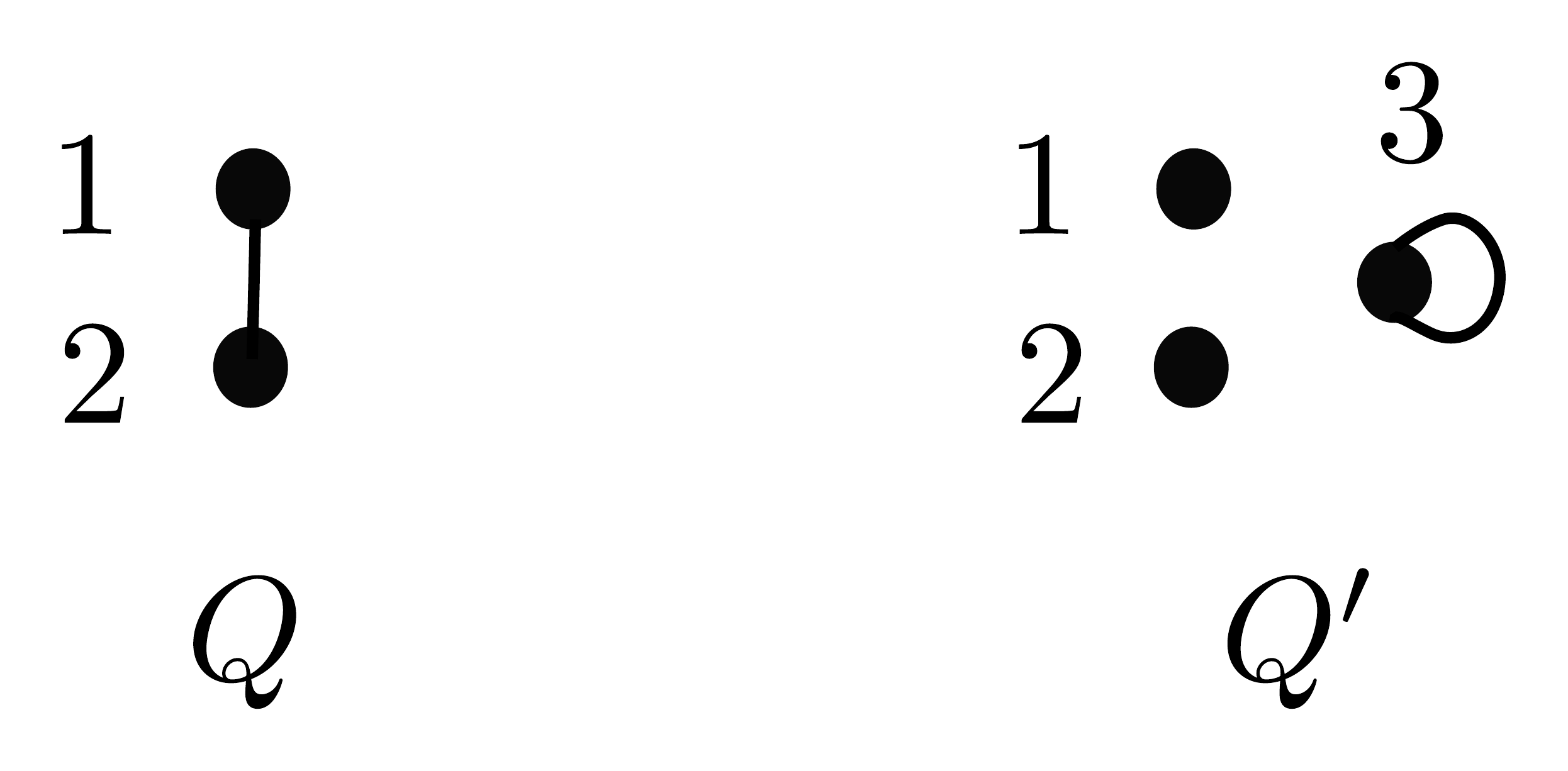}
		\caption{The quivers $Q$ and $Q'$ related by unlinking.}
		\label{fig:skein-simplest}
	\end{center}
\end{figure}

Consider now the quiver $Q$ with two nodes on the left hand side of Figure \ref{fig:skein-relation-2-3}, corresponding to basic disks as explained above. Unlinking these circles gives a new quiver $Q'$ with three nodes, as on the right hand side of Figure \ref{fig:skein-relation-2-3}. The adjacency matrix of the quiver transforms as
\be
\label{eq:link-removal}
	C = \left(
	\begin{array}{cc}
		0 & 1 \\
		1 & 0
	\end{array}
	\right) 
	\quad
	\rightsquigarrow \quad
	C' = \left(
	\begin{array}{ccc}
		0 & 0 & 0\\
		0 & 0 & 0\\
		0 & 0 & 1
	\end{array}
	\right)\,.
\ee
To see that the entries in the new quiver matrix $C'$ are as claimed, we argue as follows. In the $2\times2$ top left corner we see self-linking and linking of the old disks. Self-linking stays unchanged as the disks move, but the linking decreases by one. Since we started from self-linking zero and linking one, we end up with zeros only. The last entry on the diagonal of $C'$ is one. It corresponds to the self-crossing left from the two original positive crossings giving the linking between the two original disks. Remaining entries measure linking between the~old disks in their new position and the glued disk. There are two crosssings: one near the self-intersection of the glued disk and one near the resolved crossing. They have opposite signs and hence the linking numbers are zero. In pictures, quivers $Q$ and $Q'$ are shown in Figure \ref{fig:skein-simplest}.

We next verify that the two quivers $Q$ and $Q'$ have identical partition functions after a suitable identification of variables. We first compute the
motivic generating series and the BPS spectrum of the quiver $Q$
\be \label{eq:before-link-removal-example}
\begin{split}
	P^Q (x_1,x_2,q)&=\sum_{d_1,d_2\geq 0}(-q)^{2d_{1}d_{2}}\frac{x_1^{d_{1}}}{(q^2;q^2)_{d_{1}}}\frac{x_2^{d_{2}}}{(q^2;q^2)_{d_{2}}} \\
	&=\left(\sum_{n=0}^{\infty}\frac{x_1^n }{(q^2;q^2)_{n}} \right)\left(\sum_{m=0}^{\infty}\frac{x_2^m }{(q^2;q^2)_{m}} \right) \left(\sum_{k=0}^{\infty} \frac{(-1)^k  q^{k(k-1)}(x_1 x_2)^k}{(q^2;q^2)_{k}}  \right)\\
	&=(x_1;q^2)^{-1}_\infty (x_2;q^2)^{-1}_\infty (x_1 x_2;q^2)^{+1}_\infty\\
	&=\textrm{Exp}\left(\frac{x_1+x_2-x_1 x_2}{1-q^2}\right)\,,
\end{split}
\ee
where we used \eqref{eq:Efimov} and the following identities
\be\label{eq:3-2-identity}
	\frac{q^{2 a b}}{(q^2;q^2)_{a}  (q^2;q^2)_{b} }
	=
	\sum_{k=0}^{\textrm{min}(a,b)} \frac{q^{k^2-k} (-1)^k}{ (q^2;q^2)_{ a-k}  (q^2;q^2)_{ b-k}   (q^2;q^2)_{k} }\,,
\ee
\be\label{eq:q-pocchh-series}
\begin{split}
	(x,q^2)_\infty & = \prod_{i\geq 0} (1-x q^{2i}) = \sum_{n=0}^{\infty} 
	\frac{(-1)^n  q^{n(n-1)}}{(1-q^{2})\cdots (1-q^{2n})} x^n\,, \\
	\frac{1}{(x,q^2)_\infty} & = \prod_{i\geq 0} (1-x q^{2i})^{-1} = \sum_{n=0}^{\infty} \frac{1}{(1-q^{2})\cdots (1-q^{2n})} x^n \,.
\end{split}
\ee
Comparing \eqref{eq:before-link-removal-example} with \eqref{eq:DT-gen-function} and \eqref{P^Q=Exp}, we see that the whole BPS spectrum is just
\be\label{eq:BPS-spectrum-one-link}
	\Omega_{(1,0),0} = \Omega_{(0,1),0} = \Omega_{(1,1),0} = 1\,.
\ee

We next compute the motivic generating series of the quiver $Q'$
\be \label{eq:after-link-removal-example}
\begin{split}
P^{Q'} (x_1,x_2,x_3,q)&=\sum_{d_1,d_2,d_3\geq 0}(-q)^{d_{3}^{2}}\frac{x_1^{d_{1}}}{(q^2;q^2)_{d_{1}}} \frac{x_2^{d_{2}}}{(q^2;q^2)_{d_{2}}} \frac{x_3^{d_{3}}}{(q^2;q^2)_{d_{3}}} \\
&=(x_1;q^2)^{-1}_\infty (x_2;q^2)^{-1}_\infty (q x_3;q^2)^{+1}_\infty=\textrm{Exp}\left(\frac{x_1+x_2-qx_3}{1-q^2}\right)\,,
\end{split}
\ee
which reduces to (\ref{eq:before-link-removal-example}) for
\be\label{eq:unlinking-variables}
	x_3 =q^{-1} x_1 x_2.
\ee
We can see that the BPS spectrum of $Q'$ is
\be\label{eq:BPS-spectrum-one-link-removed}
	\Omega_{(1,0,0),0} = \Omega_{(0,1,0),0} = \Omega_{(0,0,1),1} = 1\,,
\ee
which agrees with \eqref{eq:BPS-spectrum-one-link} after relabelling.\footnote{The reader might be worried that the spin of the BPS states seems to shift, but in our conventions the~spin is given by $s+|d|-1$ so in (\ref{eq:BPS-spectrum-one-link}) and (\ref{eq:BPS-spectrum-one-link-removed}) we have two states of spin 0 and one state of spin 1.}

\subsection{Proof of invariance for general quivers: unlinking}\label{sec:Proof of invariance-unlink}

We prove the invariance of the motivic generating series under unlinking for general symmetric quivers. 
Without loss of generality we can assume that $Q$ has three nodes: two for which we change the linking and one spectator -- we can erase it or add more spectators if necessary. Therefore the adjacency matrix can be written as
\begin{equation}\label{eq:general-quiver}
	C=\left(\begin{array}{ccc}
	r & k & a\\
	k & s & b\\
	a & b & c
	\end{array}\right)\,,
\end{equation}
which gives
\begin{equation}\label{eq:Cij-before-unlinking}
\begin{split}
	P^{Q}(x_1,x_2,x_3,q)&=\sum_{d_{1},d_{2},d_{3}\geq0}(-q)^{\sum_{i,j}C_{ij}d_{i}d_{j}}\frac{x_{1}^{d_{1}}}{(q^{2};q^{2})_{d_{1}}}\frac{x_{2}^{d_{2}}}{(q^{2};q^{2})_{d_{2}}}\frac{x_{3}^{d_{3}}}{(q^{2};q^{2})_{d_{3}}}\,,\\
\sum_{i,j}C_{ij}d_{i}d_{j}&=rd_{1}^{2}+sd_{2}^{2}+cd_{3}^{2}+2(kd_{1}d_{2}+ad_{1}d_{3}+bd_{2}d_{3})\,.
\end{split}
\end{equation}

\begin{figure}[h!]
\begin{center}
\includegraphics[width=0.75\textwidth]{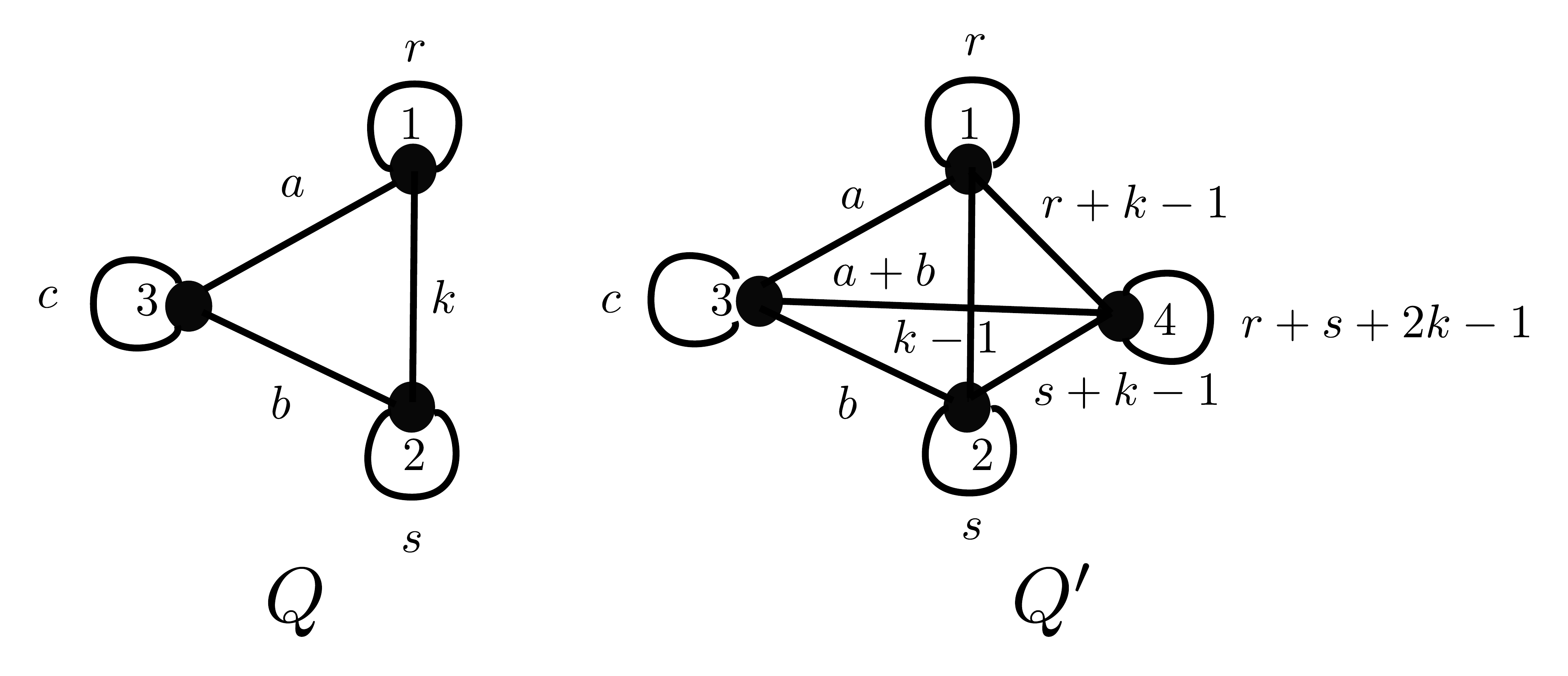}
\caption{Unlinking -- general case. Numbers next to lines and loops denote the number of pairs of arrows and the number of loops respectively.}
\label{fig:skein-generic}
\end{center}
\end{figure}

We will show that the motivic generating series of the quiver $Q'$ given by 
\begin{equation}\label{eq:Q-after-unlinking}
	C'=\left(\begin{array}{cccc}
	r & k-1 & a & r+k-1\\
	k-1 & s & b & s+k-1\\
	a & b & c & a+b\\
	r+k-1 & s+k-1 & a+b & r+s+2k-1
	\end{array}\right)
\end{equation}
is equal to $P^{Q}$ (after appropriate change of variables). We can see that the annihilation of one link is compensated by the creation of the new node which self-linking and linking with old vertices depends on initial arrows, see Figure \ref{fig:skein-generic}.

The motivic generating series of $Q'$ reads
\begin{equation}
	P^{Q'}(x_1,x_2,x_3,x_4,q)=\sum_{\delta_{1},\delta_{2},\delta_{3},\delta_{4}\geq0}
	\frac{(-q)^{\sum_{i,j}C'_{ij}\delta_{i}\delta_{j}} x_{1}^{\delta_{1}}x_{2}^{\delta_{2}}x_{3}^{\delta_{3}}x_{4}^{\delta_{4}}}
	{(q^{2};q^{2})_{\delta_{1}}(q^{2};q^{2})_{\delta_{2}}(q^{2};q^{2})_{\delta_{3}}(q^{2};q^{2})_{\delta_{4}}}\,,
\end{equation}
where 
\begin{equation}
\begin{split}
	\sum_{i,j}C'_{ij}\delta_{i}\delta_{j}= & r\delta_{1}^{2}+s\delta_{2}^{2}+c\delta_{3}^{2}+(r+s+2k-1)\delta_{4}^{2}+2(k-1)\delta_{1}\delta_{2}+
	2a\delta_{1}\delta_{3}\\
	& +2b\delta_{2}\delta_{3}+2(r+k-1)\delta_{1}\delta_{4}+2(a+b)\delta_{3}\delta_{4}+2(s+k-1)\delta_{2}\delta_{4}\,.
\end{split}
\end{equation}
After the change of variables
\be
d_{1}= \delta_{1}+\delta_{4}, \qquad d_{2}= \delta_{2}+\delta_{4}, \qquad d_{3}=  \delta_{3}, \qquad  d_{4}=  \delta_{4}, \qquad x_{4}=q^{-1}x_{1}x_{2}
\ee 
we obtain
\begin{equation}
\begin{split}
	\left. P^{Q'}(x_1,x_2,x_3,x_4,q) \right|_{x_{4}=q^{-1}x_{1}x_{2}}= & \sum_{d_{1},d_{2},d_{3}\geq0}(-q)^{\sum_{i,j}C_{ij}d_{i}d_{j}-2d_1d_2}\frac{x_{1}^{d_{1}}x_{2}^{d_{2}}x_{3}^{d_{3}}}
	{(q^{2};q^{2})_{d_{3}}}\\
 & \times\sum_{d_{4}=0}^{\textrm{min}(d_{1},d_{2})}\frac{(-1)^{d_{4}}q^{d_{4}^{2}-d_{4}}}{(q^{2};q^{2})_{d_{1}-d_{4}}(q^{2};q^{2})_{d_{2}-d_{4}}(q^{2};q^{2})_{d_{4}}}\,,
\end{split}
\end{equation}
where $\sum_{i,j}C_{ij}d_{i}d_{j}$ is given by (\ref{eq:Cij-before-unlinking}). 
Using (\ref{eq:3-2-identity}), we immediately have
\be\label{eq:unlinking-invariance}
	\left. P^{Q'}(x_1,x_2,x_3,x_4,q) \right|_{x_{4}=q^{-1}x_{1}x_{2}} = P^{Q}(x_1,x_2,x_3,q) \,,
\ee
which we wanted to show.

Note that the example from Section \ref{sec:simpleunlink} was a special case of this reasoning for 
$k=1$, $r=s=0$, and without the spectator node.

\subsection{Redundant pairs of nodes}\label{sec:redundant}
\begin{figure}[h!]
\begin{center}
\includegraphics[width=0.35\textwidth]{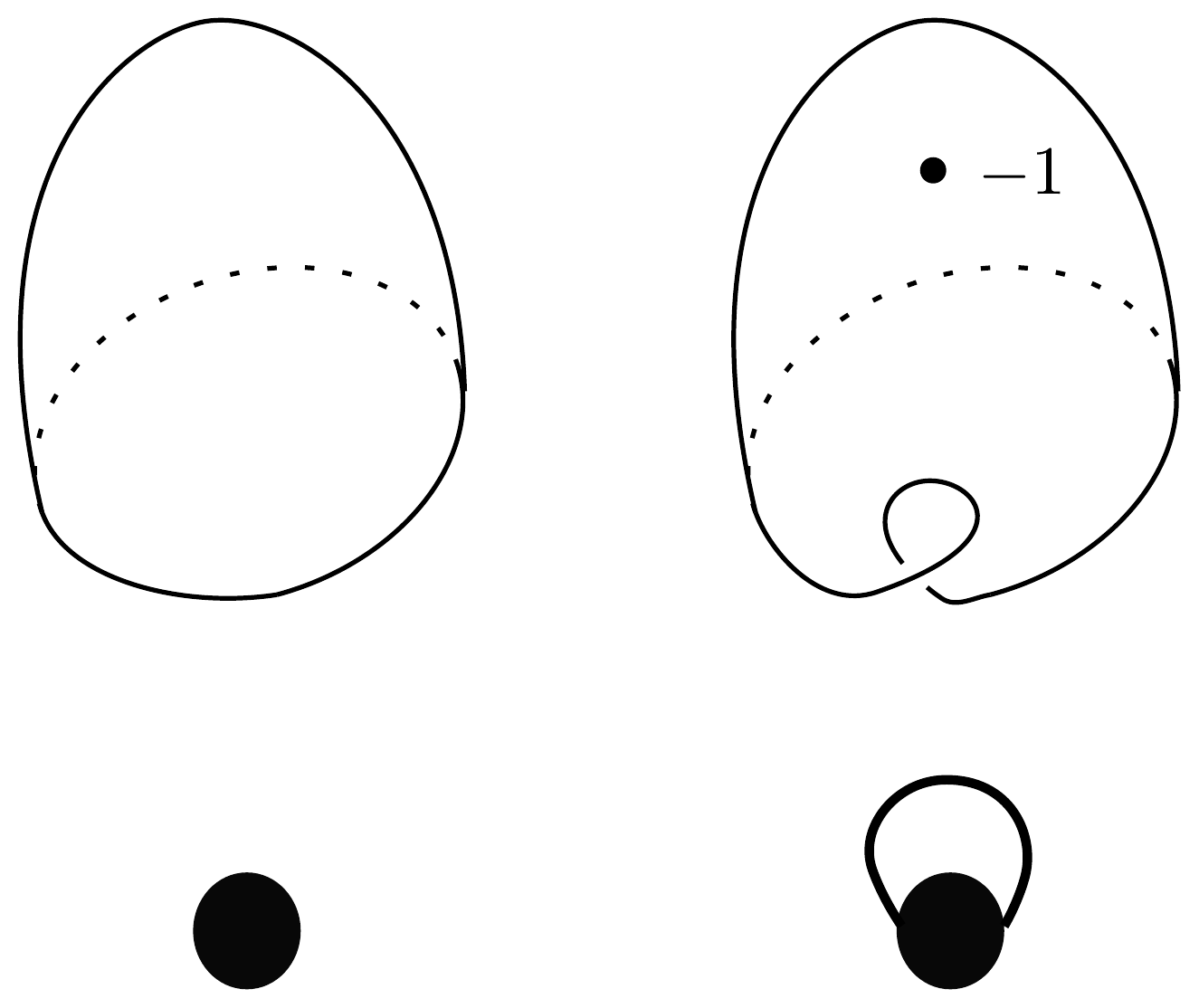}
\caption{Redundant pair of nodes.}
\label{fig:redundant-pair}
\end{center}
\end{figure}

Redundant pairs of nodes  were observed first in \cite{Kucharski:2017ogk,Kucharski:2017poe}. We start in the simplest case of the~two node quiver in Figure \ref{fig:redundant-pair}.
Note that the partition function of this quiver factorizes into 
\be\label{eq:redundant-nodes-cancellation-identity}
	\left(\sum_{d_1} \frac{x_1^{d_1}}{(q^2;q^2)_{d_1}}\right) 
	\left(\sum_{d_2} (-q)^{d_2^2}\frac{x_2^{d_2}}{(q^2;q^2)_{d_2}}\right) 
	 = (x_1;q^2)^{-1}_\infty (q x_2;q^2)_\infty\,,
\ee 
which is trivial (equals $1$) if we set $x_2 = q^{-1} x_1$.

The geometric interpretation of this quiver is the following. The first node corresponding to $x_{1}$ is a disk with no self-linking. The second one is a disk with one unit of self-linking as well as a negative shift of the 4-chain intersection compared to the first one (which leads to $x_2 = q^{-1} x_1$, see \cite{Ekholm:2018eee}), as depicted in Figure \ref{fig:redundant-pair}. 
We note that these two canceling nodes resemble the unknot nodes \cite{Kucharski:2017ogk,Kucharski:2017poe} with the important difference that the different powers of $a$ (the conifold K\"ahler modulus) are now the same, leading to cancellation of their contributions.

We next show that a redunant pair of disks that link in the same way to all other nodes does not affect the partition function. Since
\begin{equation}
\sum_{\alpha=0}^{n}(-1)^{\alpha}q^{\alpha^{2}-\alpha}\frac{(q^{2};q^{2})_{n}}{(q^{2};q^{2})_{\alpha}(q^{2};q^{2})_{n-\alpha}}=
(1;q^{2})_{n}=\begin{cases}
1 & n=0\\
0 & n\geq1
\end{cases}
\end{equation}
we can write 
\begin{equation}
\begin{split}
	1= & \sum_{n\geq0}(1;q^{2})_{n}(-q)^{a_{0}n^{2}+2(a_{1}+\ldots+a_{m})n(d_{1}+\ldots+d_{m})}\frac{x^{n}}{(q^{2};q^{2})_{n}}\\
 	 = & \sum_{n\geq0}\sum_{d_{m+1}+d_{m+2}=n}(-1)^{d_{m+2}}q^{d_{m+2}^{2}-d_{m+2}}\frac{(q^{2};q^{2})_{n}}{(q^{2};q^{2})_{d_{m+2}}(q^{2};q^{2})_{d_{m+1}}}\\
	   & \qquad \qquad \qquad \qquad\qquad\qquad\times(-q)^{a_{0}n^{2}+2(a_{1}+\ldots+a_{m})n(d_{1}+\ldots+d_{m})}\frac{x^{n}}{(q^{2};q^{2})_{n}}\\
	 = & \sum_{d_{m+1},d_{m+2}\geq0}(-q)^{a_{0}(d_{m+1}+d_{m+2})^2+d_{m+2}^{2}+2(a_{1}+\ldots+a_{m})(d_{m+1}+d_{m+2})(d_{1}+\ldots+d_{m})}\\
	   & \qquad \qquad \qquad \qquad\qquad\qquad\times \frac{x^{d_{m+1}}(q^{-1}x)^{d_{m+2}}}{(q^{2};q^{2})_{d_{m+1}}(q^{2};q^{2})_{d_{m+2}}}.
\end{split}
\end{equation}
If we multiply this unit by the motivic generating series of an arbitrary
quiver $Q$ with $m$~vertices and adjacency matrix $C$ and set
\be\label{eq:redundant-disks-variables}
	x=x_{m+1}=qx_{m+2}\,,
\ee
we obtain the motivic generating series of the new quiver $Q''$
\be
	P^{Q''}(x_1,\ldots,x_{m+2},q) = \sum_{d_1,\ldots,d_{m+2}\geq0} (-q)^{C''_{ij} d_i d_j}\prod_{i=1}^{m+2}
	\frac{x_i^{d_i}}{(q^{2};q^{2})_{d_i}}\,,
\ee
where
\begin{equation}\label{eq:redundant-pair-general}
	C''=\left(
	\begin{array}{ccccc}
		 &  &  & a_{1} & a_{1}\\
		 & C &  & \vdots & \vdots\\
		 &  &  & a_{m} & a_{m}\\
		a_{1} & \ldots & a_{m} & a_{0} & a_{0}\\
		a_{1} & \ldots & a_{m} & a_{0} & a_{0}+1
	\end{array}\right).
\end{equation}
We find that for $x_{m+1}=qx_{m+2}$ nodes $m+1$ and $m+2$ are indeed redundant and 
\be
	\left. P^{Q''}(x_1,\ldots,x_{m},x_{m+1},x_{m+2},q) \right|_{x_{m+1}=qx_{m+2}} = P^{Q}(x_1,\ldots,x_{m},q).
\ee

\subsection{Simple linking}\label{sec:simplelink}
We next consider linking instead of unlinking, as in Section \ref{sec:simpleunlink}. This case is more involved than unlinking. (Reversing the orientation of the Lagrangian would switch the roles between linking and unlinking.) We start from a basic case of two unlinked disks that correspond to a quiver $Q$ with adjacency matrix
\be
C = \left(
\begin{array}{cc}
	0 & 0 \\
	0 & 0
\end{array}
\right)\,.
\ee
The motivic generating series is
\be\label{eq:two-nodes-Q}
\begin{split}
	P^Q (x_1,x_2,q)&=\sum_{d_1,d_2\geq 0}\frac{x_1^{d_{1}}}{(q^2;q^2)_{d_{1}}} \frac{x_2^{d_{2}}}{(q^2;q^2)_{d_{2}}} \\
	&=(x_1;q^2)^{-1}_\infty (x_2;q^2)^{-1}_\infty =\textrm{Exp}\left(\frac{x_1+x_2}{1-q^2}\right)\,,
\end{split}
\ee
so the whole BPS spectrum is just
\be\label{eq:BPS-spectrum-before-linking}
\Omega_{(1,0),0} = \Omega_{(0,1),0} = 1 \,.
\ee

\begin{figure}[h!]
	\begin{center}
		\includegraphics[width=0.5\textwidth]{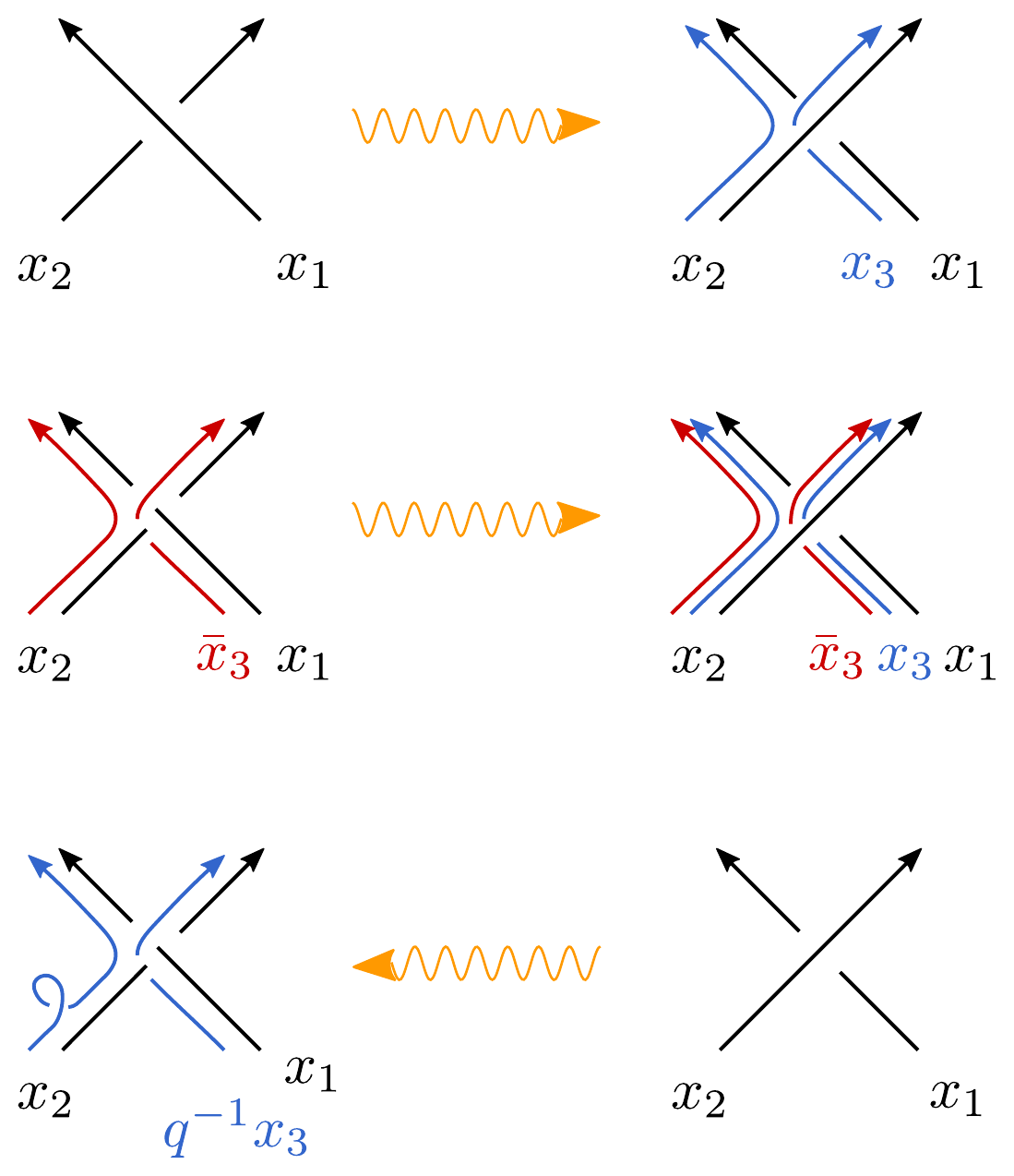}
		\caption{Starting from the standard unlinking based on the skein relation (upper picture), we add a red anti-disk on both sides (middle picture) and then the disk/anti-disk pair is annihilated, whereas the red anti-disk on the left can be exchanged for a regular blue disk with self-linking and 4-chain intersection (lower picture).}
		\label{fig:antidisk}
	\end{center}
\end{figure}

From the unlinking case in Section \ref{sec:simplelink}, we know that these two disks -- alongside a~glued disk with self-linking one -- arise from unlinking linked versions of the two disks. We would now like to run time backwards in this process. This however requires the presence of the~glued disk that we do not have. To remedy this, we create a pair of canceling glued disks and carry the one with negative orientation compared to the unlinking case to the~other side. Effectively we obtain the unlinking case amended by the presence of a disk with a negative orientation sign. This anti-disk may be exchanged for a regular disk with self-linking and 4-chain intersection, as observed in Section \ref{sec:redundant}. The geometric process is depicted in Figure \ref{fig:antidisk}.  

We need to interpret this as an adjacency matrix. To this end, we note that the anti-disk links with the two original disks exactly as the corresponding disk and hence we find that the last entries in the first two rows and the first two colums are zero. For the anti-disk the diagonal entry is again as for the disk, which means it is a one. Finally, changing the anti-disk to a disk with self-linking and 4-chain intersection decreases the total self-linking to zero and we get the following adjacency matrix 
\be
C' = \left(
\begin{array}{ccc}
	0 & 1 & 0 \\
	1 & 0 & 0 \\
	0 & 0 & 0
\end{array}
\right)\,.
\ee
The motivic generating series of $Q'$ is
\be
\begin{split}
	P^{Q'} (x_1,x_2,x_3,q)&=\sum_{d_1,d_2,d_3\geq 0}(-q)^{2d_{1}d_{2}}\frac{x_1^{d_{1}}}{(q^2;q^2)_{d_{1}}}\frac{x_2^{d_{2}}}{(q^2;q^2)_{d_{2}}}\frac{x_3^{d_{3}}}{(q^2;q^2)_{d_{3}}} \\
	&=(x_1;q^2)^{-1}_\infty (x_2;q^2)^{-1}_\infty (x_3;q^2)^{-1}_\infty (x_1 x_2;q^2)^{+1}_\infty \\
	&=\textrm{Exp}\left(\frac{x_1+x_2+x_3-x_1 x_2}{1-q^2}\right)\,,
\end{split}
\ee
which reduces to \eqref{eq:two-nodes-Q} for
\be
x_3 = x_1 x_2.
\ee

From the point of view of the BPS spectrum this identification causes a cancellation between the basic state coming from the third node and the boundstate of the two old disk in their new linked position. Consequently, the spectrum
\be\label{eq:BPS-spectrum-after-linking}
\Omega_{(1,0,0),0} = \Omega_{(0,1,0),0} = \Omega_{(0,0,1),0} =  \Omega_{(1,1,0),0} = 1
\ee
reduces to \eqref{eq:BPS-spectrum-before-linking}.

\begin{figure}[h!]
	\begin{center}
		\includegraphics[width=0.4\textwidth]{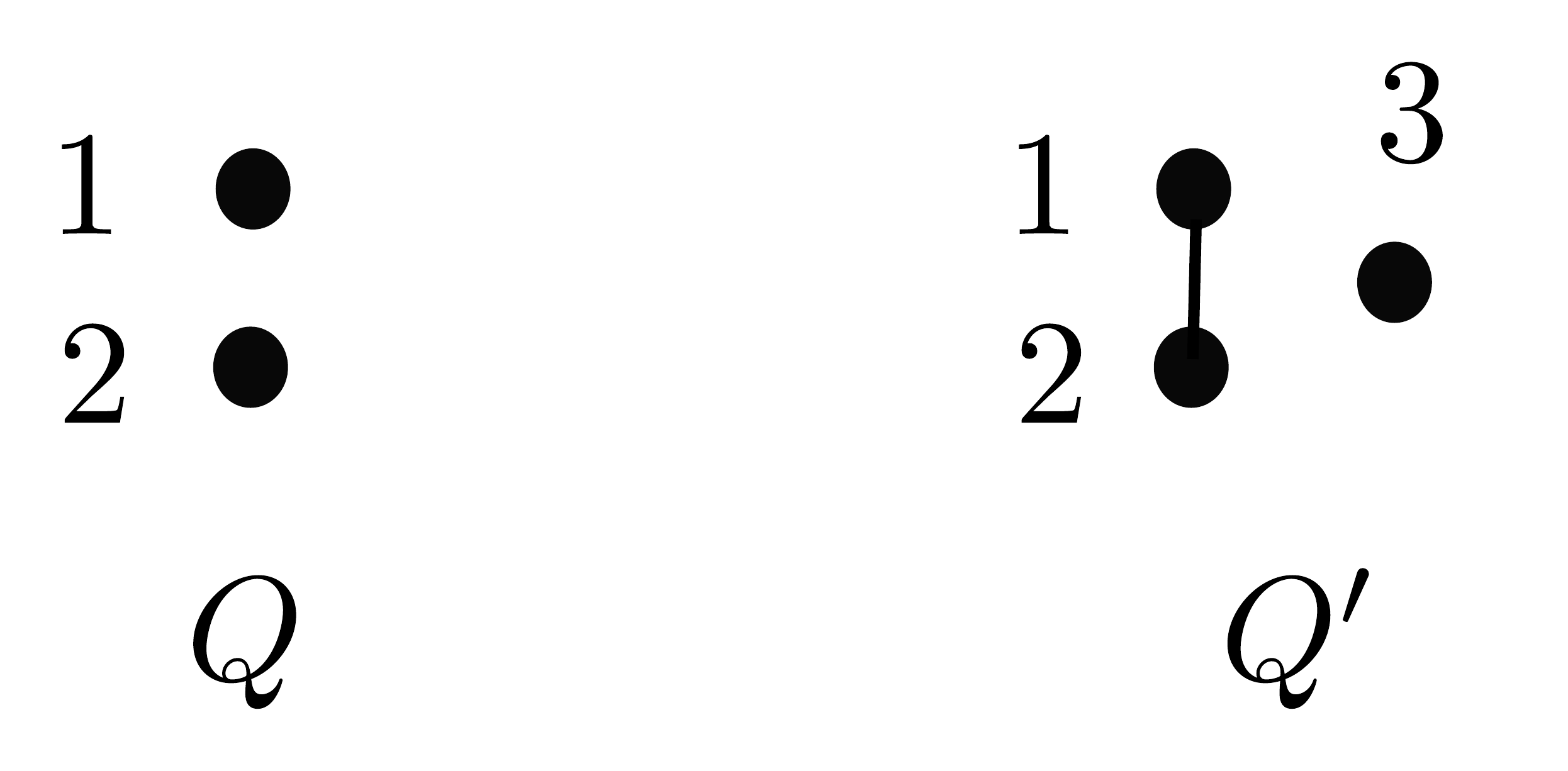}
		\caption{The quivers $Q$ and $Q'$ related by linking.}
		\label{fig:linking-simplest}
	\end{center}
\end{figure}

The quivers $Q$ and $Q'$ are presented in Figure \ref{fig:linking-simplest}.

\subsection{Proof of invariance for general quivers: linking}\label{sec:Proof of invariance-link}

We next prove the invariance under the introduction of additional linking between two nodes.
More precisely, we will show we can add a~redundant pair of nodes to $Q$ in such a way that the new quiver $Q''$ can be obtained by the unlinking of some other quiver $Q'$. This is equivalent to the statement that $Q'$ is the result of linking of $Q$ and $P^{Q'}=P^{Q}$ (after appropriate change of variables).

In analogy to Section \ref{sec:Proof of invariance-unlink} it is sufficient to focus on $Q$ given by
\begin{equation}\label{eq:general-Q-for-linking}
	C=\left(\begin{array}{ccc}
	r & k & a\\
	k & s & b\\
	a & b & c
	\end{array}\right)\,.
\end{equation}
We can enlarge it by a~redundant pair of nodes using \eqref{eq:redundant-pair-general} with $a_{1}=r+k$, $a_{2}=s+k$, $a_{3}=a+b$, $a_{0}=r+s+2k$. Then
\begin{equation}
	C''=\left(
	\begin{array}{ccccc}
		r & k & a & r+k & r+k\\
		k & s & b & s+k & s+k\\
		a & b & c & a+b & a+b\\
		r+k & s+k & a+b & r+s+2k & r+s+2k\\
		r+k & s+k & a+b & r+s+2k & r+s+2k+1
	\end{array}\right)
\end{equation}
and we know that
\begin{equation}
	\left. P^{Q''}(x_1,x_2,x_3,x_4,x_5,q) \right|_{x_{4}=qx_{5}} = P^{Q}(x_1,x_2,x_3,q)\,.
\end{equation}
On the other hand we can obtain $Q''$ by unlinking of
the quiver $Q'$ given by
\begin{equation}\label{eq:Q-after-linking}
	C'=\left(
	\begin{array}{cccc}
		r & k+1 & a & r+k\\
		k+1 & s & b & s+k\\
		a & b & c & a+b\\
		r+k & s+k & a+b & r+s+2k
	\end{array}\right)\,.
\end{equation}
Since
\be\label{eq:linking-with-redundant-nodes}
	P^{Q'}(x_1,x_2,x_3,x_4,q)=\left.P^{Q''}(x_1,x_2,x_3,x_4,x_5,q)\right|_{x_{5}=q^{-1}x_{1}x_{2}}\,,
\ee
we have 
\be\label{eq:linking-invariance}
	\left.P^{Q'}(x_1,x_2,x_3,x_4,q)\right|_{x_{4}=x_{1}x_{2}}=P^{Q}(x_1,x_2,x_3,q)\,.
\ee
Therefore if we define the linking of $Q$ given by (\ref{eq:general-Q-for-linking}) as $Q'$ given by (\ref{eq:Q-after-linking}), then (\ref{eq:linking-invariance}) guarantees the invariance of the motivic generating series under this transformation. 

\begin{figure}[h!]
\begin{center}
\includegraphics[width=0.75\textwidth]{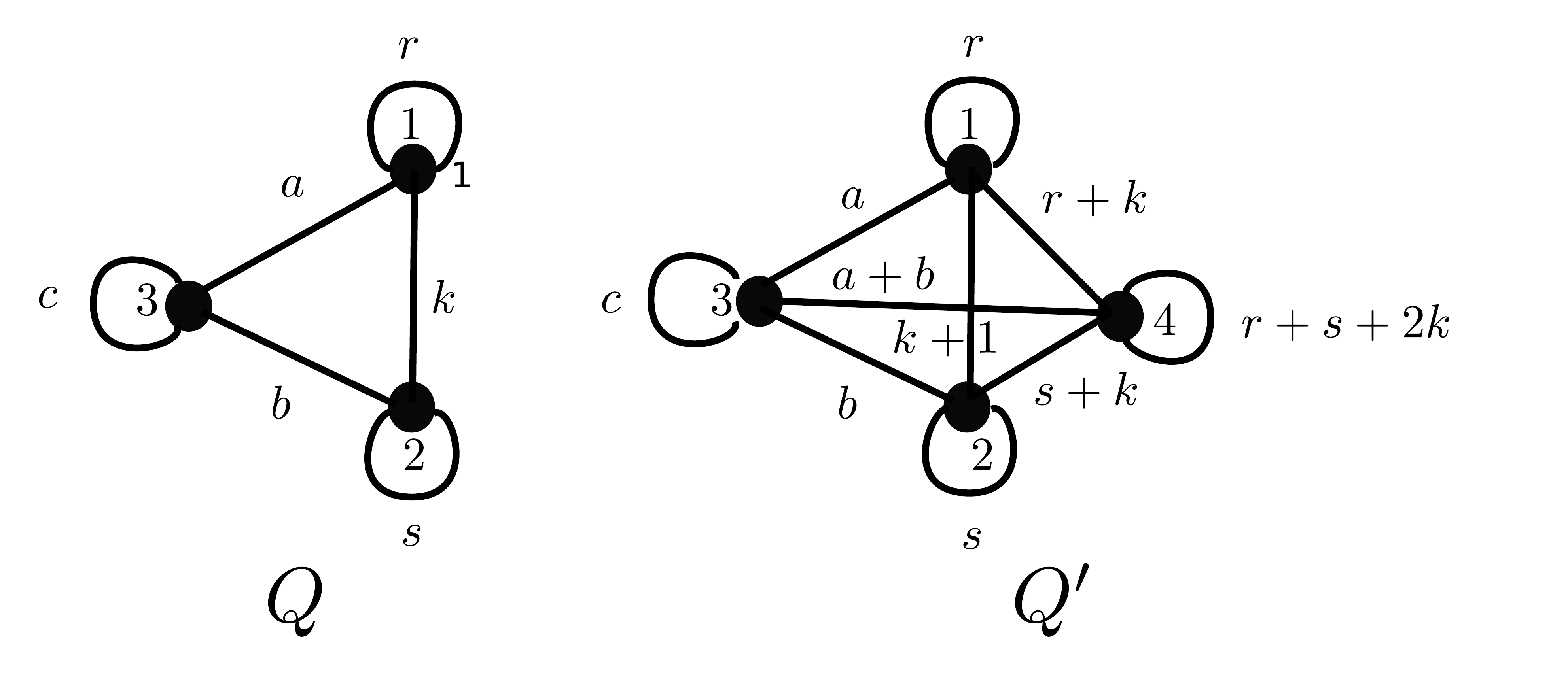}
\caption{Linking -- general case.}
\label{fig:skein-generic-add}
\end{center}
\end{figure}

The quivers $Q$ and $Q'$ are presented in Figure \ref{fig:skein-generic-add}.

The example from Section \ref{sec:simplelink} was a special case of this reasoning for 
$
r=k=s=0
$
with the spectator node erased.

\subsection{Equivalence of quivers}\label{sec:equivalence-of-quivers}

We will refer to the linking and unlinking operations introduced above, together with the~addition/removal of redundant pairs of nodes, collectively as \emph{quiver multi-cover skein relations}.
Since these relations produce an infinite number of quivers with the same partition functions (upon suitable identification of quiver variables $x_i$), we use them to define an equivalence relation on the set of quivers with variables as follows.

\begin{dfn}
Let $Q$, $Q'$ be quivers with $m$ and $m'$ nodes respectively.
We say that $Q$ and $Q'$ are equivalent under multi-cover skein relations 
\be\label{eq:skquiv}
	Q\sim Q'
\ee
if there exists a sequence of multi-cover skein relations that takes $Q$ into $Q'$ and vice versa.
\end{dfn}

If \eqref{eq:skquiv} holds, then there exist two sets of variables  $(x_1\dots x_m)$ and $(x'_1\dots x'_{m'})$, related in a specific way to each other, such that
\be
	P^Q(x_1\dots x_m) = P^{Q'}(x'_1\dots x'_{m'}) \,.
\ee

This equivalence relation contains the one defined in \cite{Kucharski:2017ogk,Kucharski:2017poe} but generates a much larger equivalence class. For example for the figure-eight knot one can find (on the ground of the~KQ correspondence) two different quivers of the same size which have the same motivic generating series. We show in Appendix \ref{sec:KRSS-equivalences} that they are related through multi-cover skein relations. 

Finally, we remark that there is another natural operation on quivers: the change of framing.
This acts on quivers by shifting the adjacency matrix by an overall integer constant $C_{ij}\to C_{ij}+f$. 
This equivalence relation is on a different footing since it does not preserve the partition function and, as explained in \cite{Ekholm:2018eee}, has a direct counterpart for generalized holomorphic curves: the curves are unchanged but the bounding chains changes and the count changes accordingly.

\section{3d $\CN=2$ dualities of multi-cover skein type}\label{sec:skein-dualities}
In the context of the knots-quivers correspondence, the generating function of symmetrically colored HOMFLY-PT polynomials \eqref{eq:HOMFLY-PT series} coincides with the  K-theoretic vortex partition function {(or, more properly, the $\IR^2\times_q S^1$ partition function)} of a 3d $\CN=2$ theory $T[L_K]$ arising on the world-volume of an M5-brane wrapped on the knot conormal $L_K$ \cite{Ooguri:1999bv, Dimofte:2010tz}.
In \cite{Ekholm:2018eee} we showed that $T[L_K]$ is dual to a~theory~$T[Q_K]$ whose structure is encoded by the quiver~$Q_K$ corresponding to the knot~$K$. 
We have argued in Section \ref{sec:skein-rel} that there is no unique quiver associated to a Lagrangian like~$L_K$, but rather an equivalence class built on the~multi-cover skein relations. This suggests the~existence of a~corresponding duality web for theories of type $T[Q_K]$. 

Furthermore, we conjectured in Section \ref{sec:general-oGW-Q} that quivers describe not only knot invariants, but also BPS spectra of open topological strings on a larger class of Lagrangians $L$ in Calabi-Yau threefolds $X$. As mentioned there, this extension of the quiver description implies a~corresponding extension of the duality between $T[L]$ and $T[Q]$. 

In this section we spell out the details of such dualities in the physical language.
We will focus entirely on quivers and the associated 3d $\CN=2$ theories of type $T[Q]$. 
The~only condition we impose on the~quiver $Q$ is that it is \emph{symmetric}, or in other words that for any pair of vertices $(i,j)$ it has an equal number of arrows $i\to j$ as in the~opposite direction~$j\to i$.
For the purpose of this section it will not matter whether such a~quiver arises from a~geometry or not. Accordingly, we will not assume any relation among the~formal variables~$x_i$ associated to nodes of~$Q$. 
In this way, all  statements we are going to make will be of rather general nature. In particular, they will automatically carry over to the~general geometric setting outlined in Section \ref{sec:general-oGW-Q}, as well as to the more specialized context of the~knots-quivers correspondence, by simply specializing variables.

\subsection{General theories of quiver type}\label{sec:general-quiver-theories}

For a given symmetric quiver $Q$ we consider a 3d $\CN=2$ theory $T[Q]$ on $\IR^2\times S^{1}$. 
This is an abelian Chen-Simons-matter theory with gauge group 
\be
	G_{\rm{gauge}} = U(1)_{\gauge,1}\times \dots\times U(1)_{\gauge,m}\,,
\ee 
where $m$ is the number of nodes in $Q$.
The matter content is a collection of chiral multiplets~$\left.\{\phi_{i}\}\right._{i=1,...,m}$, with charges $Q_{i}^{(j)}=\delta_{ij}$ under $U(1)_{\gauge,j}$. 
The flavor symmetry is maximally gauged, there are no residual axial symmetries.
On the other hand there is an abelian dual group of topological symmetries
\be
	G_{\rm{top}} = U(1)_{\top,1}\times \dots\times U(1)_{\top,m}\,.
\ee
The conserved current of $G_{\rm{top}}$ is $j \sim \star dA$, therefore conserved charges are given by the first Chern class for the gauge connection and correspond to vortex numbers $(d_1,\dots, d_m)$.
Mass parameters for $U(1)_{\top,i}$ correspond to Fayet-Iliopoulos (FI) couplings and will be denoted by~$\log x_i$.
The central charge of a vortex with global topological charge ${\mathbf d}$ is 
\be
	Z({\mathbf{d}}) = \sum_{i} d_i \log x_i = \log  {\mathbf x}^{{\mathbf d}} \,,
\ee
where $\mathbf{x} = (x_1,\dots, x_m)$ is the collection of FI couplings and $\mathbf{d} = (d_1,\dots, d_m)$ is that of vortex charges.
Finally, $T[Q]$ has mixed Chern-Simons couplings $C_{ij} \in \IZ$. 
More precisely, these are the \emph{effective} couplings related to the bare ones by 1-loop contributions of chiral multiplets \cite{Aharony:1997bx}
\be\label{eq:Quiver-CS}
	C_{ij} = \kappa_{ij} + \frac{1}{2} \sum_{k=1}^{m} Q_{i}^{(k)}Q_{j}^{(k)} = \kappa_{ij} + \frac{1}{2} \delta_{ij}\,.
\ee
At the level of a classical description, we always work on the Coulomb branch where all chirals are massive due to the VEVs acquired by vector multiplets of $G_{\gauge}$. Therefore we always work with effective Chern-Simons couplings, which must be integers.

We consider $T[Q]$ on {$\IR^2\times_q S^1$ with $q = e^{\hbar}$ parametrizing a rotation of $\IR^2$ around the $S^1$}. 
This localizes BPS vortices to the origin of $\IR^2$, and confers the latter an effective volume $\frac{1}{2\hbar}$.  (For applications to topological strings recall that $q^2=e^{g_s}$.)
The K-theoretic vortex partition function of $T[Q]$ coincides with the generating function of stable quiver representations \cite{Ekholm:2018eee}
\be\label{eq:vortex-quiver}
	\CZ^{\vort}_{T[Q]}(\mathbf{x},q) = P^{Q}(\mathbf{x},q)
\ee
where $q=e^{\hbar}$. 
Recall that the~quiver partition function is explicitly known in terms of the adjacency matrix $C_{ij}$
\be\label{eq:Efimov-bis}
	P^{Q}(\mathbf{x},q)=\sum_{d_{1},\ldots,d_{m}\geq0}(-q)^{\sum_{1\leq i,j\leq m}C_{ij}d_{i}d_{j}}\prod_{i=1}^{m}\frac{x_{i}^{d_{i}}}{(q^{2};q^{2})_{d_{i}}} \,,
\ee
therefore vortex partition functions of theories $T[Q]$ are completely under control.
Once again, let us stress that we are not imposing any constraint on the~FI~parameters $\mathbf{x}$. They are all independent.

\subsection{Semiclassical description} 

In the semiclassical limit $\hbar\to 0$, the partition function takes the universal form
\be\label{eq:bar-P-semiclassical-limit-bis}
\begin{split}
	& P^{Q}(\mathbf{x},q)
	\mathop{\longrightarrow}^{\hbar\to 0}_{q^{2 d_i}\to y_i}  
	\int \prod_{i=1}^{m}  \frac{dy_i}{y_i} \, 
	\exp \left[ \frac{1}{2\hbar} \( \tCW_{T[Q]}(\mathbf{x},\mathbf{y}) + O(\hbar) \) \right] \\
	& \tCW_{T[Q]}(\mathbf{x},\mathbf{y})
	= \sum_{i} \Li_2(y_i) + \log \((-1)^{C_{ii}}x_{i}\) \, \log y_i  + \sum_{i,j} \frac{C_{ij}}{2} \log y_i \, \log y_j\,.   
\end{split}
\ee
Here $y_i$ are fugacities for $G_\gauge$ and $\log y_i$ descend from the top components of vector multiplets via localization.
Strictly speaking, the $x_i$ appearing above are not the same as the~FI~couplings considered in Section \ref{sec:general-quiver-theories}, but are related to them by an overall rescaling of $\log x_i$
\be
	\log x_i \to 2\pi R \cdot \log x_i\,,
\ee
with $R$ being the radius of compactification of the theory on $S^1\times \IR^2$. Only after this rescaling the FI coupling $\log x_i$ becomes dimensionless and this is what appears in (\ref{eq:bar-P-semiclassical-limit-bis}). An~analogous statement applies to the relation between gauge fugacities and the top component of gauge vector multiplets.
We will generally suppress $2\pi R$ except where necessary.

The structure of the twisted superpotential therefore reflects the Lagrangian description of~$T[Q]$, where each dilogarithm corresponds to the 1-loop contribution of a~chiral multiplet with dynamical mass $\log y_i$.

\subsection{Quantum moduli space of vacua on $S^1\times\IR^2$}

In this section we highlight some properties of the quantum moduli spaces of vacua of theories of type $T[Q]$. Readers interested only in the statement on dualities induced by multi-cover skein relations may skip ahead to Section \ref{sec:3dN2-skein-dualities}.

The semiclassical description  (\ref{eq:bar-P-semiclassical-limit-bis}) is formulated on the Coulomb branch. 
On~$\IR^3$ the~Higgs branch and Coulomb branch are generically separated, joining only at singularities (although exceptions to this are known, for example in the case of non-Abelian gauge theories \cite{Aharony:1997bx}). 
The details of this picture can be however modified in several ways, for example by turning on mass deformations which can lift, partially or completely, the~Higgs branch.
Moreover when working on $\IR^2\times S^1$, BPS vortices wrapping $S^1$ produce instanton corrections for the K\"ahler potential of the order $e^{-2\pi R \cdot Z(\mathbf{d})}$. 
The effect of these is to smooth out the quantum moduli space, merging several branches together.
At the quantum level, and with a circle of finite radius, there is no invariant distinction between branches that would otherwise be separated on~$\IR^3$. 

\subsubsection{An example -- SQED}
Let us illustrate these effects through  a concrete example. To this end, we will consider a~model that is not of the type $T[Q]$ but closely related, as will become clear later on. 
We consider a $U(1)_\gauge$ gauge theory with a chiral $u$ with charge~+1 and a chiral $\tilde u$ with charge~-1. 
There is an axial symmetry $U(1)_{{\axial}}$ under which both chirals have charge +1, we may turn on a mass deformation for this with fugacity denoted by $\mu = e^{2\pi R\, m}$. 
We also include the possibility to turn on a FI coupling which corresponds to the twisted mass of the topological symmetry $U(1)_{\top}$. This model is known as $N_f=1$ SQED.

\begin{figure}[h!]
\begin{center}
\begin{subfigure}[b]{0.35\textwidth}
        \centering
        \includegraphics[height=1.2in]{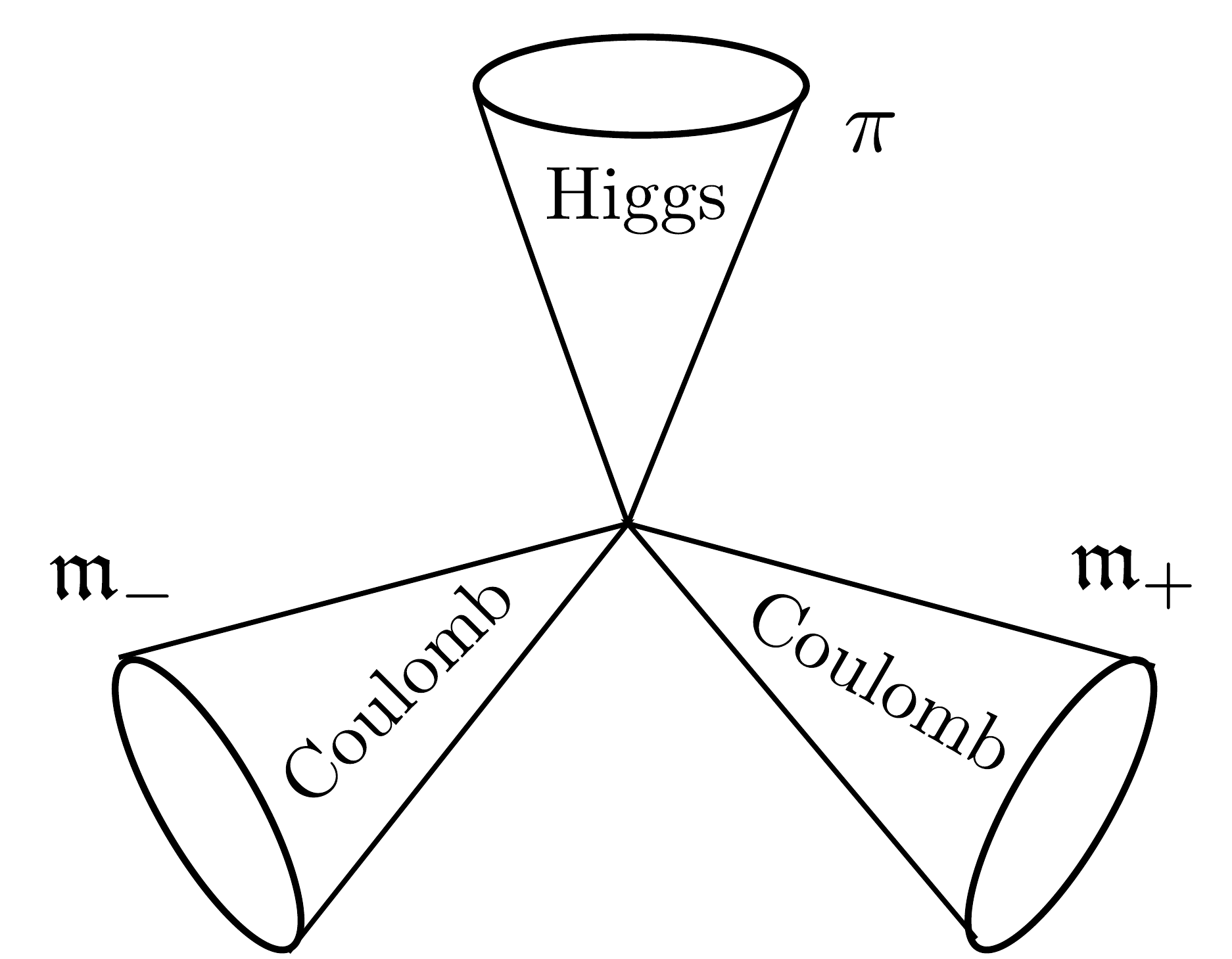}
        \caption{Theory on $\IR^3$ with $\mu=1$.}
        \label{fig:SQED-vacua-singular}
    \end{subfigure}%
    \hspace*{.1\textwidth}
    \begin{subfigure}[b]{0.35\textwidth}
        \centering
        \includegraphics[height=1.2in]{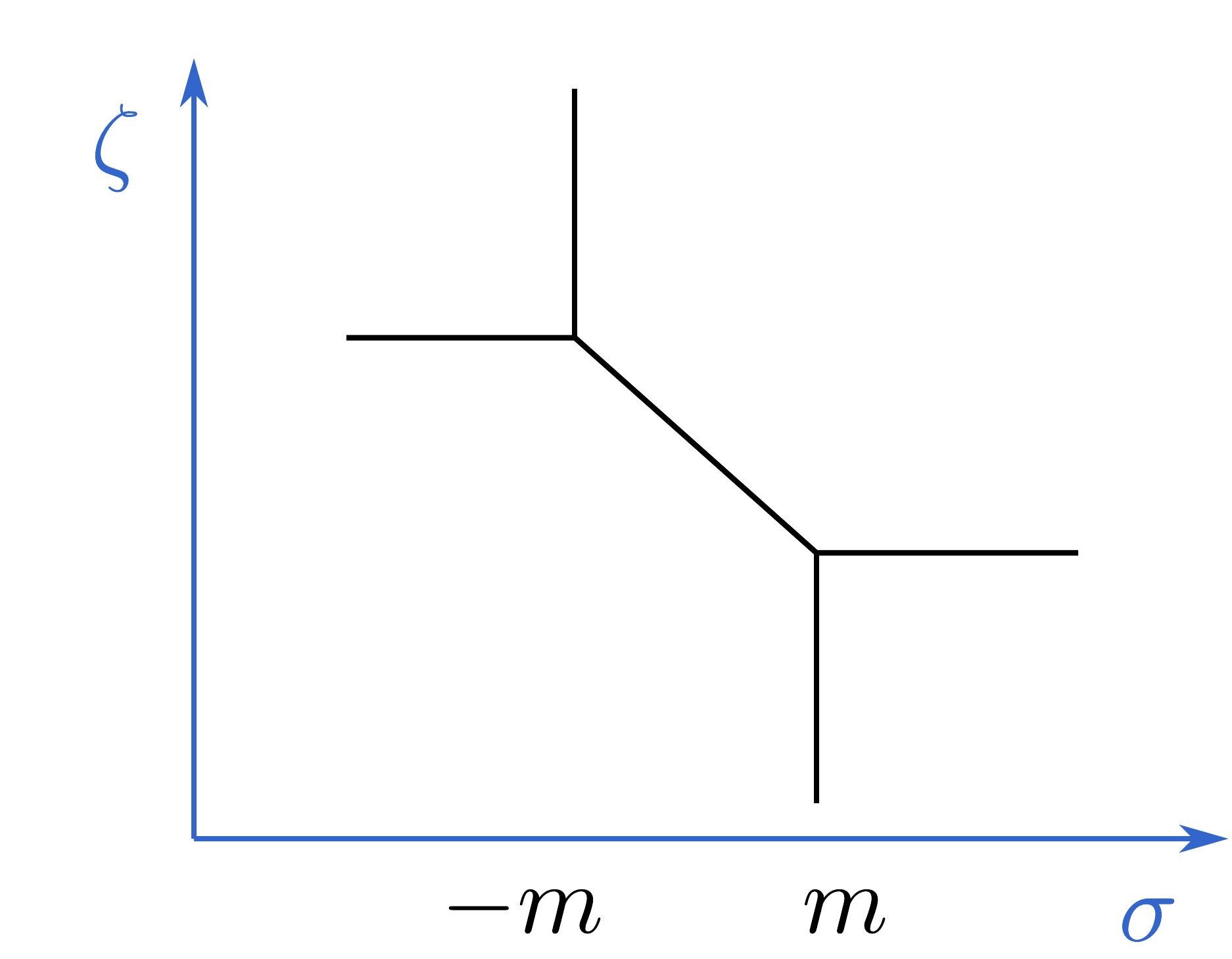}
        \caption{Theory on $\IR^3$ with $\mu\neq 1$.}
        \label{fig:SQED-vacua-massive}
    \end{subfigure}%
    \caption{Vacua of $N_f=1$ SQED on $\IR^3$.}
\label{fig:SQED-vacua}
\end{center}
\end{figure}

When this theory is considered on $\IR^3$, its moduli space of vacua is the set of minima of the potential
\be
	V_{\textrm{SQED}} = \frac{e^2}{2} \(|u|^2 - |\tilde u|^2 - \zeta\)^2 + (\sigma+m)^2 |u|^2 +(\sigma-m)^2 |\tilde u|^2 \,.
\ee
Here $\sigma$ and $\zeta$ are respectively the VEV of the top component in the gauge multiplet and the FI coupling, $e$ is the gauge coupling. 
The quantum moduli space of this theory is well-known \cite{Aharony:1997bx}.
If $\mu = 1$, it consists of a Higgs branch parameterized by the meson $\pi = u\tilde u $ for $\zeta\neq 0$ and a two-component Coulomb branch parameterized by VEVs of monopole operators $\fm_\pm$ at $\sigma>0 $ and $\sigma<0$ for $\zeta=0$. The Higgs branch has the structure of a~cone, due to the fact that the meson operator $\pi = u\tilde u$ can be assigned a gauge-invariant phase. Likewise for the gauge-invariant monopole operators, conferring the  two halves of the Coulomb branch a cone structure as well (see Figure \ref{fig:SQED-vacua-singular}).

If we turn on the axial mass $\mu = e^{2\pi R\, m}$, this breaks the Higgs branch: now $V_{\textrm{SQED}}=0$ requires either $\sigma=-m$ and $\tilde u=0$ or $\sigma=m$ and $u=0$. In both cases $\pi=u\tilde u=0$.
However it is still possible to turn on a nonzero $\zeta$: if $\sigma=-m$ and $\zeta>0$, then $u$ can be set to $|u|=\zeta^{1/2}$ to minimize the potential. Likewise for $\sigma = +m$ and $\zeta<0$ one can always take $|\tilde u|=(-\zeta)^{1/2}$ to minimize the potential.  
Overall, there are now discrete vacua for different values of $(\zeta,\sigma)$. 
The moduli space has a structure which is the one shown schematically in Figure~\ref{fig:SQED-vacua-massive}.

\begin{figure}[h!]
\begin{center}
\begin{subfigure}[b]{0.35\textwidth}
        \centering
        \includegraphics[height=1.2in]{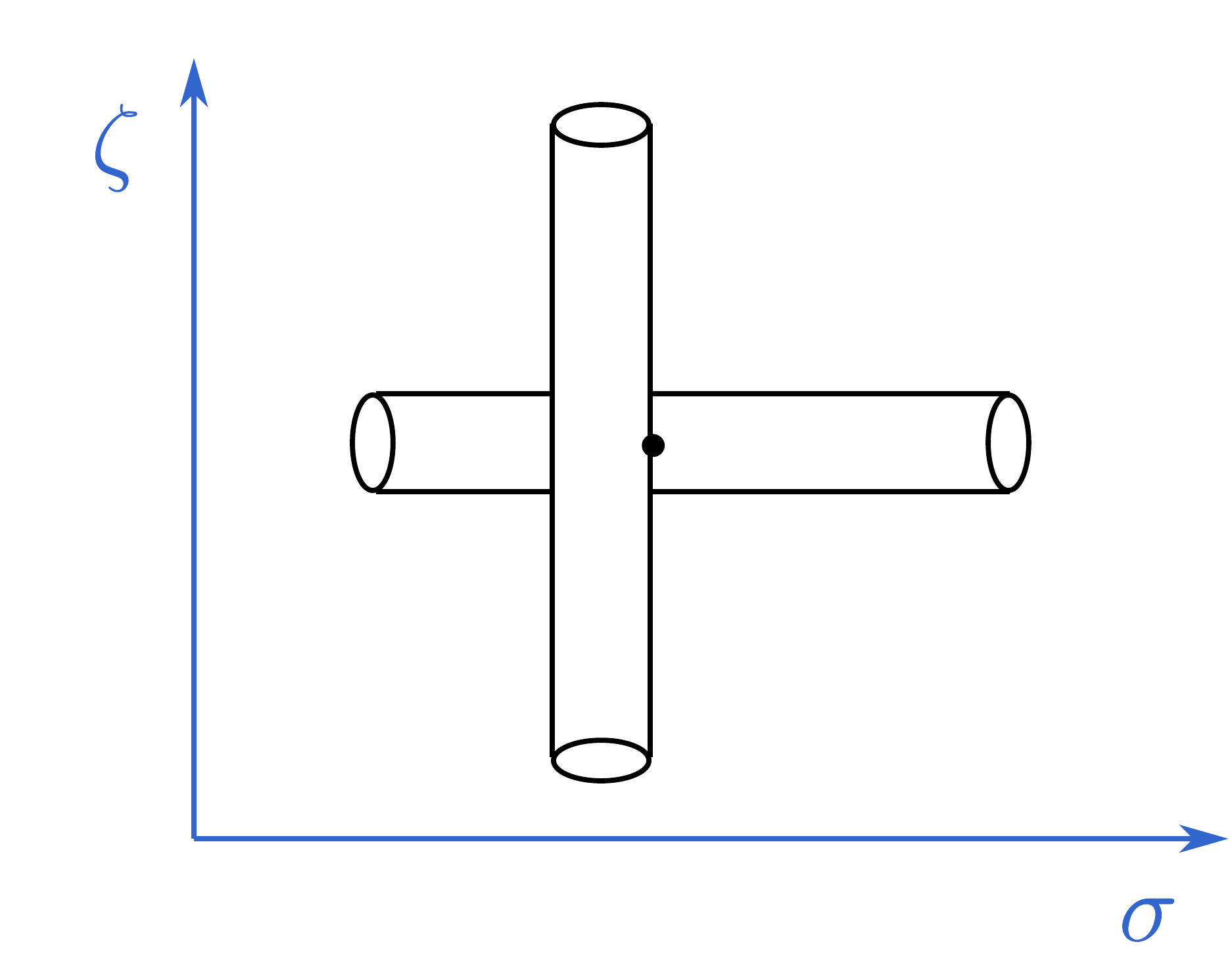}
        \caption{Theory on $\IR^2\times S^1$ with $\mu=1$.}
        \label{fig:SQED-vacua-singular-circle}
    \end{subfigure}%
    \hspace*{.1\textwidth}
    \begin{subfigure}[b]{0.35\textwidth}
        \centering
        \includegraphics[height=1.2in]{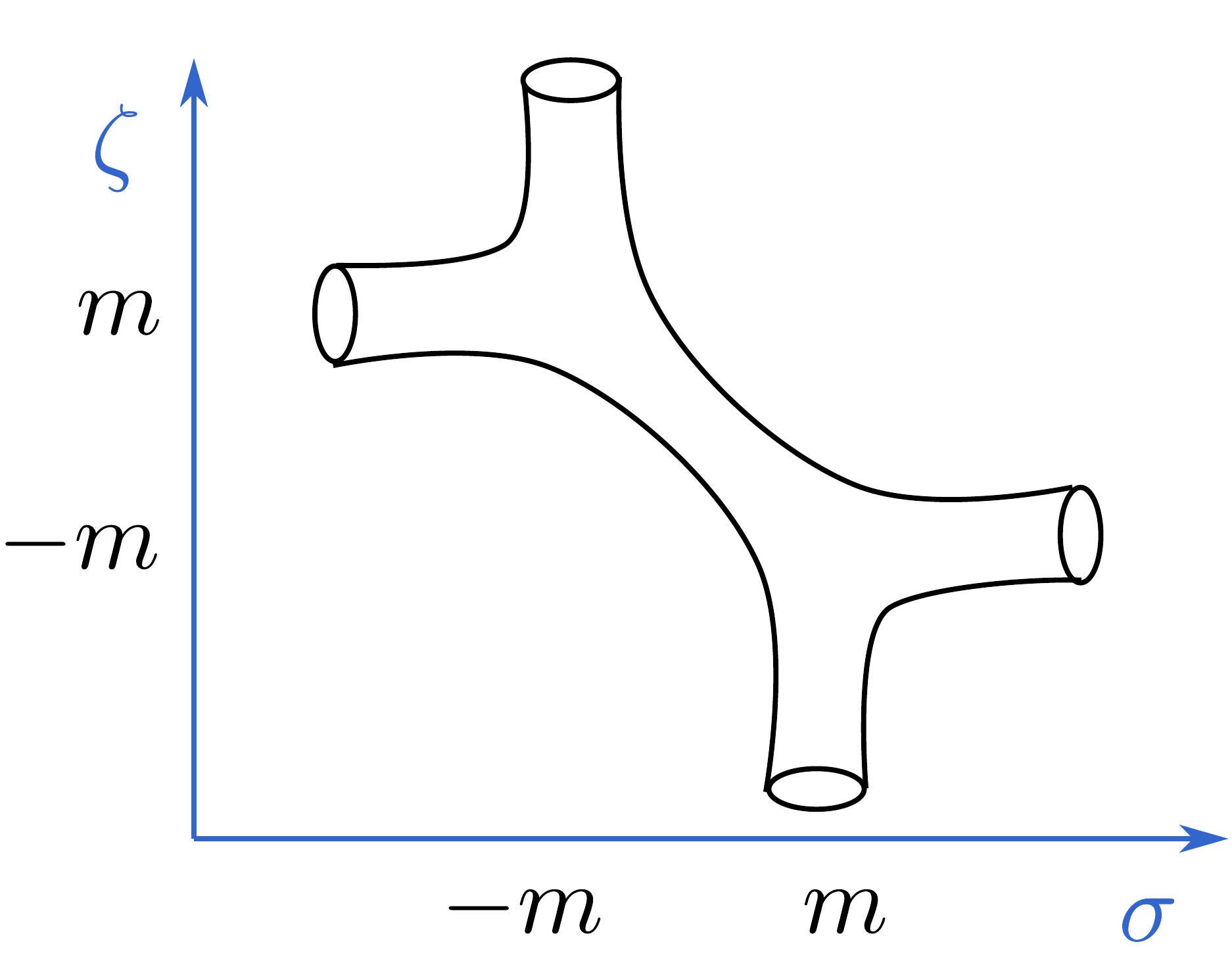}
        \caption{Theory on $\IR^2\times S^1$ with $\mu\neq 1$.}
        \label{fig:SQED-vacua-massive-circle}
    \end{subfigure}%
    \caption{Vacua of $N_f=1$ SQED on $\IR^2\times S^1$.}
    \label{fig:SQED-vacua-circle}
\end{center}
\end{figure}

When the theory is compactified on a circle of radius $R$, both $\sigma$ and $\zeta$ get complexified and it is convenient to introduce coordinates $(x,y)\in \IC^*\times \IC^*$, related to the original ones by $2\pi R \, \sigma \sim \Re \log y $ and $-2\pi R \zeta \sim \Re \log x$.
The partition function of this theory can be written down in the semiclassical limit by a mild generalization of formula (\ref{eq:bar-P-semiclassical-limit-bis})
\be\label{eq:SQED-semiclassical}
\begin{split}
	&Z(x,\mu,\hbar) \sim \\
	&\qquad\int \frac{dy}{y} \, 
	\exp \left[  \frac{1}{2\hbar} \(  \Li_2(\mu y)  + \Li_2(\mu y^{-1}) + \log(-x) \, \log y  +  \frac{1}{2} (\log y)^2 + O(\hbar) \) \right]\,. \\
\end{split}
\ee
The vacuum manifold is then 
\be\label{eq:SQED-vacua-circle}
	\mu x - x y +\mu y - 1 = 0\,.
\ee 
This is a sphere with four punctures at positions 
\be
	(x,y) \in \{  
	(0,\mu^{-1}),\ 
	(\mu^{-1}, 0),\
	(\mu, \infty),\ 
	(\infty,\mu)
	\}\,.
\ee
Noting that these position correspond exactly to the asymptotics of the vacua on $\IR^3$ in Figure \ref{fig:SQED-vacua-massive}, we deduce that the moduli spaces now has the form shown in Figure \ref{fig:SQED-vacua-massive-circle}. 
If we set $\mu=1$, the curve factorizes into two copies of $\IC^*$
 touching at the point $x=y=1$  as shown in Figure \ref{fig:SQED-vacua-singular-circle}
 \be
 	(y-1)(x-1)=0\,.
 \ee
 In the compactification from $\IR^3$ to $\IR^2\times S^1$ the asymptotics of $\sigma$ and $\zeta$ just gain a circle, but deep inside the moduli space nontrivial corrections take place. In 3d $\CN=2$ language these come from vortices wrapping the $S^1$, and they are responsible for smoothing out the~trivalent junctions of Figure \ref{fig:SQED-vacua-massive} into the smooth curve in Figure \ref{fig:SQED-vacua-massive-circle}.

\subsubsection{SQED and the theory on the unknot conormal}
The resemblance of the moduli space of vacua of the theory on a circle and the mirror curve of the resolved conifold has a simple explanation.
SQED is the worldvolume theory~$T[L]$ that arises on the toric brane $L$ in the conifold \cite{Hanany:1997vm, Aganagic:2001nx, Dimofte:2010tz}.
Incidentally, this brane \emph{essentially} coincides with the unknot conormal $L_{0_1}$ \cite{Ooguri:1999bv} and the mirror curve of $L_{0_1}$ is the augmentation curve of the unknot \cite{Aganagic:2013jpa}.
The BPS vortices of $T[L]$ descend from M2-branes wrapping holomorphic curves with boundary on $L$, resulting in the equality of the~open Gromov-Witten partition function on $L$ and the K-theoretic vortex partition function of~$T[L]$~\cite{Dimofte:2010tz}.

However, the unknot theory and the theory on the toric brane in the conifold (SQED) are almost the same, but \emph{not quite the same}. To be precise, let us compare (\ref{eq:SQED-semiclassical}) with the~twisted effective superpotential for $T[L_{0_1}]$ in \cite[eq. (5.26)]{Ekholm:2018eee}. Here we report it with $t=-1$ (to work in the unrefined case), use standard identities for dilogarithms \cite{Zagier:2007knq}, and neglect constant terms
\be
\begin{split}\label{Twisted superpotential}
	\widetilde{\mathcal{W}}_{T[L_{0_{1}}]} 
	& =\textrm{Li}_{2}\left(y\right)-\textrm{Li}_{2}\left(a^{2}y\right)+\textrm{Li}_{2}\left(a^{2}\right)+\log x\log y \\
	& =\textrm{Li}_{2}\left(y\right) + \textrm{Li}_{2}\left(a^{-2}y^{-1}\right) + \frac{1}{2} \log(-a^2 y)^{2} +\textrm{Li}_{2}\left(a^{2}\right)+\log x\log y\,.
\end{split}
\ee
Performing a~rescaling of variables $y\to\mu y$, $x\to \mu x$ and identifying $a=\mu^{-1}$ gives a~theory with matter content defined by three dilogarithms: $\Li_2(\mu y^{\pm1})$ and $\Li_2(\mu^{-2})$. 
While the~first two coincide with terms from SQED (\ref{eq:SQED-semiclassical}), the last term is an extra gauge-neutral particle with axial charge $-2$.
This particle is better reinterpreted through the identity $\textrm{Li}_{2} \left(\mu^{-2}\right)+ \frac{1}{2} \log(-\mu^{-2} )^{2}  = -\Li_2(\mu^2)$
where the minus sign, and the fact that it is gauge-neutral, suggest that we view this as a particle in a \emph{dual} theory. 
Indeed SQED theory is dual to the~XYZ~model, a~theory of three free chirals \cite{Aharony:1997bx}. One of them is the meson $\pi=u\tilde u$ which is gauge-neutral and has axial charge +2 (like the new dilogarithm). 
The other two are the monopole operators, which appear in the Gromov-Witten disk potential of the unknot (see \cite[eq. (5.30)]{Ekholm:2018eee}).

To summarize, SQED {differs} from the unknot theory: the latter features an extra neutral particle with axial charge $-2$. In the context of SQED, this particle is `swapped' into the dual XYZ model where it is identified with the meson of SQED.
This subtle difference does not affect the moduli space of vacua since the particle carries neither gauge charge nor topological charge, only the axial charge. For this reason, the moduli space of SQED coincides with that of the unknot theory. 
This is an example of {two different theories with the same moduli space of vacua}. 
Geometrically, the dilogarithm $\Li_2(a^2)$ may be interpreted as arising in the semiclassical limit from the net contribution of two multi-covers of the sphere with single units of 4-chain intersection of opposite signs (that is: replacing $a$ with $q^{\pm 1} a^2$ in (\ref{eq:fpformula}), taking the ratio, and putting $g_s\to 0$).

\subsubsection{General moduli spaces of vacua}
To conclude, let us remark on how this picture generalizes to theories of type $T[Q]$.
In fact, the SQED theory we just analyzed \emph{is} of type $T[Q]$ since it corresponds to the~unknot~\cite{Ekholm:2018eee}.
The quiver adjacency matrix in this case is
\be
	C = \(\begin{array}{cc} 0 & 0 \\ 0 & 1 \end{array}\) \,.
\ee
The moduli space of vacua of this theory is determined by the \emph{quiver $A$-polynomials} introduced in \cite{Ekholm:2018eee} (see also \cite{PSS1802,Panfil:2018faz,Smo2017}). For the matrix $C$ they are given by
\be
	A_1(\mathbf{x},\mathbf{y}) = 1 - y_1 - x_1 = 0\,,
	\qquad
	A_2(\mathbf{x},\mathbf{y}) = 1 - y_2 + x_2 y_2 = 0\,,
\ee
see \eqref{eq:quiver-A-poly}. Together with the identification of variables
\be\label{eq:KQ-vars-SQED}
	x_1 = \mu x\,,
	\qquad 
	x_2 = \mu^{-1} x\,,
	\qquad
	y_1 y_2 = \mu y\,,
\ee
they reproduce (\ref{eq:SQED-vacua-circle}). 

This brings us to another general fact about theories of type $T[Q]$: if we did not enforce the specialization of variables (\ref{eq:KQ-vars-SQED}), the moduli space of vacua would be 2-complex-dimensional, hence a complex surface rather than a complex curve.
The extra dimension is hiding in $m\sim \log \mu$ in Figure \ref{fig:SQED-vacua-circle}. In other words, the~full quantum moduli space of the~theory~$T[Q]$ would be the~total space of the~fibration of the~augmentation curve over the~complex parameter space with local coordinates $(x,\mu)\sim (x_1, x_2)$.
This is a general feature of quiver-type theories: the quantum moduli spaces of vacua of $T[Q]$ on $\IR^2\times S^1$ is an $m$-dimensional algebraic variety
\be\label{eq:TQ-moduli-space}
	\CM_{Q} := \{A^Q_i(\mathbf{x},\mathbf{y})  = 0\,, \ \ 1\leq i\leq m\}\quad  \subset\quad  \prod_{i=1}^m \IC^*_{x_i}\times \IC^*_{y_i}
\ee 
defined by the quiver A-polynomials 
\be\label{eq:quiver-A-poly}
	A^Q_i(\mathbf{x},\mathbf{y}) = 1 - y_i - (-1)^{C_{ii}} x_i \prod_{j=1}^{m} y_j^{C_{ij}} \,.
\ee
The variety $\CM_Q$ is middle-dimensional and Lagrangian with respect to the standard symplectic form on the $2m$-dimensional algebraic torus. 
In fact it is a higher-dimensional analogue of the augmentation variety (or its specialization, the A-polynomial). 
In the~context of the~KQ correspondence, or its generalization introduced in Section~\ref{sec:general-oGW-Q}, the latter would be recovered by imposing $m-1$ relations among the $x_i$ variables.

\subsection{3d $\CN=2$ multi-cover skein dualities}\label{sec:3dN2-skein-dualities}

In Section \ref{sec:skein-rel} we presented a new class of dualities among quivers.
The basic operation consists of modifying $Q$ by removing a link between two nodes and adding a new node linked in a particular way to others to obtain a new quiver $Q'$. With a suitable identification between parameters $x_i$ and $x_i'$ we then found that the partition functions of $Q$ and $Q'$ exactly match.
Due to the vortex interpretation of quiver partition functions (\ref{eq:vortex-quiver}), this duality can be translated into the language of 3d $\CN=2$ quiver type theories on $\IR^2\times_q S^1$.
This leads us to conjecture an infrared duality between the following theories:

\paragraph{Theory $T[Q]$:}
This is a theory of quiver type defined by a quiver $Q$ with $m$ nodes. The~gauge group is
\be
	G_{\rm{gauge}}^{(Q)} = U(1)_{\gauge,1}\times \dots\times U(1)_{\gauge,m}\,,
\ee 
with mixed gauge Chern-Simons couplings fixed by the quiver adjacency matrix $C_{ij}$, as in~(\ref{eq:Quiver-CS}). The mass deformations of this theory consist entirely of FI couplings $x_1,\dots, x_m$, or twisted masses for the topological symmetry group
\be
	G_{\rm{top}}^{(Q)} = U(1)_{\top,1}\times \dots\times U(1)_{\top,m}\,.
\ee

\paragraph{Theory $T[Q']$:}
This is a theory of quiver type defined by a quiver $Q'$ with $m+1$ nodes. The gauge group is
\be
	G_{\rm{gauge}}^{(Q')} = U(1)_{\gauge,1}\times \dots\times U(1)_{\gauge,m+1}\,,
\ee 
with mixed gauge Chern-Simons couplings fixed by the quiver adjacency matrix $C'_{ij}$.
The quiver $Q'$ is related to $Q$ by deletion of a link between nodes $a$ and $b$. 
Therefore $C'_{ab}=C'_{ba} = C_{ab} - 1$, while  $C'_{ij} = C_{ij}$ for all other $(i,j)\neq (a,b),(b,a)$ and $i,j\leq m$. 
In addition, $C'_{ij}$ also encodes mixed gauge Chern-Simons couplings for the new gauge group, labeled by $i=m+1$.
Its mass deformations consist entirely of FI couplings $x'_1\dots x'_{m+1}$.
This theory also has a~monopole potential 
\be\label{eq:monopole-potential}
	W_{Q'} = \fm_{m+1} \fm_{a} \fm_{b} \,,
\ee 
where $\fm_i$ are monopole operators with charges 
\be
	\begin{array}{c|ccc}
	& \fm_{a} & \fm_{b} & \fm_{m+1}  \\
	\hline
	U(1)_{\gauge,i} & 0 & 0 & 0 \\
	\hline
	U(1)_{\top,a} & -1 & 0 & 0 \\
	U(1)_{\top,b} & 0 & -1 & 0 \\
	U(1)_{\top,m+1} & 0 & 0 & 1 \\
	U(1)_{\top,i\neq a,b,m+1} & 0 & 0 & 0 \\
	\end{array}
\ee

\paragraph{Evidence:}

The monopole potential (\ref{eq:monopole-potential}) breaks the topological symmetry group of theory $T[Q']$, reducing its rank by one
\be
	G_{\rm{top}}^{(Q')} = U(1)_{\top,1}\times \dots\times U(1)_{\top,m}\,.
\ee
In fact the potential enforces
\be\label{eq:FI-new}
	x'_{m+1} = q^{-1} x'_a x'_b \,,
\ee
and we claim that the duality between $T[Q]$ and $T[Q']$ holds with (\ref{eq:FI-new}) supplemented by 
\be\label{eq:FI-other}
	x'_i = x_i \qquad i=1,\dots, m \,.
\ee
The fact that we can match global continuous symmetries of the two theories is already a~good piece of evidence for the duality. Let us mention that, although the FI couplings of the first $m$ coincide, this is generally not the case for the gauge fugacities. Later we will see examples of this.

Another piece of evidence for this duality includes the equality of K-theoretic vortex partition functions. This follows from  (\ref{eq:vortex-quiver}) and (\ref{eq:unlinking-invariance}) 
\be\label{eq:duality-vortex}
	\CZ^{\vort}_{T[Q']}(\mathbf{x}',q)
	=
	\CZ^{\vort}_{T[Q]}(\mathbf{x},q)
\ee
provided (\ref{eq:FI-new}) and (\ref{eq:FI-other}) hold.

Moreover, it follows from the semiclassical limit $\hbar \to 0$ of (\ref{eq:duality-vortex}) that $T[Q]$ and $T[Q']$ have the same quantum moduli space of vacua. 
However, the dimensions of $\CM_{Q}$ and $\CM_{Q'}$, defined as in (\ref{eq:TQ-moduli-space}), do not seem to match: $\dim_{\IC}\CM_{Q'} = \dim_{\IC} \CM_{Q} +1$. 
The equation we need to supply is a relation for the gauge fugacities. Motivated by the geometric interpretation in terms of holomorphic disks, we supply in fact two equations:
\be\label{eq:gauge-fugacities}
	y_a = y_a' y_{m+1}' \,, \qquad y_b = y_b' y_{m+1}'\,.
\ee
With these, the algebraic varieties are equivalent:
\be
	\CM_{Q} \simeq \CM_{Q'}\,.
\ee

The geometric interpretation of (\ref{eq:gauge-fugacities}) is rather simple: when the multi-cover skein relation in Figure \ref{fig:skein-relation-2-3} is applied, the meridian holonomy of basic disks $D_a$ and $D_b$ is broken up into that of the unlinked disks $D'_{a},D'_{b}$ plus that of the (now basic) boundstate $D'_{m+1}$.
This counting is based on the interpretation of meridian holonomies as the effective result of the infinite towers of multi-coverings of basic disks \cite{Ooguri:1999bv}. 
The multi-cover skein relation reorganizes these towers and these changes of variables simply follow.
On the other hand, algebraically imposing these two equations is nontrivial, since it potentially overconstrains the problem. Their consistency is predicted by the geometric picture, below it will be verified in some examples. 
In order to describe these operations at the level of moduli spaces appropriately, we need a framework of Lagrangian correspondences, which will be reviewed in Section \ref{sec:quantum-gluing}.

\subsection{A basic example: pentagon duality}\label{sec:pentagon-duality}
The fundamental multi-cover skein duality relates the quivers with adjacency matrices
\be
	C = \(\begin{array}{cc}
	0 & 1 \\ 1 & 0
	\end{array}\)
	\qquad
	C' = \(\begin{array}{ccc}
	0 & 0 & 0 \\ 0 & 0 & 0 \\ 0 & 0 & 1
	\end{array}\)\,.
\ee
The corresponding quivers are 
\be
	Q=\raisebox{-3pt}{\includegraphics[width=.01\textwidth]{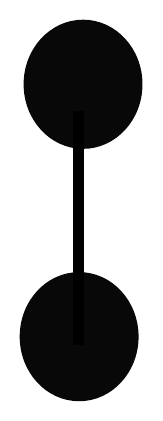}}
	\qquad 
	Q'=\raisebox{-3pt}{\includegraphics[width=.03\textwidth]{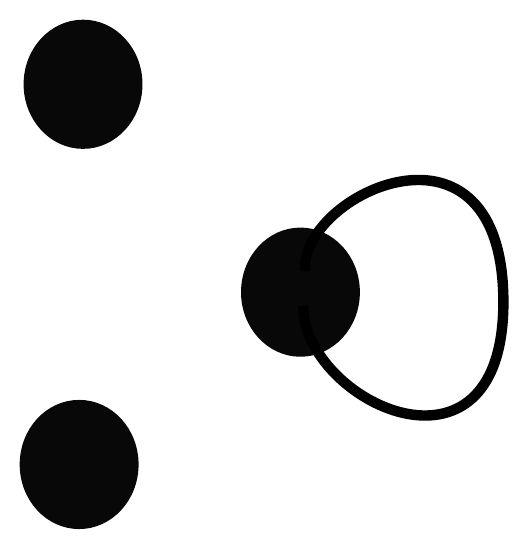}}\,.
\ee 
We proved the equality of the partition functions of $Q$ and $Q'$ in Section \ref{sec:simpleunlink}.
Here we discuss the description of the corresponding 3d $\CN=2$ theories, and the corresponding semi-classical picture.

\paragraph{Theory $T[Q]$:}
The gauge group is
\be
	G_{\rm{gauge}}^{(Q)} = U(1)_{\gauge,1}\times U(1)_{\gauge,2}\,,
\ee 
with effective mixed gauge Chern-Simons coupling $C_{12}=1$. 
Both FI couplings  $x_1, x_2$ are turned on, the corresponding topological symmetry group is
\be
	G_{\rm{top}}^{(Q)} = U(1)_{\top,1}\times U(1)_{\top,2}\,.
\ee
The semiclassical limit of the K theoretic vortex partition function is
\be\label{eq:2-node-partition-function}
\begin{split}
	& Z_Q(x_1,x_2,\hbar) \sim 
	\int \frac{dy_1}{y_1}\, \frac{dy_2}{y_2} \, e^{\frac{1}{2\hbar} \tCW_Q} \,,\\
	& \tCW_Q = 
	\Li_2(y_1)  + \Li_2(y_2) + \log x_1 \, \log y_1 + \log x_2 \, \log y_2  +  \log y_1 \log y_2 + O(\hbar)\,. 
\end{split}
\ee
The vacuum manifold of this theory is
\be\label{eq:2-node-vacua}
	\CM_Q:\{ 1-y_1 -x_1 y_2 = 0\, , \,\,1-y_2 -x_2 y_1 = 0\} \subset (\IC^*\times \IC^*)^2\,.
\ee

\paragraph{Theory $T[Q']$:}
The gauge group is
\be
	G_{\rm{gauge}}^{(Q')} = U(1)'_{\gauge,1}\times U(1)'_{\gauge,2} \times U(1)'_{\gauge,3}\,,
\ee 
the only nonzero effective gauge Chern-Simons coupling is $C'_{33}=1$.
The FI couplings of this theory are $x'_1,x'_2,x'_3$.
Finally, this theory has a monopole potential 
\be\label{eq:monopole-potential-2}
	W_{Q'} = \fm_{1} \fm_{2} \fm_{3} \,,
\ee 
where $\fm_i$ are monopole operators with charges 
\be
	\begin{array}{c|ccc}
	& \fm_{1} & \fm_{2} & \fm_{3}  \\
	\hline
	U(1)'_{\gauge,i} & 0 & 0 & 0 \\
	\hline
	U(1)'_{\top,1} & -1 & 0 & 0 \\
	U(1)'_{\top,2} & 0 & -1 & 0 \\
	U(1)'_{\top,3} & 0 & 0 & 1 \\
	\end{array}
\ee
This potential enforces $x'_3 = x'_1 x'_2$ at the classical level.
Taking this into account, the~semiclassical limit of the K theoretic vortex partition function is
\be\label{eq:3-node-partition-function}
\begin{split}
	& Z_{Q'}(x'_1,x'_2,\hbar) \sim 
	\int \frac{dy'_1}{y'_1}\, \frac{dy'_2}{y'_2}\, \frac{dy'_3}{y'_3} \, e^{\frac{1}{2\hbar} \tCW_{Q'}} \,,\\
	& \tCW_{Q'} = 
	\Li_2(y'_1)  + \Li_2(y'_2) + \Li_2(y'_3) + \frac{1}{2} \( \log y'_3\)^2  \\
	& \phantom{\tCW_{Q'}  } + \log x'_1 \, \log y'_1 + \log x'_2 \, \log y'_2+ \log(- x'_1x'_2) \, \log y'_3  + O(\hbar) \,.
\end{split}
\ee
The vacuum manifold of this theory is
\be
	\CM_{Q'}:\{ 1-y'_1 -x'_1  = 0\, ,\,\, 1-y'_2 -x'_2 = 0\, , \,\, 1-y'_3 + x'_1x'_2\, y'_3 = 0\} \subset (\IC^*\times \IC^*)^3\,.
\ee
Let us check the equivalence of the vacuum manifolds.
Solving for $y'_3$ gives $y_3' = (1-x_1'x_2')^{-1}$. Then we use the map (\ref{eq:FI-other}) to set $x_1'=x_1, x_2'=x_2$.
Next we solve for $y_1', y_2'$ and use the~map (\ref{eq:gauge-fugacities}) to obtain 
\be
	y_1 = y_1'y_3' = \frac{1-x_1}{1-x_1 x_2}\,,
	\qquad
	y_2 = y_2'y_3' = \frac{1-x_2}{1-x_1 x_2}\,.
\ee
It can be easily checked that this agrees with the description of $\CM_Q$ in (\ref{eq:2-node-vacua}).

\subsection{Relation to other known dualities}\label{sec:relation-known-dualities}
In general, multi-cover skein dualities of 3d $\CN=2$ theories {appear to} give new relations.
However, in special cases, multi-cover skein dualities coincide with known dualities of 3d $\CN=2$ theories. One example is the SQED-XYZ `mirror symmetry'.

\subsubsection{Pentagon duality and SQED -- XYZ mirror symmetry}
Let us consider the pentagon duality illustrated above.
We start from Theory $T[Q]$:
taking the saddle point with respect to $y_1$ in \eqref{eq:2-node-partition-function} localizes the integral to $y_1 = {1-x_1 y_2}$:
\be
\begin{split}
	& 
	\int \frac{dy_2}{y_2}  
	e^{\frac{1}{2\hbar}
	\(
	\Li_2(1- x_1 y_2)+ \Li_2(y_2) 
	+ \log (1-x_1 y_2)\log y_2 + \log x_1 \log (1-x_1 y_2) + \log x_2 \log y_2
	\)}
	\\
	& \qquad=
	e^{\frac{1}{2\hbar}
	\(
	+\frac{1}{2} [\log(-x_1)]^2 + \log x_2 \log x_1^{-1/2}
	\)} \\
	&\qquad\qquad \times \int \frac{dy}{y}  
	e^{\frac{1}{2\hbar}
	\(
	\Li_2(x_1^{-1/2} y^{-1})+ \Li_2(x_1^{-1/2}y) 
	+ \log y\, \log (-x_1^{1/2} x_2) + \frac{1}{2} (\log y)^2 
	\)}\,,
\end{split}
\ee
where we introduced the effective (or shifted) gauge fugacity $y = y_2  x_1^{1/2}$, and used standard dilogarithm identities. The resulting integrand is exactly that of SQED theory. If we identify $(y, x_1^{-1/2}, x_2 \, x_1^{1/2} )$ as the fugacities of $U(1)_{\gauge} \times U(1)_{\axial} \times U(1)_{\top}$,  this integral coincides precisely with  \eqref{eq:SQED-semiclassical}.

Next we can check what happens on the other side. In Theory $T[Q']$ we can directly perform the integrals \eqref{eq:3-node-partition-function} to get
\be
	 \exp\left[\frac{1}{2\hbar}\(
	\Li_2(x_1^{-1})+\Li_2(x_2^{-1})+\Li_2(x_1 x_2)
	+ \dots
	\)\right]\,,  
\ee
where ellipses refer to usual squares of logarithms, which can be computed using standard identities for $\Li_2$.
This signals the presence of three chirals with the following charges 
\be
	\begin{array}{c|cc}
		& U(1)_{\axial} & U(1)_{\top} \\
		\hline
		\pi & 2 & 0 \\
		\fm_{-} & -1 & -1 \\
		\fm_+ & -1 & 1
	\end{array}
\ee
These correspond to the meson and the two monopole operators in the XYZ dual description of SQED (compare, for example, charge assignments with those in \cite[Section 3]{Dimofte:2011ju}). 
Therefore after integrating out some of the gauge fugacities, the pentagon multi-cover skein duality is related to SQED-XYZ mirror symmetry.

\subsubsection{Beyond the pentagon}
So far we have focused on a single example of multi-cover skein duality: the pentagon.
General multi-cover skein dualities are harder to describe in terms of known 3d $\CN=2$ dualities, in particular since they generate infinite sets of dual theories. 

\begin{figure}[h!]
\begin{center}
\includegraphics[width=0.95\textwidth]{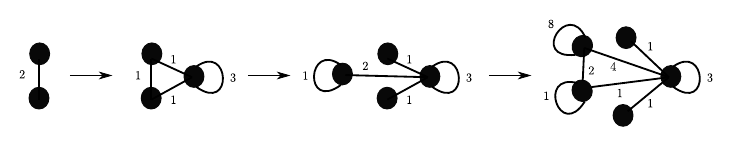}
\caption{Chain of dualities obtained by unlinking. Numbers next to edges denote multiplicity.}
\label{fig:2-node-2-link-dualities}
\end{center}
\end{figure}

As an example, consider the quiver with two nodes and two pairs of arrows:
\be
	Q = \raisebox{-1pt}{\includegraphics[width=0.05\textwidth]{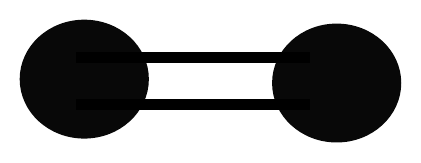}}\,,
	\qquad 
	C = \(\begin{array}{cc}
	0 & 2 \\ 2 & 0
	\end{array}\)\,.
\ee
The gauge group is once again
\be
	G_{\rm{gauge}}^{(Q)} = U(1)_{\gauge,1}\times U(1)_{\gauge,2}\,,
\ee 
however now the effective mixed gauge Chern-Simons coupling is $C_{12}=2$. 
Both FI couplings  $x_1, x_2$ are turned on, the corresponding topological symmetry group is
\be
	G_{\rm{top}}^{(Q)} = U(1)_{\top,1}\times U(1)_{\top,2}\,.
\ee
The semiclassical limit of the K-theoretic vortex partition function is
\be\label{eq:2-node-theory}
\begin{split}
	& Z_Q(x_1,x_2,\hbar) \sim 
	\int \frac{dy_1}{y_1}\, \frac{dy_2}{y_2} \, e^{\frac{1}{2\hbar} \tCW_Q} \,,\\
	& \tCW_Q = 
	\Li_2(y_1)  + \Li_2(y_2) + \log x_1 \, \log y_1 + \log x_2 \, \log y_2  + 2 \log y_1 \log y_2 + O(\hbar)\,, 
\end{split}
\ee
and the vacuum manifold of the theory is
\be\label{eq:2-node-2-link-vacua}
	\CM_Q:\{ 1-y_1 -x_1 y_2^2 = 0\, , \,\, 1-y_2 -x_2 y_1^2 = 0\} \subset (\IC^*\times \IC^*)^2\,.
\ee
Applying multi-cover quiver skein dualities successively we obtain more and more complicated theories. The first few in the family are shown in Figure \ref{fig:2-node-2-link-dualities}. 

As we continue with this operation the gauge theory description becomes more complicated, involving larger gauge groups, more matter fields, more Chern-Simons couplings, and more monopole potential terms. 
The dualities produced by multi-cover skein relations can be quite nontrivial: given one of these more complicated theories, it would be very hard to guess that it admits a simple dual such as (\ref{eq:2-node-theory}). 
{
It would also be important to determine whether this chain of dualities, and more generally multi-cover skein dualities, can be obtained by combination of known 3d $\CN=2$ dualities, such as described in \cite{Dimofte:2011ju}. 
}

\section{Operator-valued partition functions, wall-crossing, and multi-cover skein relations}\label{sec:wall-crossing}

We have shown above that there is a whole family of quivers associated to a knot. They are generated by creation and destruction of quiver links accompanied by addition of suitable nodes. 
From the viewpoint of counts of holomorphic curves each node corresponds to a basic disk. The multi-cover skein relation induces a change in the set of basic disks that generate the BPS spectrum without changing the partition function which counts all generalized holomorphic curves with boundary on $L_K$.

In this section we will give a more quantitative description of the change in the set of basis disks 
at the level of rather explicit formulas for partition functions that make it more manifest which holomorphic curves are basic, which ones are boundstates, which ones are multi-covers, etc.
For this purpose we will introduce an appropriate formalism which  leads to interesting connections to work on wall-crossing by Kontsevich and Soibelman \cite{Kontsevich:2008fj}.

\subsection{Quantum torus algebra}\label{sec:QTA}

The partition function of quiver representations obeys functional identities associated to \emph{quantum quiver A-polynomials}
\be	
	\hat A_{i}(\mathbf{x}, \mathbf{y}) P^{Q}(\mathbf{x},q) = 0 \,.
\ee
These arise as straightforward quantizations of the classical quiver A-polynomials as explained in \cite{Ekholm:2018eee}.\footnote{We have been informed by H.~Larraguivel, D.~Noshchenko, M. Panfil, and P.~Sulkowski that quantum quiver A-polynomials have been independently obtained in their upcoming work which focuses on the~topological recursion.} More precisely, if $C$ is the adjacency matrix of the quiver, the general formula for its quantum quiver A-polynomial reads  
\be\label{eq:quantum-Ai}
	\hat A_{i}(\mathbf{x}, \mathbf{y}) = 1 - \hat y_i -  \hat x_i (-q\hat y_i)^{C_{ii}}\prod_{j\neq i} \hat y_j^{C_{ij}} \,.
\ee
The operators $\hat x_i $ and $\hat y_i$ are defined by
\be
\begin{split}
	\hat x_i  f(x_1,\ldots, x_m,y_1,\ldots, y_m) &= x_i \, f(x_1,\ldots, x_m,y_1,\ldots, y_m)\,,\\
	\hat y_i  f(x_1,\ldots, x_m,y_1,\ldots, y_m) &= f(x_1,\ldots,q^2 x_i,\ldots, x_m,y_1,\ldots, y_m)\,,\\
\end{split}
\ee
They generate a quantum torus algebra
\be\label{eq:QTA}
\begin{split}
	\hat x_i \hat x_j = \hat x_j \hat x_i\,,
	&\qquad
	\hat y_i \hat y_j = \hat y_j \hat y_i\,,	
	\\
	\hat y_i  \hat x_j &= q^{2\delta_{i,j}}\ \hat x_j  \hat y_i\,.
\end{split}
\ee

For knot conormals $L_{K}$ this is the algebra that arises by deformation-quantization on the moduli space of flat abelian connections on $L_K\setminus \left.\{L_{i}\}\right._{i=1,...,m}$, 
the knot conormal where we excise the tubular neighborhood $L_i$ of the boundary of each basic disk. 
In the~semiclassical limit ($q\to 1$) $\hat x_i$ and $\hat y_i$ tend to longitude and meridian on $T_i^2=\partial L_i$. This is also consistent with the identification $y_i\sim q^{2 d_i}$ in the semiclassical limit of $P^{Q}(\mathbf{x},q) $, see~\cite{Ekholm:2018eee}.

\subsection{Assembling a quiver}\label{sec:assembly}
In this section we explain how to write the partition function of any symmetric quiver as a simple product in non-commutative variables in the quantum torus algebra. 
Consider a~symmetric quiver $Q$ and suppose that we wish to add to it a new node labeled by $0$ to obtain another quiver $Q'$. Let~$\ell$ be the~number of loops on the new node and let $v_i$ be the number of links between the zeroth node and the $i$-th node of $Q$. 
The quiver partition function changes as follows 
\be
\begin{split}
	P^{Q'}(x_0,\mathbf{x},q) 
	& = \sum_{d_0,\mathbf{d}} (-q)^{\mathbf{d}\cdot C\cdot \mathbf{d} + 2 d_0\,  \mathbf{v}\cdot \mathbf{d} +\ell \, {d_0}^2} \frac{\mathbf{x}^{\mathbf{d}} x_0^{d_0}}{(q^2;q^2)_{\mathbf{d}} (q^2;q^2)_{d_0}}
	\\
	& = 
	\sum_{d_0} (-q)^{\ell\, {d_0}^2} \frac{x_0^{d_0}}{ (q^2;q^2)_{d_0}}
	\sum_{\mathbf{d}}(-q)^{\mathbf{d}\cdot C\cdot \mathbf{d} } \ \frac{(-q)^{2 d_0\, \mathbf{v}\cdot \mathbf{d} } \mathbf{x}^{\mathbf{d}} }{(q^2;q^2)_{\mathbf{d}}}
	\\
	& = \left[\sum_{d_0\geq 0}  \frac{(-q)^{\ell\, {d_0}^2}}{ (q^2;q^2)_{d_0}}
	\ \hat x_0^{d_0}\ 
	\Big(\prod_i {\hat y_i}{}^{v_i} \Big)^{d_0}\right] P^{Q}(\mathbf{x},q) 
\end{split}
\ee
where $\mathbf{d}\cdot C\cdot \mathbf{d}=\sum_{i,j}C_{ij}d_i d_j,\;$ $\mathbf{v}\cdot \mathbf{d}=\sum_{i}v_i d_i,\;$ $\mathbf{x}^{\mathbf{d}}=\prod_{i}x_{i}^{d_{i}}$, and $(q^2;q^2)_{\mathbf{d}}= \prod_{i}(q^2;q^2)_{d_{i}}$.

Now notice that  
\be\label{eq:self-framing}
	(\hat x_0 \hat y_0^k)^n = \hat x_0^n \hat y_0^{nk} q^{(n^2-n)k}
\ee
(see Section \ref{sec:selflinkfactor} for a geometric interpretation) and recall the definition of the  quantum dilogarithm: 
\be
	\Psi_{q}(\xi) := 
	\sum_{n=0} \frac{q^{n}}{(q^2,q^2)_n} \xi^n \,.
\ee
Then the addition of a node to the quiver (as above) corresponds to the~action of the~following $q$-difference operator:
\be
	P^{Q'}(x_0,\mathbf{x},q)  = 
	\Psi_q\left((-1)^{\ell} \, q^{\ell-1} \, \hat x_0 \, \hat y_0^{\ell} \prod_{i} \hat y_{i}^{v_{i}}\right)
	P^{Q}(\mathbf{x},q)  \,.
\ee

By iteration, one may construct the partition function of any quiver in this way, starting from the empty quiver $Q=\varnothing$ with $P^{\varnothing}(q)=1$, and adding all nodes with appropriate linking data successively. 
If we define\footnote{Note that this is similar, but not identical, to the expressions appearing in the $\hat A_i$.}
\be\label{eq:X-vars}
	X_i = (-1)^{C_{ii}} \, q^{C_{ii}-1} \, \hat x_i \, \hat y_i^{C_{ii}} \,\prod_{j<i} {\hat y_j}{\ }^{C_{ij}} 
\ee
for $i=1,\dots,m$, we get the following compact expression for the quiver partition function:
\be\label{eq:P-finite-product}
	\boxed{
	\IP^{Q} = 
	\Psi_q( X_m )\cdot \Psi_q( X_{m-1} )\ \cdot\ \ldots\ \cdot \ \Psi_q( X_{1} )\,.
	}
\ee
More precisely, $\IP^Q$ is an operator in the quantum torus algebra that encodes the quiver partition function.  There is one quantum dilogarithm $\Psi_q$ for each node of the quiver, and the variables $X_i$ are non-commutative. In fact
\be
	X_i X_j = q^{2 A_{ij}} X_j X_i\,,
\ee
where $A_{ij}$ is a skew-symmetric matrix
\be\label{eq:A-from-C}
	A_{ij} = \left\{\begin{array}{lr}
	C_{ij} \qquad & (i>j) \\
	0 \qquad & (i=j)\\
	-C_{ij}\qquad & (i<j) 
	\end{array}
	\right.\,.
\ee
We point out that we made a choice of ordering of the quiver nodes. Different orderings give different definitions of $X_i$ as well as different presentations \eqref{eq:P-finite-product}. However, they are all equivalent and involve the same number of dilogarithms equal to the number of nodes in $Q$.

\subsubsection{Normal ordering}
The quiver partition function written as in (\ref{eq:P-finite-product}) stands in striking contrast to the factorization that defines the motivic DT invariants (\ref{P^Q=Exp}). In both cases the partition function is a product of $q$-Pochhammers, however in (\ref{eq:P-finite-product})  the product is \emph{finite} and all powers are equal to $-1$, whereas in (\ref{P^Q=Exp}) the factorization typically involves infinitely many nonzero motivic DT invariants.

The relation between the two can be described by a simple operation that we call the~\emph{normal ordering}. Given a formal series in $\hat{\mathbf{x}}$ and $\hat{\mathbf{y}}$, the normal ordering is defined as the operation of reordering each monomial so that all $\hat y_i$ are brought  \emph{to the right} and removed. Since $\hat y_{i}$ act as the identity on the constant function $1$, this just corresponds to the result of acting by the operator on the function $1$.

Applying the normal ordering to \eqref{eq:P-finite-product} results in a formal series that coincides by definition with $P^Q(\mathbf{x},q)$, as written in \eqref{eq:Efimov}, with factorization that yields \eqref{P^Q=Exp}.

\subsubsection{Self-linking}\label{sec:selflinkfactor}
Let us comment on the geometric interpretation of the $q$-shift induced by the addition of loops on single nodes, accounted by formula \eqref{eq:self-framing} through the quantum torus algebra. 
Loops on a node correspond to `self-linking' of the basic disk dual to that node. 
Geometrically, this can be thought of as a local kink of the disk boundary, with a compensating 4-chain intersection of the opposite sign \cite{Ekholm:2018eee}. 
For the basic disk these two give canceling powers of $q$.
However, multi-covers counted by $x_i^{d_i}$ picks up a power of $q^{n^2}$ for $d_i=n$, because with the kink the disk boundary must cross (a copy of) itself $n^2$ times, see Section~\ref{sec:general-oGW-Q}. Apart from this, the multi-cover also pick up $n$ intersections with the 4-chain. The~combination of  these effects explains the factor $q^{n^2-n}$ in \eqref{eq:self-framing}.

\subsection{Review of wall-crossing}\label{sec:WC-R}
We briefly recall the basic setup of the Kontsevich-Soibelman wall-crossing formula \cite{Kontsevich:2008fj}.
Let $\Gamma$ be a Poisson lattice endowed with a skew-symmetric integral pairing $\langle\, \cdot\, ,\, \cdot\, \rangle$.
We define the quantum torus algebra $\IC[\Gamma]$ by 
\be\label{eq:QTA-KS}
	X_{\gamma} X_{\gamma'} 
	= q^{\langle\gamma,\gamma'\rangle} X_{\gamma+\gamma'}\,.
\ee
Note that this implies $X_{\gamma} X_{\gamma'} = q^{2\langle\gamma,\gamma'\rangle} X_{\gamma'} X_{\gamma} $.

Let $Z\in {\rm Hom}(\Gamma,\IC)$ be the \emph{central charge} homomorphism that associates  $\gamma \mapsto Z_{\gamma}\in \IC $. We denote by $\CB$ the space of such homomorphisms and by $u$ a point in $\CB$. Therefore $u$~fixes a choice of $Z_{\gamma}$ for all $\gamma$, in particular if fixes the relative partial ordering of $\arg Z_{\gamma}$. 

The \emph{BPS spectrum at $u$} is encoded by a collection of Laurent polynomials $\Omega(\gamma,q,u)\in \IZ[q,q^{-1}]$. We denote by $\Omega_j(\gamma,u)$ the coefficient of $(-q)^{j}$.
Let us fix a sector $\sphericalangle$ of the unit circle and consider
\be
	\IU_{\sphericalangle}(u) : = \prod^{\curvearrowleft}_{Z_\gamma\in \sphericalangle} \Psi_q( (-1)^{j+1} q ^j  X_{\gamma} )^{\Omega_{j}(\gamma,u)} 
\ee
where the product is taken over all BPS states with charge $\gamma$ whose central charge $Z_\gamma$ has phase within this sector with increasing ordering of $\arg Z$ from right to left.

We can now state the content of the wall-crossing formula. 
Let $u_0,u_1\in \CB$ be two points connected by a smooth path $u(t)\subset\CB$, such that no $Z_\gamma$ crosses the boundary of $\sphericalangle$ if $\Omega(\gamma,u(t))\neq 0$. 
However, the phase-ordering of central charges within $\sphericalangle$ may reshuffle arbitrarily along the path.
Then
\be
	\IU_{\sphericalangle}(u_1) = \IU_{\sphericalangle}(u_0)\,.
\ee
This turns out to fix entirely $\Omega(\gamma,q,u_1)$ in terms of $\Omega(\gamma,q,u_0)$.

The most basic example of a wall-crossing formula involves a rank-two lattice $\gamma_1\IZ\oplus\gamma_2\IZ$ with $\langle\gamma_2,\gamma_1\rangle = 1$. 
Let $u_0$ correspond to $\arg Z_{\gamma_1}<\arg Z_{\gamma_2}$ and $u_1$ to the opposite ordering. Given the BPS spectrum $\Omega(\gamma_1,q,u_0)= \Omega(\gamma_2,q,u_0) =1$, the wall-crossing formula  
\be\label{eq:pentagon}
	\Psi_{q}(-X_{\gamma_2}) \Psi_{q}(-X_{\gamma_1}) = \Psi_{q}(-X_{\gamma_1}) \Psi_{q}(-X_{\gamma_1+\gamma_2}) \Psi_{q}(-X_{\gamma_2}) 
\ee
predicts the BPS spectrum at $u_1$, namely $\Omega(\gamma_1,q,u_1)= \Omega(\gamma_2,q,u_1) = \Omega(\gamma_1+\gamma_2,q,u_1) =1$, corresponding to the factorization on the right hand side.

\subsection{Wall-crossing as multi-cover skein relations: the pentagon}\label{sec:WC-S}

Let us now return to the basic example of link removal studied in detail in Section \ref{sec:simpleunlink}.
The~two equivalent quivers, related by application of skein relations, are depicted in Figure~\ref{fig:skein-simplest}. 

The first quiver consists of two nodes with one link between them. Let us assemble this quiver as explained in Section \ref{sec:assembly}. 
Variables (\ref{eq:X-vars})  and their algebra (\ref{eq:A-from-C}) in this case are:
\be\label{eq:2-node-QTA}
\begin{split}
	X_1 &= q^{-1}\hat  x_1\,,
	\qquad 
	X_2 = q^{-1} \hat x_2 \hat y_1\,,
	\\
	&\quad  X_1 X_2 = q^{-2} X_2 X_1\,.
\end{split}
\ee
By  (\ref{eq:P-finite-product}) the partition function is therefore
\be\label{eq:2-node-quantum}
	\IP^{\ \includegraphics[width=.01\textwidth]{figures/two-node-simplest.pdf}} 
	= \Psi_q(X_2) \Psi_q(X_1)\,.
\ee

The second quiver of Figure \ref{fig:skein-simplest} has three nodes, no links among them, and one loop on the third node. 
The non-commutative variables and their algebra are now
\be
\begin{split}
	X'_1 = q^{-1} \hat x_1\,,
	\qquad 
	X'_2 &= q^{-1} \hat x_2 \,,
	\qquad 
	X'_3 = - \hat x_3 \hat y_3\,,
	\\
	X'_i X'_j &= X'_j X'_i\,.
\end{split}
\ee
By (\ref{eq:P-finite-product}) the partition function can be expressed as
\be
	\IP^{\ \includegraphics[width=.03\textwidth]{figures/three-node-simplest.pdf}} 
	= \Psi_q(X_1') \Psi_q(X_3') \Psi_q(X_2')\,,
\ee
where we have reshuffled the arguments using the fact that  of $X_i'$ mutually commute.
Recall from (\ref{eq:unlinking-variables}) that $x_3 = q^{-1} x_1 x_2$. 
We can view the above partition function as the~normal-ordered version of 
\be\label{eq:3-node-quantum}
\begin{split}
	\IP^{\ \includegraphics[width=.03\textwidth]{figures/three-node-simplest.pdf}} 
	& = \Psi_q(q^{-1}\hat x_1) \Psi_q(- q^{-1} \hat x_1 \hat x_2 \hat y_1) \Psi_q(q^{-1} \hat x_2 \hat y_1) \\
	& = \Psi_q(X_1) \Psi_q(-q \, X_1 X_2) \Psi_q(X_2) \,.
\end{split}
\ee
Here we simply inserted $\hat y_1$ inside the last dilogarithm (which does nothing upon normal ordering), and traded $\hat y_3$ for $\hat y_1$ in the second factor. 
This latter modification is also allowed since upon normal ordering it provides the same $q$-power as $\hat x_3 \hat y_3$, thanks to the~simultaneous presence of $\hat x_1$.

The multi-cover quiver skein relation guarantees that 
\be\label{eq:quiver-pentagon}
	\IP^{\ \includegraphics[width=.01\textwidth]{figures/two-node-simplest.pdf}} 
	=
	\IP^{\ \includegraphics[width=.03\textwidth]{figures/three-node-simplest.pdf}} \,.
\ee
More precisely, the multi-cover quiver skein relations gives this statement at the level of representation theory of symmetric quivers. This means that one first applies normal-ordering to each side of \eqref{eq:quiver-pentagon} and after that identifies variables as in \eqref{eq:unlinking-variables}. Here, we promoted this statement to an operator identity valued in the quantum torus algebra.

Identifying $X_i = -X_{\gamma_i}$ and using $q X_{\gamma_1}X_{\gamma_2}= X_{\gamma_1+\gamma_2}$ it is clear that (\ref{eq:quiver-pentagon}) is nothing but the pentagon identity (\ref{eq:pentagon}).
This is a basic example of the following more general principle, that will be further illustrated below:
\begin{center}
\fbox{
Skein relations on symmetric quivers generate wall-crossing identities.}
\end{center}

The emergence of the wall-crossing formalism here is strongly reminiscent of another setting in which BPS states arise from holomorphic curves wrapped by M2 branes in the~context of class $\mathcal{S}$ theories \cite{Gaiotto:2009hg, Gaiotto:2012rg}.
The analogy with the present work is quite tight in some ways.
On the one hand it was pointed out by \cite{Alim:2011kw, Alim:2011ae} that quivers compute 4d $\CN=2$ BPS spectra.
It was then observed in \cite{Gabella:2017hpz} that in the~context of class $\mathcal{S}$ theories the nodes of those quivers correspond to basic holomorphic disks arising from edges of BPS~graphs. 
Boundstates of basic disks generate the whole BPS spectrum.
The counterpart of $\IP^Q$ is the Kontsevich-Soibelman invariant, or motivic spectrum generator. 
Just like the former is determined by the linking data of basic disks, it was shown in \cite{Longhi:2016wtv} that the motivic spectrum generator is likewise encoded by the linking data (more precisely the BPS graph) of the corresponding set of basic disks. 
{
In this vein, the expression (\ref{eq:P-finite-product}) for the $\IR^2\times_qS^1$ partition function is also  reminiscent of conjectural relations between motivic spectrum generators and Schur indices of 4d $\CN=2$ theories \cite{Cecotti:2010fi, Iqbal:2012xm, Cordova:2015nma}. A possible interpretation of this may be obtained via a coupled 3d-4d system such as those considered in \cite{Cecotti:2011iy, Dimofte:2013lba}.\footnote{We thank the anonymous referee for drawing this point to our attention.}
}

\subsection{Operator form of the multi-cover skein relation}\label{sec:skein-higher}
In this section we reformulate the multi-cover skein relation for quivers, see Section \ref{sec:equivalence-of-quivers}, in the operator language introduced above. 

We start with unlinking and linking. Consider a pair of disks with linking number $k$ corresponding to two nodes of a quiver $Q$ with $m-1$ nodes. Write the partition function of $Q$ in the product form as in \eqref{eq:P-finite-product}
\be
\IP^{Q}=\Psi_{q}(X_{m})\cdot\ldots\cdot\Psi_{q}(X_{4})\Psi_{q}(X_{2})\Psi_{q}(X_{1}),
\ee
with the last factors $\Psi_{q}(X_{2})\Psi_{q}(X_{1})$ corresponding to the two nodes in the pair. Perform either the unlinking multi-cover skein move ($k\to k-1$) or the linking multi-cover skein move ($k\to k+1$) on the two nodes in the pair to obtain a new quiver $Q'$. 
The partition functions of $Q$ and $Q'$ are equal, as explained in Sections \ref{sec:Proof of invariance-unlink} and \ref{sec:Proof of invariance-link}, but the factorization transforms as follows:   
\begin{align}\label{eq:multi-cover-skein-identity}
\IP^{Q} &= \Psi_{q}(X_{m})\cdot\ldots\cdot\Psi_{q}(X_{4})\Psi_{q}(X_{2})\Psi_{q}(X_{1})\\\notag
&=\Psi_{q}(X_{m})\cdot\ldots\cdot\Psi_{q}(X_{4})\Psi_{q}(X_{1}')\Psi_{q}(X_{3}')\Psi_{q}(X_{2}')=\IP^{Q'}\,.
\end{align}
In order to understand the relation between variables, let start from $\hat x_{j},\hat y_{j}$ -- the fundamental operators associated to the torus boundary of a tubular neighborhood of the boundary of the $j^{\rm th}$ disk. Then in the \emph{unlinking} case we have
\begin{align}\label{eq:chvarunlink} 
\hat x_1'=\hat x_1,\qquad  \hat   x_2'=\hat x_2,\qquad   \hat  x_3' &= q^{-1} \hat x_1 \hat x_2,\qquad
\hat y_1 = \hat y_1'\hat y_3',\qquad  \hat y_2 = \hat y_2'\hat y_3',\\\notag
X_1' =X_1 \hat y_2^{k-1},\qquad  X_3'&=-q^{2k-1} X_1 X_2  \hat y_2^{k-1},\qquad  X_2' = X_2,
\end{align}
and in the \emph{linking} case
\begin{align}\label{eq:chvarlink}
\hat x_1'=\hat x_1,\qquad \hat x_2'=\hat x_2,\qquad \hat x_3' &= \hat x_1 \hat x_2,\qquad \hat y_1 = \hat y_1' \hat y_3,\qquad 
\hat y_2 = \hat y_2' \hat y_3,\\\notag
X_1'=X_1 \hat y_2'{}^{k+1} \hat y_3'{}^{k} \,,\qquad
X_3'&= q^{2k+1} X_1 X_2 \,,\qquad
X_2' = X_2\,.
\end{align}
We give a detailed derivation of these formulas in Appendix \ref{app:skein-formulae}. 

Consider next the case of redundant nodes. Here we add two new nodes to a quiver $Q$ with $m-2$ nodes and produce a new quiver $Q'$ without changing the partition function. Then
\begin{align}\label{eq:multi-cover-skein-identity2}
\IP^{Q} &= \Psi_{q}(X_{m})\cdot\ldots\cdot\Psi_{q}(X_{3})\\\notag
&=\Psi_{q}(X_{m})\cdot\ldots\cdot\Psi_{q}(X_{3})\Psi_{q}(X_{2})\Psi_{q}(X_{1})=\IP^{Q'},
\end{align}
where 
\begin{equation}\label{eq:chvarredundant}
\hat x_{2}=q\hat x_{1},\qquad \hat y_{2}=\hat y_{1}^{-1},\qquad X_{2}=\hat x_{2}\prod_{j=3}^{m}\hat y_{j}^{l_{j}},\qquad X_{1}=q^{-1}X_{2}\hat y_{1}.
\end{equation}
This is a straightforward consequence of the discussion in Section \ref{sec:redundant}.

From the viewpoint of disks, \eqref{eq:multi-cover-skein-identity} expresses how multi-coverings of basic disks get reorganized when basic disks undergo boundary crossings and \eqref{eq:multi-cover-skein-identity2} when they undergo birth/death (pair production/annihilation).  
Each quantum dilogarithm, taken alone, counts multi-covers of a single disk without taking into account linking and the quantum torus algebra encodes the generalized curves produced from these multi-covers.

\subsection{Quantum torus algebra and holomorphic curve counting}\label{sec:QT-S}

The relation between wall-crossing identities and skein relations arises naturally once we write the quiver partition function as an ordered product of quantum dilogarithms valued in the quantum torus algebra.
In this section we show that this way of expressing the~quiver partition function contains more information. In particular, it encodes a consistent description of the spectrum of basic disks and of their boundstates for each of the two quivers appearing on either side of the pentagon relation \eqref{eq:quiver-pentagon}. By \emph{consistent} we mean that this description reflects precisely the occurrence of the unlinking by multi-cover skein relation from the left hand side to the right hand side.

To see how this works, we expand both sides of \eqref{eq:quiver-pentagon} 
\be\label{eq:2-node-QTA-expansion}
\begin{split}
	\IP^{\ \includegraphics[width=.01\textwidth]{figures/two-node-simplest.pdf}}
	& = 
	1 + X_1 \frac{q}{1-q^2}  + X_2 \frac{q}{1-q^2} \\
	&
	{\color{red} + X_2 X_1 \frac{q^2}{(1-q^2)^2}}
	+ X_1^2 \frac{q^2}{(1-q^2)(1-q^4)}
	+ X_2^2 \frac{q^2}{(1-q^2)(1-q^4)}
	\\
	&
	{\color{blue}
	+ X_2^2 X_1 \frac{q^3}{(1-q^2)^2(1-q^4)}
	}
	+X_2 X_1^2 \frac{q^3}{(1-q^2)^2(1-q^4)}
	 +\dots
\end{split}
\ee
\be\label{eq:3-node-QTA-expansion}
\begin{split}
	\IP^{\ \includegraphics[width=.03\textwidth]{figures/three-node-simplest.pdf}} 
	& = 
	1 + X_1 \frac{q}{1-q^2}  + X_2 \frac{q}{1-q^2} \\
	&
	{\color{red}
	\underbrace{- X_1 X_2 \frac{q^2}{(1-q^2)}}_{(0,0,1)}
	\underbrace{+ X_1 X_2 \frac{q^2}{(1-q^2)^2}}_{(1,1,0)}
	}
	+ X_1^2 \frac{q^2}{(1-q^2)(1-q^4)}
	+ X_2^2 \frac{q^2}{(1-q^2)(1-q^4)}
	\\
	&
	\underbrace{-X_1^2 X_2 \frac{q^3}{(1-q^2)^2}}_{(1,0,1)}
	\underbrace{+X_1^2 X_2 \frac{q^3}{(1-q^2)^2(1-q^4)}}_{(2,1,0)}
	\\
	&
	{\color{blue}
	\underbrace{-X_1 X_2^2 \frac{q^3}{(1-q^2)^2}}_{(0,1,1)}
	\underbrace{+X_1 X_2^2 \frac{q^3}{(1-q^2)^2(1-q^4)}}_{(1,2,0)}
	}
	+\dots
\end{split}
\ee
where we included labels $(n_1,n_2,n_{3})$ to keep track of the origin of each term in the product of expansions of the three quantum dilogarithms. 
This is important since each dilogarithm corresponds to a node and therefore to a basic holomorphic disk.
It is easy to check that the two sides match using the non-commutative product rule \eqref{eq:2-node-QTA}.

Now recall that $q^2 = e^{g_s}$, and that powers of $g_s$ correspond to Euler characteristics of the generalized holomorphic curves counted by the partition function. Compare terms in red:
\be\label{eq:skein-basic-disks-QTA}
	\underbrace{
	 { X_2 X_1 \frac{q^2}{(1-q^2)^2}}
	 }_{
	 \text{two linked disks}
	 }
	  = 
	 {
	\underbrace{- X_1 X_2 \frac{q^2}{(1-q^2)}}_{(0,0,1) = \text{fused disks}}
	\underbrace{+ X_1 X_2 \frac{q^2}{(1-q^2)^2}}_{(1,1,0) = \text{disjoint disks}}\,.
	}
\ee
On the left we have two linked disks since the coefficient diverges like $(g_s)^{-2}$ and since this term comes from the quiver where each node is a basic disk linked to the other one.
On the right we also have a term (the second one) which coefficient diverges like $(g_s)^{-2}$ but this one comes from the product of monomials in  $\Psi_1(X_1)$ and $\Psi_1(X_2)$ in different order. We interpret it as corresponding to two disjoint disks coming from unlinking.
The first term on the right instead comes from the expansion of $\Psi_1(-q X_1 X_2)$: this node arises by fusing the~two disks according to the multi-cover skein relations, as also evident from \eqref{eq:unlinking-variables}. Another confirmation that this term corresponds to a single fused disk comes from the fact that its coefficient diverges like $(g_s)^{-1}$.
Therefore identity (\ref{eq:skein-basic-disks-QTA}), valid in the quantum torus algebra, matches the multi-cover skein relation \eqref{eq:skein-relations-brane} on the basic disks shown in Figure \ref{fig:skein-relation-2-3}. (The apparent $q$-power mismatch will be clarified in Section \ref{sec:skein-higher}.)

\begin{figure}[h!]
\begin{center}
\includegraphics[width=0.75\textwidth]{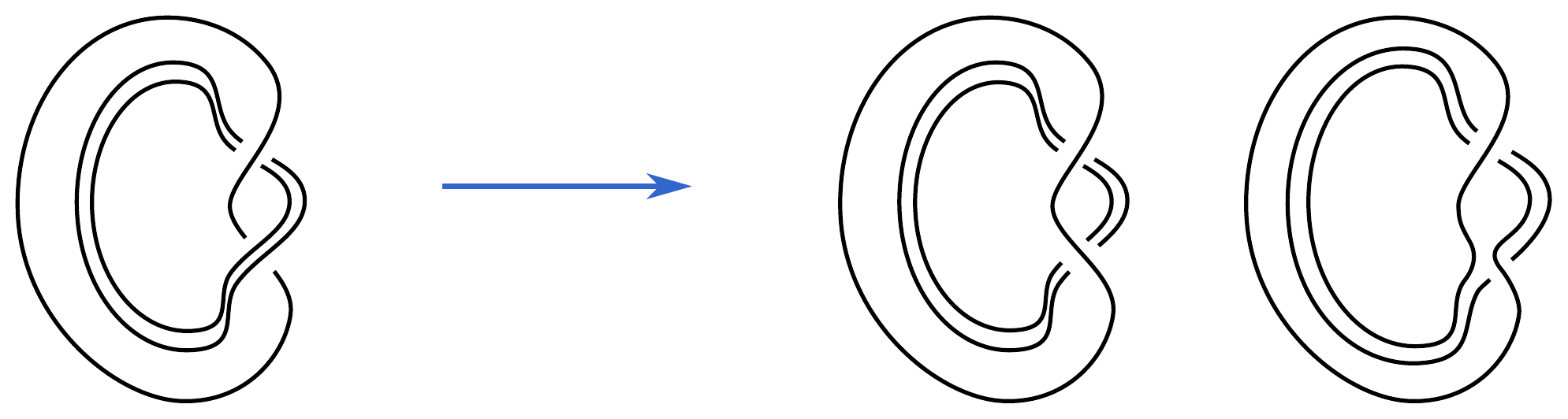}
\caption{Skein relation for multi-covers.}
\label{fig:skein-lvl3}
\end{center}
\end{figure}

The generating function of quiver representations contains much more information than this. We illustrate it here by looking at the simplest multi-covers. Consider the terms in blue:
\be\label{eq:skein-lvl3}
\begin{split}
	X_2^2 X_1 \frac{q^3}{(1-q^2)^2(1-q^4)} &
	\qquad 
	\text{disk 1 linked to two copies of disk 2}
	\\
	= & \\
	\underbrace{-X_1 X_2^2 \frac{q^3}{(1-q^2)^2}}_{(0,1,1)}
	& 
	\qquad \text{disk 3=`1+2' and one copy of disk 2}
	\\
	\underbrace{+X_1 X_2^2 \frac{q^3}{(1-q^2)^2(1-q^4)}}_{(1,2,0)}
	&
	\qquad \text{disk 1 and two copies of disk 2, all unlinked}
\end{split}
\ee
The interpretation of each of these terms is evident again by keeping track of their labeling $(n_1,n_2,n_3)$ and the power of $g_s^{-\chi}$.
Once again, we observe that the identity (\ref{eq:skein-lvl3}) is nothing but a way to write down the multi-cover skein relation depicted in Figure \ref{fig:skein-lvl3}.

The close parallel between quantum torus algebra and skein relations goes on to all orders in the identity \eqref{eq:quiver-pentagon}. Reformulating the quiver partition function in non-commutative form leads to extra information comparing to the usual form of the partition function~\eqref{eq:Efimov}.
Expanding the non-commutative version of the partition function order by order encodes exactly how holomorphic curves are obtained as boundstates of basic disks and their covers, and the rules of quantum torus algebra reproduce precisely the identities predicted by multi-cover skein relations.
The key ingredient is equation \eqref{eq:quiver-pentagon} -- an identity of wall-crossing type which relates different presentations of the form \eqref{eq:P-finite-product} corresponding to distinct `phases' for the~ensemble of holomorphic curves.

\subsection{Beyond the pentagon}\label{sec:beyond}

The pentagon relation (\ref{eq:pentagon}) is only the most basic example of a wall-crossing formula.
For instance, if we change the pairing $\langle\gamma_2,\gamma_1\rangle = 2$, then the formula reads 
\be\label{eq:WCI2}
\begin{split}
	\Psi_{q}(-X_{\gamma_2}) \Psi_{q}(-X_{\gamma_1}) 
	 = &
	\Psi_{q}(-X_{\gamma_1})  \Psi_{q}(-X_{\gamma_1+(\gamma_1+\gamma_2)}) 
	\cdot\ldots\cdot \Psi_{q}(-X_{\gamma_1+n(\gamma_1+\gamma_2)})\cdot\ldots  \\
	& \ldots\cdot
	\Psi_{q}(q^{-1} X_{\gamma_1+\gamma_2})^{-1} \Psi_{q}(q X_{\gamma_1+\gamma_2})^{-1}\cdot\ldots
	\\
	&\ldots\cdot  \Psi_{q}(-X_{\gamma_2+n(\gamma_1+\gamma_2)}) \cdot\ldots\cdot 
	\Psi_{q}(-X_{\gamma_2+(\gamma_1+\gamma_2)}) 
	\Psi_{q}(-X_{\gamma_2}) \,.
\end{split}
\ee
For $\langle\gamma_2,\gamma_1\rangle > 2$ the formula becomes much more complicated and there is an~interesting structure in the motivic DT invariants appearing on the right hand side \cite{2003math......4193R, 2009arXiv0909.5153G, Galakhov:2013oja}.
Howerver, the universal feature of wall-crossing formulas is that they always take the form of products of quantum dilogarithms with integer powers and with arguments valued in a suitable quantum torus algebra.

It follows that any identity of this type can be interpreted, through \eqref{eq:P-finite-product}, as a relation between two quiver partition functions (which are recovered by applying normal ordering).
These will in general feature a different number of nodes and therefore should be related by appropriate multi-cover skein relations. 

We note that multi-cover skein relations and wall crossing formulas encode information in different ways. 
For example, consider the quiver \includegraphics[width=0.05\textwidth]{figures/two-node-two-links.pdf}.
This corresponds to the~wall-crossing identity \eqref{eq:WCI2} in the sense outlined above.
Here, the~wall-crossing identity immediately leads to infinitely many nodes, whereas the~quiver multi-cover skein relation of link removal increases the number of nodes by one at a time, as illustrated in Figure \ref{fig:2-node-2-link-dualities}.
Thus the multi-cover skein relation follows the different phases of the ensemble of holomorphic curves  more closely than standard wall-crossing identities.

The relation between wall-crossing and multi-cover skein is very interesting and should be systematically studied, we leave this to future work.

\subsection{Quantum gluing of 3-manifolds along tori}\label{sec:quantum-gluing}

The quiver-assembling construction of the partition function can be given an interpretation  in terms of gluing together 3-manifolds along tori.
In the case under consideration, the~basic building block is a solid torus $S^{1}\times D^{2}$ with a Wilson line inserted on its central circle $S^{1}\times\{0\}$. Such solid tori can be glued together into a system of linked Wilson lines through the formalism leading to formula \eqref{eq:P-finite-product} which expresses the quantum partition function associated to the resulting 3-manifold. This is similar to well-known constructions in Floer theory related to the Atiyah-Floer conjecture, see e.g. \cite{woodward,FukayaAF}.

To understand the geometric interpretation of \eqref{eq:P-finite-product}, we start by recalling the general geometric setup in Section \ref{sec:general-oGW-Q}.
Each quiver node is a basic holomorphic disk with boundary on a Lagrangian $L$. 
Each factor  $\Psi_q$ in the formula for $\IP^Q$ is associated to such a disk and specifically  accounts for all higher-genus multi-covers.
We can think of building a~3-manifold $L$ as follows. We start with $L=\IR^2\times S^1$, without any Wilson lines (no disks). Its partition function is just $\IP^{\varnothing}=1$ and operators $\hat x$ and $\hat y$, corresponding to the longitude and the meridian at infinity, arise from the quantization of $U(1)$ Chern-Simons theory on the solid torus.

Each factor $\Psi_q$ modifies the geometry by cutting out a small solid torus around the~central curve of $L$ and gluing in a new solid torus with a Wilson line along its central circle.
Such operations change the partition function. Furthermore, if we glue in several unlinked parallel Wilson lines, the partition function changes in the obvious way, the $\Psi_q$-factors commute and the whole spectrum of generalized holomorphc curves consists of multi-covers of the basic disks.

This becomes more involved when the disk boundaries are linked, as there are non-trivial generalized holomorphic curves (bound states).
The quantum torus algebra introduced by the variables \eqref{eq:X-vars} keeps track of multi-cover linking and successive multiplication of $\Psi_q$-factors with non-commutative arguments 
correctly produce the partition function of all bound states.

\subsubsection{Semi-classical limit, disk potentials, and Lagrangian correspondences}\label{sec:lagcorr}
We give a geometric interpretation of the discussion about $A$-polynomials and associated disk potentials in \cite[Section 3]{Ekholm:2018eee} in our current setup. In the next section we discuss how this generalizes to the full partition function.

Let $L\approx S^{1}\times \R^{2}$ and write $T_{\infty}$ for the ideal torus boundary of $L$. 
Cut out a tubular neighborhoods of two disk boundaries in $L$. Write $N_{j}$, $j=1,\dots,m$, for the neighborhoods of the disk boundaries and $T_j$ for their boundary tori. Consider $L^{\ast}=L\setminus (\bigcup_{j}N_{j})$. Flat connections on $L^{\ast}$ have a (complexified) phase space determined by the boundary $\partial L^* = \bigcup_{j}T_j\cup T_{\infty}$:
\be
	\CP_\tot = \prod_{j}\CP_{j} \times \CP_\infty = \(\IC^* \times \IC^*\)^{m+1}\,.
\ee
This comes with coordinates $(x,y)=(e^{\xi},e^{\eta})$ on~$\CP_{\infty}$ and coordinates $(x_{j},y_{j})=(e^{\xi_{j}},e^{\eta_{j}})$ on~$\CP_{j}$. We also have symplectic forms $d\xi\wedge d\eta$ and $d\xi_{j}\wedge d\eta_{j}$ corresponding to intersections of longitudes and meridians thought of as ideal boundaries of bounding chains and projections $\pi_{j}$ and $\pi_{\infty}$ to factors.

Homology relations between longitudes and meridians give a Lagrangian subvariety $\CL_{\tot}\subset \CP_{\tot}$ defined by the $m+1$ equations
\[ 
\xi=\xi_{1}+\sum_{k}C_{1k}\eta_{k}=\dots=\xi_{m}+\sum_{k}C_{mk}\eta_{m},
\qquad\eta=-\sum_{j}\eta_{j},
\]  
where linking of disks boundaries is measured by $C_{ij}=C_{ji}$ and self-linking by $C_{jj}$. Here the negative signs on $\sum_{j}\eta_{j}$ come from viewing $T_{j}$ as the boundary of $L^{\ast}$ rather than $N_{j}$. Note that acting by the exponential on these relations gives monodromy relations for flat $U(1)$-connections on $L^{\ast}$. 

We now consider the disk potential counting generalized holomorphic disks that are combinations of multiple covers of the basic disks. In $N_{j}$ we have 
\be\label{eq:building-block}
	\Psi_q(q^{-1} x_{j}) = \sum_{n\geq 0} \frac{x_{j}^n}{(q^2;q^2)_n} \sim \exp\(-\frac{1}{g_{s}} \Li_2(x) + \dots\)\,,
\ee
therefore the disk potential is $W = -\Li_2(x_{j})$ and the semi-classical moduli space $\CL_{j}\subset\CP_{j}$ is given by
\be
	y_{j} = \exp\(\frac{\partial W}{\partial \xi_{j}}\) = 1-x_{j} \,.
\ee
For a geometric model, think of the toric Lagrangian brane of  $\IC^3$ \cite{Aganagic:2001nx}. 

To compute the disk potential of $L$ we reinterpret the reasoning in \cite{Ekholm:2018eee}: The disk potential of $L$ is obtained by transporting the product Lagrangian $\prod_{j}\CL_{j}\subset \prod_{j}\CP_{j}$, given by the individual disk potentials in $N_{j}$, through the Lagrangian correspondence $\CL_{\tot}$. In other words, we define the Lagrangian $\CL_{\infty}\subset\CP_{\infty}$ as
\begin{equation}\label{eq:diskpotential}
\CL_{\infty}=\pi_{\infty}(((\CL_{1}\times\CL_{2})\times\CP_{\infty})\cap\CL_{\tot})
\end{equation}  
and then the disk potential $W$ of $L$ is the local defining function $y=\frac{\partial W}{\partial x}$ of $\CL_{\infty}$. 

\subsubsection{The quantized Lagrangian correspondence}
In this section we give a conjectural interpretation of the operator formula \eqref{eq:P-finite-product} for the~quiver partition function in the spirit of Section \ref{sec:lagcorr}. We use notation as there and give an interpretation in terms of the D-model \cite{Aganagic:2013jpa}. At the full quantum level we first consider the ambient space complex symplectic space $\CP=\prod_{j}\CP_{j}$ with the Lagrangian~$\CL=\prod_{j}\CL_{j}$ in it. The D-model is the A-model topological string in $\CP$ with a Lagrangian brane on $\CL$ and a coisotropic space filling brane. The wave function of this D-model is simply the product
\[ 
\Psi=\prod_{j}\Psi_{q}(x_{j}).
\] 
The above discussion about Lagrangian correspondences suggests that one should view   \eqref{eq:P-finite-product} as the result of carrying the Lagrangian $\CL$ and the space filling brane along $\CP$ via~$\CL_{\tot}$ at the quantum level to get a D-model in $\CP_{\infty}$, which is then the usual B-model with wave function given by the operator form of the quiver partition function.

\appendix

\section{Multi-cover skein relations beyond disks}\label{app:disk-annulus}

In Section \ref{sec:skein-rel} we showed how the usual skein relation on basic disks extend to the multi-cover skein relation, which relates two disks to three, and which counts all generalized holomorphic curves coming from multi-covers of the disks before and after gluing/crossing. Here the orientation of the moduli space of the glued disk played a role and gave rise to different quiver relations for linking and unlinking. To revert the unlinking we needed to introduce a disk/anti-disk pair and use the linking skein relation. 

In this section we show on the example of the annulus that there will not be a simple two-to-three curve multi-cover skein for higher genus curves. Our approach to the annulus is to write it as a combination of disks and then use the multi-cover skein that we already know. One could approach curves of all genera in this way and obtain (finite) wall-crossing formulas. It would be very interesting to understand these formulas from a mathematical perspective using obstruction bundles near embedded nodal curves. 

\begin{figure}[h!]
	\begin{center}
		\includegraphics[width=0.4\textwidth]{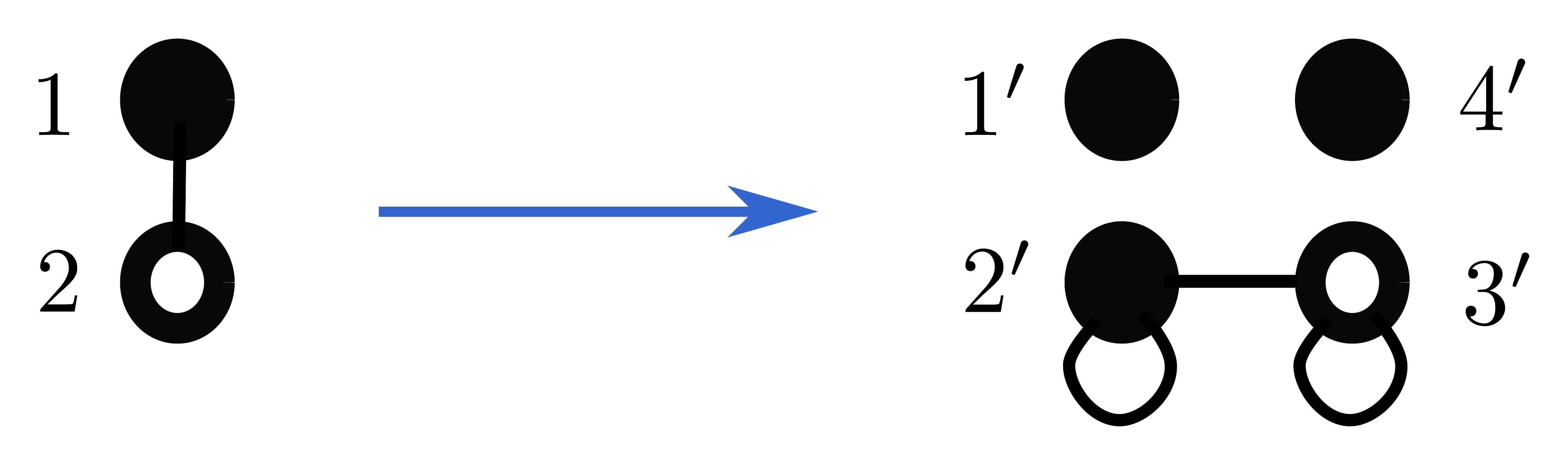}
		\caption{Disk-annulus multi-covering skein relation.}
		\label{fig:DA}
	\end{center}
\end{figure}

To derive the formula we first observe that a single annulus can be expressed in terms of two disks with opposite 4-chain intersection. Geometrically, the annulus appear when we glue a constant disk in the Lagrangian to the 1-parameter family of holomorphic curves intersecting it generically. Counting multicovers we have:
\[ 
\exp\left(\sum_{d}\frac{1}{d}\frac{(qx)^{d}}{q^{d}-q^{-d}}\right)
=\exp\left(\sum_{d}\frac{1}{d}\frac{(q^{-1}x)^{d}}{q^{d}-q^{-d}}\right)\cdot\exp\left(\sum_{d}\frac{1}{d}x^{d}\right)
\] 
In our treatment below we will rewrite this as
\[ 
\exp\left(\sum_{d}\frac{1}{d}\frac{(qx)^{d}}{q^{d}-q^{-d}}\right)\cdot\exp\left(-\sum_{d}\frac{1}{d}\frac{(q^{-1}x)^{d}}{q^{d}-q^{-d}}\right)
=\exp\left(\sum_{d}\frac{1}{d}x^{d}\right),
\] 
and then replace the anti-disk factor using redundant pairs. More precisely, we compute as follows.
Trading the annulus for a pair of disks with shifted 4-chain intersections, as described above, we get the partition function of a disk-annulus (d.a.) pair that links once:
\be
\begin{split}
	\IP^{\text{d.a.}}
	& \equiv \frac{1}{1-\hat x_2 \hat y_1}  (\hat x_1;q^2)^{-1}_\infty 
	= \Psi_q(q^2 X_2)^{-1} \Psi_q(X_2)  \Psi_q(X_1), 
\end{split}
\ee
where the variables are as in \eqref{eq:X-vars}, $X_2 X_1 = q^2 X_1 X_2$.
We then use the pentagon identity between \eqref{eq:2-node-quantum} and \eqref{eq:3-node-quantum}, which we write in two ways:
\be
\begin{split}
\Psi_q(X_2)\Psi_q(X_1) &= \Psi_q(X_1)\Psi_q(-qX_1X_2)\Psi_q(X_2)\\
\Psi_q(X_2)^{-1} \Psi_q(X_1) &= \Psi_q(X_1)\Psi_q(X_2)^{-1} \Psi_q(-qX_1X_2)^{-1}\,.
\end{split}
\ee
Using these we rewrite the disk-annulus partition function as follows
\be
\begin{split}
	\IP^{\text{d.a.}}
	& = {\Psi_q(q^2 X_2)}^{-1}  \Psi_q(X_1)\Psi_q(-qX_1X_2)\Psi_q(X_2) \\
	& =  \Psi_q(X_1)\Psi_q(q^2 X_2)^{-1} \Psi_q(-q^3X_1X_2)^{-1}     \Psi_q(-qX_1X_2)\Psi_q(X_2) 
\end{split}
\ee
Next we trade the multi-covers of disks encoded by the third and fourth factor for an~annulus (the reverse of what was done for the original annulus):
\be
\begin{split}
	\IP^{\text{d.a.}}
	& = \Psi_q(X_1)\Psi_q(q^2 X_2)^{-1}\frac{1}{1 + q^2 X_1X_2} \Psi_q(X_2)\,. 
\end{split}
\ee
After normal ordering this becomes
\be
\begin{split}
	\IP^{\text{d.a.}}
	& \rightsquigarrow 
	\sum_{d_1\dots d_4} 
	(-q)^{d_2^2 + d_3^2 + 2 d_2 d_3}
	\frac{(x_1)^{d_1}}{(q^2;q^2)_{d_1}} 
	\frac{(q x_2)^{d_2}}{(q^2;q^2)_{d_2}} 
	(q^{-1} x_1 x_2)^{d_3}
	\frac{x_2^{d_4}}{(q^2;q^2)_{d_4}}\,.
	\\
\end{split}
\ee
This has the form of a (generalized) quiver partition function, involving three disks and one annulus.
Variables of the new quiver are related to the old ones by
\be
x_1' = x_1\,,
\qquad
x_2' = q x_2\,,
\qquad
x_3'=  q^{-1} x_1 x_2\,,
\qquad
x_4' = x_2\,,
\ee
see Figure \ref{fig:DA}.

Geometrically, the process of unlinking the disk and annulus can be described as follows. First the annulus was replaced by two disks (nodes $2'$ and $4'$). 
One of them (node~$4'$) has no intersection with the 4-chain, the other one (node $2'$) has a positive 4-chain intersection.
We then create an annulus (node $3'$) by combining two disks, this is clearly a~boundstate of the original disk and annulus. 
Both the new annulus and the disk corresponding to node $2'$ have a unit of self-linking. 
In addition, the new annulus has a negative unit of 4-chain intersection and links with the disk encoded by node $2'$.

For comparison, we consider what a two-to-three term ansatz to wall-crossing would give in this case. 
We have
\be\label{eq:PDA}
\begin{split}
	\IP^{\text{d.a.}} & \equiv \frac{1}{1-\hat x_2 \hat y_1}  (\hat x_1;q^2)^{-1} 
	\rightsquigarrow \sum_{d_1 d_2} q^{2 d_1 d_2} x_2^{d_2}   \frac{x_1^{d_1}}{(q^2;q^2)_{d_1}}\,.
\end{split}
\ee

Applying the naive skein relation to the basic objects and then taking their multi-covering partition functions, one gets a new copy of the disk and the annulus (now mutually unlinked), as well as a new annulus arising from their boundstate and carrying a self-intersection. Denoting the holonomy of the boundstate by $x_3=\sigma q^\alpha x_1 x_2$ ($\sigma=\pm1$ and $\alpha\in \IZ$ are to be determined) we can write the partition function as
\be\label{eq:daa-naive}
\begin{split}
	\IP^{\text{d.a.a.}} & \equiv
	\frac{1}{1-\hat x_2}(\hat x_1;q^2)^{-1} \sum_{d_3} (\hat  x_3 \hat y_3)^{d_3} \\
	& \rightsquigarrow
	\sum_{d_1, d_2} x_2^{d_2} \frac{x_1^{d_1}}{(q^2;q^2)_{d_1}} \sum_{d_3} q^{d_3(d_3-1)} (\sigma q^\alpha x_1 x_2)^{d_3} \,.
\end{split}
\ee
Matching with quadratic terms in (\ref{eq:PDA}) fixes $\sigma=-1$, $\alpha=0$.
Nevertheless, higher terms will not match. For example the terms of the order $x_1 x_2^2$ are
\be\label{eq:daa-missing}
\IP^{\text{d.a.}} \supset \frac{x_1 x_2^2 \ q^4}{1-q^2}\, \neq \,
 \frac{x_1 x_2^2}{1-q^2} - x_1 x_2^2 \subset \IP^{\text{d.a.a.}} \,.
\ee
We learn from this that the basic skein relation does not carry over to a multi-covering formula for annuli. The correct formula is written in quiver language in Figure \ref{fig:DA}.

\section{Nonuniqueness of the quiver for a given knot -- $4_1$ example}\label{sec:KRSS-equivalences}

The invariance properties of the quiver partition function under linking and unlinking, shown in Section \ref{sec:skein-rel}, turn out to explain neatly some puzzling observations. For example for the figure-eight knot one can find two quivers of
the same size which have the same motivic generating series.\footnote{Note that quivers in this appendix correspond to the \emph{reduced}
normalization. This property translates automatically to the unreduced normalization as well, but in that case quivers would be very big.}  One is
given by
\cite{Kucharski:2017ogk}
\begin{equation}
C_{4_{1}}=\left[\begin{array}{ccccc}
0 & 0 & -1 & 0 & -1\\
0 & 2 & 0 & 1 & {\color{red}-1}\\
-1 & 0 & -1 & {\color{red}0} & -2\\
0 & 1 & {\color{red}0} & 1 & -1\\
-1 & {\color{red}-1} & -2 & -1 & -2
\end{array}\right],\label{eq:41quiver}
\end{equation}
the second differs only by a permutation of 4 entries (which cannot
be obtained by vertices' relabelling)
\begin{equation}
\tilde{C}_{4_{1}}=\left[\begin{array}{ccccc}
0 & 0 & -1 & 0 & -1\\
0 & 2 & 0 & 1 & {\color{red}0}\\
-1 & 0 & -1 & {\color{red}-1} & -2\\
0 & 1 & {\color{red}-1} & 1 & -1\\
-1 & {\color{red}0} & -2 & -1 & -2
\end{array}\right].\label{eq:42quiver'}
\end{equation}
We can obtain $\tilde{C}_{4_{1}}$ from $C_{4_{1}}$ by unlinking and the
\emph{inverse }of unlinking. In order to see it, let us relabel vertices
of $Q_{4_{1}}$such that 
\begin{equation}
C_{4_{1}}\sim\left[\begin{array}{ccccc}
-1 & {\color{red}0} & -2 & 0 & -1\\
{\color{red}0} & 1 & -1 & 1 & 0\\
-2 & -1 & -2 & {\color{red}-1} & -1\\
0 & 1 & {\color{red}-1} & 2 & 0\\
-1 & 0 & -1 & 0 & 0
\end{array}\right].
\end{equation}
Now we apply the unlinking for the first two nodes (top left corner
of the matrix) with the~remaining three being spectators. In the notation
from the Section \ref{sec:Proof of invariance-unlink} we have
\[
r=-1,\; s=1,\; k=0\,,
\]
so the unlinking $k\rightarrow k-1$ leads to
\begin{equation}
C_{4_{1}}\sim\left[\begin{array}{cccccc}
-1 & {\color{red}-1} & -2 & 0 & -1 & -2\\
{\color{red}-1} & 1 & -1 & 1 & 0 & 0\\
-2 & -1 & -2 & {\color{red}-1} & -1 & -3\\
0 & 1 & {\color{red}-1} & 2 & 0 & 1\\
-1 & 0 & -1 & 0 & 0 & -1\\
-2 & 0 & -3 & 1 & -1 & -1
\end{array}\right].
\end{equation}
Now we can relabel vertices again to have
\begin{equation}
C_{4_{1}}\sim\left[\begin{array}{cccccc}
-2 & {\color{red}-1} & -2 & -1 & -1 & -3\\
{\color{red}-1} & 2 & 0 & 1 & 0 & 1\\
-2 & 0 & -1 & {\color{red}-1} & -1 & -2\\
-1 & 1 & {\color{red}-1} & 1 & 0 & 0\\
-1 & 0 & -1 & 0 & 0 & -1\\
-3 & 1 & -2 & 0 & -1 & -1
\end{array}\right].
\end{equation}
This matrix matches the structure of (\ref{eq:Q-after-unlinking})
for
\[
r=-2,\; s=2,\; k=0\,,
\]
so it can be simplified (by the inverse of unlinking) to
\begin{equation}
C_{4_{1}}\sim\left[\begin{array}{ccccc}
-2 & {\color{red}0} & -2 & -1 & -1\\
{\color{red}0} & 2 & 0 & 1 & 0\\
-2 & 0 & -1 & {\color{red}-1} & -1\\
-1 & 1 & {\color{red}-1} & 1 & 0\\
-1 & 0 & -1 & 0 & 0
\end{array}\right].
\end{equation}
Relabelling again we obtain
\begin{equation}
C_{4_{1}}\sim\left[\begin{array}{ccccc}
0 & 0 & -1 & 0 & -1\\
0 & 2 & 0 & 1 & {\color{red}0}\\
-1 & 0 & -1 & {\color{red}-1} & -2\\
0 & 1 & {\color{red}-1} & 1 & -1\\
-1 & {\color{red}0} & -2 & -1 & -2
\end{array}\right]=\tilde{C}_{4_{1}}\,,
\end{equation}
therefore quivers given by (\ref{eq:41quiver}) and (\ref{eq:42quiver'})
are in the same equivalence class, as expected.

\section{Details of generalized multi-cover skein identities}\label{app:skein-formulae}

Here we fill in the details on the variables appearing in the general multi-cover skein identity~(\ref{eq:multi-cover-skein-identity}). Let us start with the case in which $Q'$ is obtained from $Q$ by unlinking.
We assume that disk $1$ has $s$ units of self-linking, disk 2 has $r$ units, and both have arbitrary amounts of linking with other basic disks. We suppress factors of $\hat y_j$ for $j\neq 1,2,3$ that would arise from linking to other nodes of the quiver, these can be simply inserted into our formulas as necessary.
Then, according to conventions set out in (\ref{eq:X-vars}), for $Q$ we have:
\be\label{eq:Q-vars-before}
	X_1 = (-1)^s q^{s-1} \hat x_1 \hat y_1^s\,,
	\qquad
	X_2 = (-1)^r q^{r-1} \hat x_2 \hat y_2^r \hat y_1^k\,,
\ee
and for $Q'$:
\be
\begin{split}
	X_1' & = (-1)^s q^{s-1} \hat x_1' \hat y_1'{}^s \hat y_2'{}^{k-1} \hat y_3'{}^{s+k-1}\,, \\
	X_2' & = (-1)^r q^{r-1} \hat x_2' \hat y_2'{}^r \,,\\
	X_3' & = (-1)^{r+s+2k-1} q^{r+s+2k-2} \hat x_3' \hat y_3'{}^{r+s+2k-1} \hat y_2'{}^{r+k-1}\,.
\end{split}
\ee
Recall from (\ref{eq:unlinking-invariance}) and (\ref{eq:gauge-fugacities})  that 
\be
\begin{split}
x_1'=x_1\,, \qquad &x_2'=x_2\,, \qquad x_3' = q^{-1} x_1 x_2\,,\\ 
y_1 = y_1'y_3'&\,,\qquad y_2 = y_2'y_3'\,.
\end{split}
\ee
This implies that
\be
	X_1'=X_1 \hat y_2^{k-1}\,,\qquad
	X_3'=-q^{2k-1} X_1 X_2  \hat y_2^{k-1}\,,\qquad
	X_2' = X_2\,,
\ee
where we inserted `by hand' a factor of $\hat y_1'{}^{s+k}$ into $X_3$ and a factor $\hat y_3'{}^{r}$ into $X_2$ since they are innocuous in (\ref{eq:multi-cover-skein-identity}) due to ordering (recall a similar trick in (\ref{eq:3-node-quantum})).
As claimed, this reduces the multi-cover skein identity (\ref{eq:multi-cover-skein-identity}) to the pentagon identity (\ref{eq:quiver-pentagon}) for $k=1$.

If there are additional `spectator' nodes, their $X_j$ variables remain unchanged. In notation from (\ref{eq:Q-after-unlinking}), this can be understood as follows. After unlinking one would need to modify $X_j$ by removing factors of $\hat y_1^a, \hat y_2^b$ and replacing them with $\hat y_1'{}^{a}\hat y_2'{}^{b}\hat y_3'{}^{a+b}$.
But due to (\ref{eq:gauge-fugacities}) this operation is trivial.

Next we consider linking. 
Similarly to the previous case, we assume that disks $1$ and~$2$ have~$s$ and~$r$ units of self-linking respectively, and arbitrary amounts of linking with other basic disks. We still suppress factors of $\hat y_j$ for $j\neq 1,2,3$ which can be inserted into our formulas if necessary.
Then, according to conventions set out in (\ref{eq:X-vars}), for $Q$ we have the~same variables as in (\ref{eq:Q-vars-before}), while for $Q'$ we now have
\be
\begin{split}
	X_1' & = (-1)^s q^{s-1} \hat x_1' \hat y_1'{}^s \hat y_2'{}^{k+1} \hat y_3'{}^{s+k}\,, \\
	X_2' & = (-1)^r q^{r-1} \hat x_2' \hat y_2'{}^r\,, \\
	X_3' & = (-1)^{r+s+2k} q^{r+s+2k-1} \hat x_3' \hat y_3'{}^{r+s+2k} \hat y_2'{}^{r+k}\,.
\end{split}
\ee
Recall from (\ref{eq:linking-invariance}) that 
\be
x_1'=x_1\,, \qquad x_2'=x_2\,, \qquad x_3' = x_1 x_2 \,. 
\ee
For $y_i$ variables we need a~bit more care.
Let us focus on (\ref{eq:linking-with-redundant-nodes}): here we have an equivalence between the quiver $Q'$ where two nodes have one additional units of linking, and a~quiver~$Q''$ which has yet an~extra node which is `dual' to the one created by linking (they form a~redundant pair of nodes). Since $Q'$ and $Q''$ are related by standard unlinking, we can immediately infer that 
\be
	y_1' = y_1'' y_5'' = y_1'' y_{4}''{}^{-1}\,, 
	\qquad 
	y_2' = y_2'' y_5'' = y_2'' y_{4}''{}^{-1}\,,
\ee
where we used (the semiclassical limit of) (\ref{eq:redundant-nodes-cancellation-identity}) to claim that $y_4'' y_5''=1$ for the redundant pair (as should be obvious from the definition of such a pair). 
Returning to the case considered here, we map $y_1'' \mapsto y_1$, $y_2'' \mapsto y_2$, $y_4'' \mapsto y_3$, while obviously $y_1', y_2'$ are already the~correct labels as considered here. This implies
\be
	y_1 = y_1' y_3 \,,
	\qquad
	y_2 = y_2' y_3 \,.
\ee
Therefore we can reexpress $X'_i$ variables in terms of $X_i$ as follows:
\be
	X_1'=X_1 \hat y_2'{}^{k+1} \hat y_3'{}^{k} \,,\qquad
	X_3'= q^{2k+1} X_1 X_2 \,,\qquad
	X_2' = X_2\,,
\ee
where we inserted `by hand' a factor of $\hat y_1'{}^{s+k}$ into $X_3$ and a factor $\hat y_3'{}^{r}$ into $X_2$ since they are innocuous in (\ref{eq:multi-cover-skein-identity}) due to ordering. 
As a check, for $k=0$ this reduces the multi-cover skein identity precisely to the expected formula for the case studied in Section \ref{sec:simplelink}.
As for the case of unlinking, the same argument shows that $X_j$ variables of spectator nodes do not change.

\newpage
\bibliography{biblio}{}
\bibliographystyle{JHEP}

\end{document}